\documentclass[11pt,a4paper]{article}
\pdfoutput=1

\usepackage{amsmath,amssymb,mathrsfs}
\usepackage{subfigure}
\usepackage{url}

\usepackage{graphicx}
\usepackage{hyperref}
\usepackage{bm}

\usepackage{cite}

\numberwithin{equation}{section}

\allowdisplaybreaks[4]

\DeclareMathOperator{\Tr}{Tr}

\begin{document}

\renewcommand{\subfigcapskip}{5pt}

\renewcommand{\theequation}{\thesection.\arabic{equation}}

\def\thefootnote{\fnsymbol{footnote}}

%%%%%%%%%%%%%%%%%%%%%%%%%%%%%%%%%%%%%%%%%%%%%%%%%%
%%%%%%%%%%%%%%%%%%%%%%%%%%%%%%%%%%%%%%%%%%%%%%%%%%

\hfill{} MPP-2019-130

\vspace{1.0 truecm}

\begin{center}

{\textbf{\LARGE
Stellar cooling anomalies 
and variant axion models
}}

\bigskip

\vspace{0.5 truecm}

{\bf
Ken'ichi Saikawa$^1$\footnote{saikawa@mpp.mpg.de, current affiliation: Institute for Theoretical Physics, Kanazawa University, Japan} and 
Tsutomu T. Yanagida$^{2,3}$\footnote{tsutomu.tyanagida@ipmu.jp}
} \\[5mm]

\begin{tabular}{lc}
&\!\! {$^1$ \em Max-Planck-Institut f\"ur Physik (Werner-Heisenberg-Institut),}\\
&{\em F\"ohringer Ring 6, D-80805 M\"unchen, Germany}\\[.4em]
&\!\! {$^2$ \em T. D. Lee Institute and School of Physics and Astronomy,}\\
&{\em Shanghai Jiao Tong University, Shanghai 200240, China}\\[.4em]
&\!\! {$^3$ \em Kavli Institute for the Physics and Mathematics of the Universe (Kavli IPMU),}\\
&{\em UTIAS, WPI, The University of Tokyo, Kashiwa, Chiba 277-8568, Japan}\\[.4em]
\end{tabular}

\vspace{1.0 truecm}

{\bf Abstract}
\end{center}

\begin{quote}
A number of observations of stellar systems show a mild preference for 
anomalously fast cooling compared with what predicted in the standard theory,
which leads to a speculation that there exists an additional energy loss mechanism
originated from the emission of axions in stars.
We revisit the global analysis of the stellar cooling anomalies by adopting conservative assessments on several systematic uncertainties
and find that the significance of the cooling hints becomes weaker
but still indicates a non-vanishing axion-electron coupling at around 2.4\,$\sigma$.
With the revised analysis results, we explore the possibility that such excessive energy losses are interpreted in the framework of
variant axion models, which require two Higgs doublets and flavor-dependent Peccei-Quinn charge assignments.
These models resolve two fundamental issues faced in the traditional KSVZ/DFSZ models 
by predicting a sizable axion coupling to electrons required to explain the cooling anomalies
and at the same time providing a solution
to the cosmological domain wall problem.
We also find that a specific structure of the axion couplings to electrons and nucleons
slightly relaxes the constraint from supernova 1987A
and enlarges viable parameter regions compared with the DFSZ models.
It is shown that good global fits to the observational data are obtained for axion mass ranges 
of $0.45\,\mathrm{meV} \lesssim m_a \lesssim 30\,\mathrm{meV}$, 
and that the predicted parameter regions can be probed in the forthcoming helioscope searches.
\end{quote}

\thispagestyle{empty}

\newpage

\tableofcontents

\renewcommand{\thepage}{\arabic{page}}
\renewcommand{\thefootnote}{\arabic{footnote}}
\setcounter{footnote}{0}

%%%%%%%%%%%%%%%%%%%%%%%%%%%%%%%%%%%%%%%%%%%%%%%%%%
\section{Introduction}
\label{sec:Introduction}
\setcounter{equation}{0}
%%%%%%%%%%%%%%%%%%%%%%%%%%%%%%%%%%%%%%%%%%%%%%%%%%

Extreme environment realized in stars can be regarded as a good laboratory to test fundamental physics~\cite{Raffelt:1996wa}.
The consideration of stellar evolution particularly leads to quite strong constraints on light weakly interacting particles.
A leading example of such a low mass particle is the axion~\cite{Weinberg:1977ma,Wilczek:1977pj}, which is a Nambu-Goldstone boson
emerging from the spontaneous breaking of hypothetical global U(1) Peccei-Quinn (PQ) symmetry
introduced to provide a solution to the strong CP problem of quantum chromodynamics (QCD)~\cite{Peccei:1977hh}.
If the scale of the PQ symmetry breaking is sufficiently high, interactions of axions with ordinary matter are quite weak.
This fact implies that axions produced in the core of a star can easily escape from the system, providing a new energy loss mechanism.
Such an exotic energy loss can affect the evolution of stars, which allows us to investigate the property of axions 
by using the results from astrophysical observations.

Recently, it has been pointed out that observations of stars in several different evolutionary stages
show systematic trends to prefer excessive energy losses over the standard cooling scenario,
and that such cooling anomalies can be explained just by 
adding new cooling channels due to axion emission~\cite{Raffelt:2011ft,Ringwald:2015lqa,Giannotti:2015dwa,Giannotti:2015kwo,Giannotti:2016hnk,Giannotti:2017hny}.
Understanding such astrophysical phenomena is not straightforward, of course, and the results should be taken carefully.
However, at this point we cannot discard the possibility of interpreting them as {\it hints} of the existence of axions.
According to the analysis in Ref.~\cite{Giannotti:2017hny}, there is a more than 3\,$\sigma$ preference for the axion interpretation.\footnote{The significance of the hints becomes weaker
if we adopt a more conservative approach to include systematic uncertainties (see Sec.~\ref{sec:implications_for_models}).}

The hints obtained from the stellar cooling anomalies can be used as a guide for
next generation axion search experiments [see e.g.~Ref.~\cite{Irastorza:2018dyq} for reviews].
In particular, the proposed helioscope, International Axion Observatory (IAXO)~\cite{IAXOloe,Armengaud:2014gea}, 
will probe a broad parameter space with improved sensitivities
and have a potential to cover a large fraction of the region motivated by the 
solution to the stellar cooling anomalies~\cite{Armengaud:2019uso}.

It should be noted that the preceding argument on the axion interpretation of the stellar cooling anomalies
might remain inadequate when we consider cosmological issues of the axion models.
It is known that the axion models suffer from a constraint from isocurvature fluctuations in the cosmic microwave background~\cite{Linde:1985yf,Seckel:1985tj},
which arises if the PQ symmetry is broken before or during inflation and never restored afterwards.
This constraint becomes severe unless the inflationary energy scale is sufficiently low.
For instance, if the scale of the PQ symmetry breaking is of order $\sim 10^9\,\mathrm{GeV}$, which is preferred by the stellar cooling hints~\cite{Giannotti:2017hny},
the Hubble scale during inflation must be lower than $H_{\rm inf}\lesssim 10^6\,\mathrm{GeV}$ in order to avoid large isocurvature fluctuations~\cite{Kobayashi:2013nva}.\footnote{The isocurvature constraint
also depends on the abundance of cold axions produced by the vacuum re-alignment mechanism~\cite{Preskill:1982cy,Abbott:1982af,Dine:1982ah},
and the limit $H_{\rm inf}\lesssim 10^6\,\mathrm{GeV}$ corresponds to the case where the fraction of the energy density of cold axions is about 1\,\% of the total dark matter abundance.
The constraint becomes tighter if the axion fraction gets larger.}
A simple way to avoid this isocurvature constraint is to assume that the PQ symmetry is broken after inflation.
However, in this case domain walls are created around the epoch of the QCD phase transition~\cite{Sikivie:1982qv},
leading to disastrous consequences if they are stable~\cite{Zeldovich:1974uw}.
This cosmological domain wall problem poses a serious concern for the popular Dine-Fischler-Srednicki-Zhitnitsky (DFSZ) models~\cite{Zhitnitsky:1980tq,Dine:1981rt},
which provide good fits to the stellar cooling hints~\cite{Giannotti:2017hny}.
On the other hand, in the Kim-Shifman-Vainshtein-Zakharov (KSVZ) models~\cite{Kim:1979if,Shifman:1979if},
which are regarded as another well-motivated class of axion models, 
the axion interaction with electrons is too small to explain the observed data~\cite{Giannotti:2017hny},
though they can avoid the domain wall problem straightforwardly.
This situation is alleviated in a KSVZ-like axion/majoron model~\cite{Shin:1987xc,Ballesteros:2016euj,Ballesteros:2016xej}, 
but in this case a sufficiently large radiative correction to the axion-electron coupling is required,
which is likely to violate the condition of perturbativity.

In this paper, we point out that both of the above astrophysical and cosmological issues can be naturally addressed in the framework
of variant axion models, which were proposed first by Peccei, Wu and Yanagida~\cite{Peccei:1986pn} and by Krauss and Wilczek~\cite{Krauss:1986wx}.
These models are constructed based on two Higgs doublets and flavor-dependent PQ charge assignments for the quark fields.
Although the first version of the variant axion models was excluded as 
it is ``visible" like the original Weinberg-Wilczek model~\cite{Weinberg:1977ma,Wilczek:1977pj}
(i.e.~the PQ scale is assumed to be as low as the electroweak scale),
it is possible to make it ``invisible"~\cite{Geng:1988nc,Hindmarsh:1997ac} by adding a singlet complex scalar field
such that the PQ scale becomes arbitrary high like all other viable axion models.\footnote{Extensions of the variant axion models were proposed recently 
in e.g.~Refs.~\cite{Ema:2016ops,Calibbi:2016hwq,Bjorkeroth:2018dzu,Bjorkeroth:2018ipq}, which are characterized by flavor-changing axion couplings.
A potential problem for such models is that they generically suffer from
a severe constraint coming from the non-observation of rare decays $K^+ \to \pi^+ + a$.
We point out that in the simplest framework of the variant axion models
this constraint can be straightforwardly avoided if only one of the up-type quarks has a nonzero PQ charge.}
The important feature of this framework is that we can arrange the model such that it leads to a unique vacuum 
in the low energy effective theory,
which evades the domain wall problem~\cite{Geng:1988nc}.
Furthermore, it can make the axion-electron coupling naturally large 
as it appears in the tree level, in a similar manner to the DFSZ framework.
This feature is contrasted to the axion/majoron model, 
where the electron coupling is induced only radiatively.

The remaining part of the paper is organized as follows.
In Sec.~\ref{sec:stellar_cooling}, we review the observational results of cooling anomalies
and previous attempts to interpret them in terms of additional cooling channels induced by axions.
The global fits to the astrophysical data are revisited with slightly different assessments on the systematic uncertainties.
In Sec.~\ref{sec:variant_axions}, we construct the variant axion models 
and derive axion couplings to the Standard Model (SM) particles.
In Sec.~\ref{sec:interpretation_of_anomalies}, we show the results of global fits to the data of stellar cooling anomalies,
specifying the preferred parameter regions. We also compare these predictions with the sensitivities of future helioscope searches such as IAXO.
Section~\ref{sec:conclusion} is devoted to discussion and conclusions.
In Appendix~\ref{app:analysis_method}, we summarize the method of the global analysis.
In Appendix~\ref{app:yukawa_interactions}, we describe the structure of the Higgs sector and Yukawa interactions in the variant axion models
in order to derive typical parameter ranges compatible with requirements of perturbativity of Yukawa interactions.

%%%%%%%%%%%%%%%%%%%%%%%%%%%%%%%%%%%%%%%%%%%%%%%%%%
\section{Axions and stellar energy losses}
\label{sec:stellar_cooling}
\setcounter{equation}{0}
%%%%%%%%%%%%%%%%%%%%%%%%%%%%%%%%%%%%%%%%%%%%%%%%%%

Low mass weakly interacting particles such as axions can affect the time scale of the evolution of stars, since
they can escape from the system and take away a considerable amount of energy.
Since the existence of such exotic cooling channels changes various astrophysical observables,
we can constrain the properties of hypothetical particles by comparing the observational results 
with the prediction of the standard cooling scenario~\cite{Raffelt:1996wa}.
Furthermore, results of some recent analyses even show a mild preference for the existence of excessive energy losses.
In this section, we revisit the hints of excessive energy losses reported in previous literatures
and their possible interpretations in terms of axions.
We take a different approach to deal with several systematic uncertainties,
shedding light on challenging aspects of the data analysis.

%%%%%%%%%%%%%%%%%%%%%%%%%%%%%%%%%%%%%%%%%%%%%%%%%%
\subsection{Horizontal and red giant branch stars in globular clusters}
\label{sec:HB/RGB}
%%%%%%%%%%%%%%%%%%%%%%%%%%%%%%%%%%%%%%%%%%%%%%%%%%

A strong indicator to measure the cooling rate of stars in different evolutionary stages is provided by the observation of globular clusters.
Globular clusters are gravitationally bound systems composed of about $10^6$ stars,
which show a characteristic distribution in the color-magnitude diagram according to a certain evolutionary phase.
In particular, the ratio of the number of stars in the horizontal branch (HB) to that of those in the red giant branch (RGB), called the $R$-parameter,
is used to estimate the relative time scale for different evolutionary stages associated with these branches.
The existence of axions can affect the value of the $R$-parameter, 
since they can be efficiently produced via the Primakoff process $\gamma + Ze \to Ze + a$ in the HB stars but not in the RGB stars,
changing the relative amount of time spent on two different branches.
Hence the additional cooling due to the emission of axions in the HB stars could be observed as a decrease in the value of $R$.

Based on the above argument, we can constrain the magnitude of the axion-photon coupling $g_{a\gamma}$, which is given by
\begin{align}
\mathcal{L}_{a\gamma} = -\frac{1}{4}g_{a\gamma}aF_{\mu\nu}\tilde{F}^{\mu\nu}, \qquad g_{a\gamma} \equiv \frac{\alpha}{2\pi f_a}C_{a\gamma},
\label{g_agamma_def}
\end{align}
where $F_{\mu\nu}$ is the electromagnetic (EM) field strength, $\tilde{F}^{\mu\nu}$ its dual,
$f_a$ the axion decay constant, $\alpha$ the fine structure constant, and $C_{a\gamma}$ a dimensionless coefficient.

Axions can also interact with electrons through the following coupling,
\begin{align}
\mathcal{L}_{ae} = -ig_{ae}a\overline{e}\gamma_5 e, \qquad g_{ae} \equiv \frac{C_{ae}m_e}{f_a},
\label{g_ae_def}
\end{align}
where $C_{ae}$ is a dimensionless coefficient and $m_e$ the electron mass.
In this case, axions can be produced in the core of red giants
through bremsstrahlung off electrons $e + Ze \to Ze + e + a$,
which cools the core and causes a delay of helium ignition.
The delay of helium ignition leads to a larger core mass, which affects the value of $R$-parameter.
Therefore, the information on the $R$ can be used to probe not only the axion-photon coupling
but also the axion-electron coupling~\cite{Giannotti:2015kwo}.

The prediction for the effect of the axion emission on the $R$-parameter is given by~\cite{Giannotti:2015kwo}
\begin{align}
R^{\rm th} = 7.33Y +0.02 -0.095\sqrt{21.86+21.08g_{10}} -1.61\delta\mathcal{M}_c - 0.067\alpha_{26},
\label{R_theory}
\end{align}
where $g_{10} = g_{a\gamma}\times 10^{10}\,\mathrm{GeV}$, $\alpha_{26}=10^{26} \times g_{ae}^2/4\pi$,
$Y$ is the helium mass fraction in the globular clusters, and
\begin{align}
\delta\mathcal{M}_c = 0.024M_{\odot}\left(\sqrt{4\pi\alpha_{26}+(1.23)^2}-1.23-0.921\alpha^{0.75}_{26}\right)
\label{deltaMc_theory}
\end{align}
is the shift of the core mass due to the the axion emission during the RGB phase~\cite{Viaux:2013hca,Viaux:2013lha}.

In Ref.~\cite{Ayala:2014pea}, observational value $R^{\rm obs} = 1.39 \pm 0.03$ was obtained and compared to a theoretical prediction to derive a bound on $g_{a\gamma}$.
This analysis was revised in Ref.~\cite{Giannotti:2015kwo} by adding the dependence on $g_{ae}$ in Eq.~\eqref{R_theory}.
As the model depends not only on $g_{a\gamma}$ and $g_{ae}$ but also on the helium mass fraction $Y$,
the results of such analyses rely on the assumption about the value of $Y$.
The authors of Ref.~\cite{Ayala:2014pea} preferred to use a value of $Y = 0.2535 \pm 0.0036$, which corresponds to
the weighted average for the helium abundance measured at different oxygen abundance (O/H) estimated in Ref.~\cite{Aver:2013wba}.
Note that this value is somewhat larger than the primordial helium abundance $Y_p$, which is obtained by extrapolating 
the observational results to $\mathrm{O/H} \to 0$.
As noted in Ref.~\cite{Ayala:2014pea}, such different assumptions on $Y$ lead to different bounds on the axion couplings.
It was found that there is a mild preference for a non-vanishing value of $g_{a\gamma}$ when the average value $Y = 0.2535 \pm 0.0036$ is adopted,
while such a hint is diminished for the primordial value $Y_p$.

The claim of Ref.~\cite{Ayala:2014pea} is that one should not use $Y_p$ extracted at zero metallicity but the actual helium abundance in the globular clusters,
since the chemical composition there should be different from the primordial one.
In this sense, the average value of $Y$ can be regarded as representative of the actual helium mass fraction in the globular clusters.
In the analysis performed in this paper, we follow this suggestion and adopt $Y = 0.2535 \pm 0.0036$ in the global fits.

Equation~\eqref{R_theory} was derived in Ref.~\cite{Giannotti:2015kwo} based on the analysis methods
developed in Refs.~\cite{Ayala:2014pea,Straniero:2015nvc}.
In these analyses, the stellar evolution models were computed via a 1D hydrostatic code FUNS~\cite{Straniero:2014},
which was developed based on the FRANEC code~\cite{Straniero:2005hc}.
In Ref.~\cite{Dominguez:1999qr}, behavior of the intermediate mass stars
in the pre-asymptotic giant branch phase computed by means of the FRANEC code was compared to
other evolutionary codes, and it was found that there is a $40\textendash50\%$ difference in the evaluation of the helium-burning lifetime
between different models.
Although we cannot directly apply this estimate to the present case of the FUNS code, 
it should be kept in mind that Eq.~\eqref{R_theory} could be subject to a similar amount of uncertainty.

The effect of stellar rotation is also a possible source of uncertainty in Eq.~\eqref{R_theory}.
Rotation could modify the structure of a star, which results in a change in its chemical composition.
The effect of rotation was studied in Ref.~\cite{Piersanti:2013rya} with a modified version of the FRANEC code. 
It was shown that a nonzero rotation velocity leads to a shorter helium-burning lifetime, which implies a smaller value of $R$.
In the models considered in Ref.~\cite{Piersanti:2013rya}, the variation of the helium-burning lifetime due to the effect of rotation
is about $3\%$ in low metallicity environments.
In this paper, we include it as an additional systematic uncertainty of $R$ when we perform the global analysis.

%%%%%%%%%%%%%%%%%%%%%%%%%%%%%%%%%%%%%%%%%%%%%%%%%%
\subsection{Tip of the red giant branch}
\label{sec:RGBTip}
%%%%%%%%%%%%%%%%%%%%%%%%%%%%%%%%%%%%%%%%%%%%%%%%%%

As mentioned in the previous subsection, the additional cooling due to the axion bremsstrahlung off electrons
can lead to a delay of helium ignition in the core of red giants.
As a result, the core becomes more massive and brighter at helium ignition, and such effects
can be observed as a tip of the RGB (TRGB) in the color-magnitude diagram of a globular cluster.
This observable can be used to constrain the axion-electron coupling $g_{ae}$.

In Refs.~\cite{Viaux:2013hca,Viaux:2013lha}, the color-magnitude diagram of the globular cluster M5 was studied, and
the observational value of $M_{I,\mathrm{TRGB}}^{\rm obs} = -4.17 \pm 0.13\,\mathrm{mag}$ for the TRGB brightness was obtained.
This value was compared to the theoretical prediction given by~\cite{Viaux:2013lha}
\begin{equation}
M_{I,\mathrm{TRGB}}^{\rm th} = -4.03 -0.25\left((g_{13}^2+0.93)^{0.5}-0.96-0.17 g_{13}^{1.5}\right) + \delta M_{I,\mathrm{TRGB}}^{\rm th}\,\mathrm{mag},
\label{MI_TRGB_theory}
\end{equation}
where $g_{13} = 10^{13}\times g_{ae}$, 
and a shift $\delta M_{I,\mathrm{TRGB}}^{\rm th} = 0.039\,\mathrm{mag}$ was introduced since the ranges of systematic errors in
$M_{I,\mathrm{TRGB}}^{\rm th}$ were not symmetric.
A number of possible sources of systematic uncertainties in the theoretical prediction $M_{I,\mathrm{TRGB}}^{\rm th}$
were investigated and summarized in Ref.~\cite{Viaux:2013hca}.
Among them, relatively important ones are the effect of mass loss [$+(0.022\textendash 0.035)\,\mathrm{mag}$],
the equation of state ($-0.0045$ to $+0.0242\,\mathrm{mag}$),
the treatment of convection ($\pm 0.056\,\mathrm{mag}$),
and the color transformations and bolometric corrections [$\pm(0.08+0.02 g_{13})\,\mathrm{mag}$].

As in the case of the $R$-parameter described in Sec.~\ref{sec:RGBTip}, 
the stellar rotation may also affect the results of $M_{I,\mathrm{TRGB}}$.
However, we can assume that such an effect is negligible in this case from the following reason.
According to Ref.~\cite{Piersanti:2013rya}, the variation of the core mass due to the stellar rotation at the TRGB phase is just about $0.35\%$
in low metallicity environments.
The analysis in Ref.~\cite{Viaux:2013lha} indicates that a change in the core mass by $3\%$ corresponds to the error of $\Delta M_{I,\mathrm{TRGB}} = \pm 0.002\,\mathrm{mag}$,
which is much smaller than other dominant sources of uncertainties.
Therefore, we expect that the stellar rotation would only have a minor effect in the estimation of $M_{I,\mathrm{TRGB}}$.

In Refs.~\cite{Viaux:2013hca,Viaux:2013lha}, all the systematic errors in $M_{I,\mathrm{TRGB}}^{\rm th}$ were 
convolved to a Gaussian statistical error, whose magnitude was estimated by adding them in quadrature.
With this estimate of the theoretical error, the comparison with the observational value
gives a bound $|g_{ae}| < 4.3 \times 10^{-13}$ (95\% CL) with a small hint that 
the agreement between observations and theoretical predictions improves at a nonzero value of $g_{ae}$.
However, the validity of this procedure of converting systematic uncertainties into statistical uncertainties is not obvious.
In the analyses performed in this paper, 
we take a different approach to deal with the error in $M_{I,\mathrm{TRGB}}^{\rm th}$.

%%%%%%%%%%%%%%%%%%%%%%%%%%%%%%%%%%%%%%%%%%%%%%%%%%
\subsection{White dwarf luminosity function}
\label{sec:WDLF}
%%%%%%%%%%%%%%%%%%%%%%%%%%%%%%%%%%%%%%%%%%%%%%%%%%

The hint for non-vanishing value of $g_{ae}$ was also pointed out in the context of the cooling of white dwarfs (WDs).
The cooling rate of WDs can be inferred from their luminosity function (WDLF), 
which represents the number density of WDs per luminosity interval.
The dominant process of the axion emission is given by the bremsstrahlung off electrons, which hastens the cooling of WDs.
This additional cooling leads to some modifications of the WDLF.
The result of the analysis in Ref.~\cite{Bertolami:2014wua} shows that 
a non-vanishing value of $|g_{ae}| \sim 1.4 \times 10^{-13}$ is marginally preferred,
although it excludes higher values of $g_{ae}$ and sets a bound $|g_{ae}| < 2.1\times 10^{-13}$ (95\% CL).

As noted in Ref.~\cite{Bertolami:2014wua}, there are two obstacles that prevent us from drawing credible conclusions
from the analysis of the WDLF.
One is the uncertainties in the stellar formation rates (SFR).
In Ref.~\cite{Bertolami:2014wua}, theoretical WDLF curves were derived with an assumption of a constant SFR,
while possible time variations of the SFR can affect the estimation of the theoretical WDLF.
The other is magnitude-dependent selection effects (survey incompleteness), which could distort the observed WDLF.

In Ref.~\cite{Isern:2008nt}, it was claimed that the shape of the bright part of the WDLF is 
almost independent of the SFR and that its effect can be absorbed in the overall normalization.
The analysis performed in Ref.~\cite{Rowell:2013vsa} with the aim of the ``inversion" from the WDLF to the SFR
(i.e. deriving the time dependence of the SFR from the observed WDLF)
appears to confirm this claim, as only a fainter part ($M_{\rm bol} \gtrsim 12$) of the WDLF is sensitive to the SFR.
Since the bound on the axion-electron coupling was derived by only using the luminosity range $3\le M_{\rm bol} \le 12.5$~\cite{Bertolami:2014wua},
we expect that the possible time variation in the SFR would have only a minor effect in the shape of the WDLF of interest.

The more important and serious issue is the uncertainties due to survey incompleteness.
The data of the ``observed" WDLF used in Ref.~\cite{Bertolami:2014wua} were constructed in Ref.~\cite{Bertolami:2014noa}
by considering two data sets obtained from different surveys:
One is the WDLF determined by the Sloan Digital Sky Survey (SDSS)~\cite{Harris:2005gd,Krzesinski:2009},
and the other is that determined by the SuperCOSMOS Sky Survey (SSS)~\cite{Rowell:2011}.
It was found that these two WDLFs do not agree with each other within their error bars.
This inconsistency might be attributed to the incompleteness of the SSS-WDLF, which could be
not uniform at all magnitude bins, biasing its shape~\cite{Bertolami:2014noa}.

Understanding the origin of the discrepancy is challenging, and further independent WDLFs are
likely to be needed to deal with possible systematics.
To compromise over this issue, in Ref.~\cite{Bertolami:2014noa}
two WDLFs are unified into an averaged WDLF, whose error bras are estimated from
differences between the SDSS-WDLF and the SSS-WDLF.
This procedure can be regarded as an intuitively reasonable guess for the systematic uncertainty,
and we use it in the analysis performed in this paper.
However, it should be noted that real systematic uncertainties can be larger than this estimate and probably highly correlated,
and this situation must be reviewed once further observational data are available.

%%%%%%%%%%%%%%%%%%%%%%%%%%%%%%%%%%%%%%%%%%%%%%%%%%
\subsection{White dwarf variables}
\label{sec:WDV}
%%%%%%%%%%%%%%%%%%%%%%%%%%%%%%%%%%%%%%%%%%%%%%%%%%

Another observable to infer the cooling rate of WDs is the period decrease of WD variables.
The time scale of the change of the pulsation period of WD variables can be related to the cooling rate, and
it can be used to probe the exotic cooling due to the axion emission~\cite{Isern:1992gia}.
The observations of several WD variables show that the observed rate of the period decrease
is substantially faster than what predicted in
the standard theory.
The reported hints of the cooling excess include the observations of 
G117-B15A~\cite{Corsico:2012ki}, R548~\cite{Corsico:2012sh}, PG1351+489~\cite{Corsico:2014mpa,Battich:2016htm}, and L19-2~\cite{Corsico:2016okh}.
These results lead to a speculation that there exits some anomalous energy loss due to the axion-electron bremsstrahlung.

The rate of the period decrease of the pulsating WD depends on its mass and effective temperature.
In the study of PG 1351+489~\cite{Corsico:2014mpa,Battich:2016htm}, 
theoretical prediction for the rate of the period change was given by considering two approaches,
the asteroseismological model, in which the mass and effective temperature are fixed such that it reproduces the observed pulsation periods,
and the spectroscopic model, in which the mass and effective temperature are chosen to be spectroscopically inferred values.
It was found that the values of the surface gravity and effective temperature found by the asteroseismological model
are not consistent with those given by spectroscopic measurements.
Due to this discrepancy, the bounds on the axion couplings become different according to which model is assumed~\cite{Battich:2016htm}.

The above model dependence can be regarded as a systematic uncertainty in the theoretical estimate of the rate of the period change of PG 1351+489,
which we include in the analysis performed in this paper.
Note that a similar caution should be given for other WD variables.
In the study of L19-2~\cite{Corsico:2016okh}, 
the results of asteroseismological models agreed with spectroscopic inferences,
and hence such a model uncertainty is not the issue in that case.

%%%%%%%%%%%%%%%%%%%%%%%%%%%%%%%%%%%%%%%%%%%%%%%%%%
\subsection{Supernova 1987A and neutron stras}
\label{sec:SN_and_NS}
%%%%%%%%%%%%%%%%%%%%%%%%%%%%%%%%%%%%%%%%%%%%%%%%%%

Several observational results described above motivate us to study the possibility to interpret the cooling anomalies in terms of the axion emission.
However, there are further caveats to be kept in mind.
One is the fact that the axion couplings to nucleons can be constrained by 
considering the energy loss of the supernova (SN) 1987A~\cite{Raffelt:1987yt,Turner:1987by,Mayle:1987as}.
The additional cooling channel due to nucleon bremsstrahlung $N + N \to N + N + a$ shortens the neutrino pulse duration from the SN,
and the observed neutrino signals lead to the bounds on the axion-nucleon couplings $g_{aN}$, which are given by
\begin{equation}
\mathcal{L}_{aN} = -ig_{aN}a\overline{N}\gamma_5 N, \qquad g_{aN} \equiv \frac{C_{aN}m_N}{f_a},
\label{g_aN_def}
\end{equation}
where $C_{aN}$ is a dimensionless coefficient, $m_N$ the mass of a nucleon,
with $N=p,n$ representing either a proton or a neutron.

We note that there is subtlety in deriving the axion bound from SN 1987A observations. 
This is mainly related to the difficulty in evaluating the axion production rate in the dense nuclear medium from the first principles.
It was pointed out that the axion emission can be suppressed by many-body and multiple-scattering effects~\cite{Raffelt:1991pw,Raffelt:1993ix},
while the degree of suppression is determined by the nucleon spin-fluctuation rate that is hard to estimate in the dense medium
because of the complication of nuclear physics~\cite{Raffelt:1996wa}.
The results of the numerical studies~\cite{Keil:1996ju,Fischer:2016cyd} using an axion emission rate with many-body effects
estimated in Refs.~\cite{Janka:1995ir,Sigl:1995ac} imply the following bound~\cite{Giannotti:2017hny}
\begin{equation}
g_{ap}^2 + g_{an}^2 < 3.6 \times 10^{-19}. \label{SN_bound}
\end{equation}

It should be emphasized that the above constraint should be taken as an indicative result rather than a sharp bound
as the nature of the axion emission is not completely understood and SN simulations do not take account of all necessary physics.
More detailed study is warranted to obtain a conclusive bound on the axion-nucleon couplings from SN observations.
Recently, the SN 1987A bound has been reviewed in Ref.~\cite{Chang:2018rso} by applying several updated nuclear physics calculations,
which yields a factor $\sim 5$ weaker limit on $f_a$ compared with the ``canonical" bound for the KSVZ models
$f_a \gtrsim 4\times 10^8\,\mathrm{GeV}$ from Ref.~\cite{Raffelt:2006cw}. 
A more recent study performed in Ref.~\cite{Carenza:2019pxu} shows a similar relaxation of the SN 1987A bound,
but the degree of reduction is milder than Ref.~\cite{Chang:2018rso}, amounting to a factor $\sim 2$ weaker limit on $f_a$ compared with the canonical bound.
The discrepancy between two results may originate from the fact that the corrections from different nuclear physics effects
are treated separately as multiplicative factors in Ref.~\cite{Chang:2018rso}
whereas in Ref.~\cite{Carenza:2019pxu} these effects are combined in a consistent manner.
However, these results are still not conclusive, as they do not take account of the feedback from the axion emission on the SN.
The bound would be further modified if the effect of the axion feedback is consistently included in a full SN simulation~\cite{Carenza:2019pxu}.

Constraints on the axion-nucleon couplings also arise from the cooling of neutron stars (NSs).
In NSs the axion emission via Cooper pair-breaking-formation (PBF) process could be important in addition to the nucleon bremsstrahlung
process, if baryons in the core of the NS undergo a transition to superfluid state~\cite{Keller:2012yr}.
Simulations of NS cooling by including the axion emission via PBF process and comparison with the data from NS observations
were performed in Refs.~\cite{Sedrakian:2015krq,Sedrakian:2018kdm}, which gives bounds on the axion decay constant
$f_a \gtrsim (5\textendash 10)\times 10^7\,\mathrm{GeV}$ for the KSVZ models and $f_a \gtrsim (5\textendash 15)\times 10^7\,\mathrm{GeV}$ for the DFSZ models.
These results are obtained by taking account of the average temperature of the NS in the Cassiopeia A (Cas A) SN remnant rather than its transient behavior, 
and they can be regarded as conservative limits.
On the other hand, some attempts focusing on the cooling rate of the Cas A NS were made
in Refs.~\cite{Leinson:2014ioa,Hamaguchi:2018oqw}, whose results turned out to be controversial.
In Ref.~\cite{Leinson:2014ioa}, it was argued that a slight extra cooling due to the axion emission is required to explain the observed fast cooling
of the Cas A NS, which shows a preference for a non-vanishing value of the axion-neutron coupling amounting to $f_a/C_{an} \sim 2.5\times 10^9\,\mathrm{GeV}$.
Such a hint was not confirmed in the analysis performed in Ref.~\cite{Hamaguchi:2018oqw},
which instead yields bounds of $f_a \gtrsim 5(7)\times 10^8\,\mathrm{GeV}$ for the KSVZ (DFSZ) models.
In addition, the study of the axion emission from a different NS in HESS J1731-347~\cite{Beznogov:2018fda} 
also disagrees with the hint reported in Ref.~\cite{Leinson:2014ioa}, giving even a stronger bound $f_a/C_{an} \gtrsim 3.4\times 10^9\,\mathrm{GeV}$.
However, given a limited understanding of the cooling of NSs, these results should be taken cautiously.
Moreover, as for the cooling of the Cas A NS, the observational data themselves are inconclusive~\cite{Elshamouty:2013nfa,Posselt:2013xva,Posselt:2018xaf}.
Because of these controversies, we do not include the results of observations of the NS cooling in the analyses performed in this paper.

%%%%%%%%%%%%%%%%%%%%%%%%%%%%%%%%%%%%%%%%%%%%%%%%%%
\subsection{Implications for axion models}
\label{sec:implications_for_models}
%%%%%%%%%%%%%%%%%%%%%%%%%%%%%%%%%%%%%%%%%%%%%%%%%%

Except for the ambiguities related to the cooling of SN and NS, 
various observations show a mild preference for additional cooling.
Although individual anomalies are not quite significant, if combined, 
they lead to a clear systematic tendency of excessive energy losses.
The interpretation in terms of several particle physics models including
neutrino anomalous magnetic moments, minicharged particles, hidden photons, 
and axion-like particles (ALPs) was discussed in Ref.~\cite{Giannotti:2015kwo}.
It turned out that axion/ALPs coupled to electrons are perfectly fit to all the anomalies while other candidates are inadequate for the explanation.
The combined analysis of the hints from WD, HB, and RGB stars performed in Ref.~\cite{Giannotti:2017hny} 
indicates non-vanishing couplings of axion/ALPs with electrons and photons.

\begin{figure}[htbp]
\centering
\includegraphics[width=110mm]{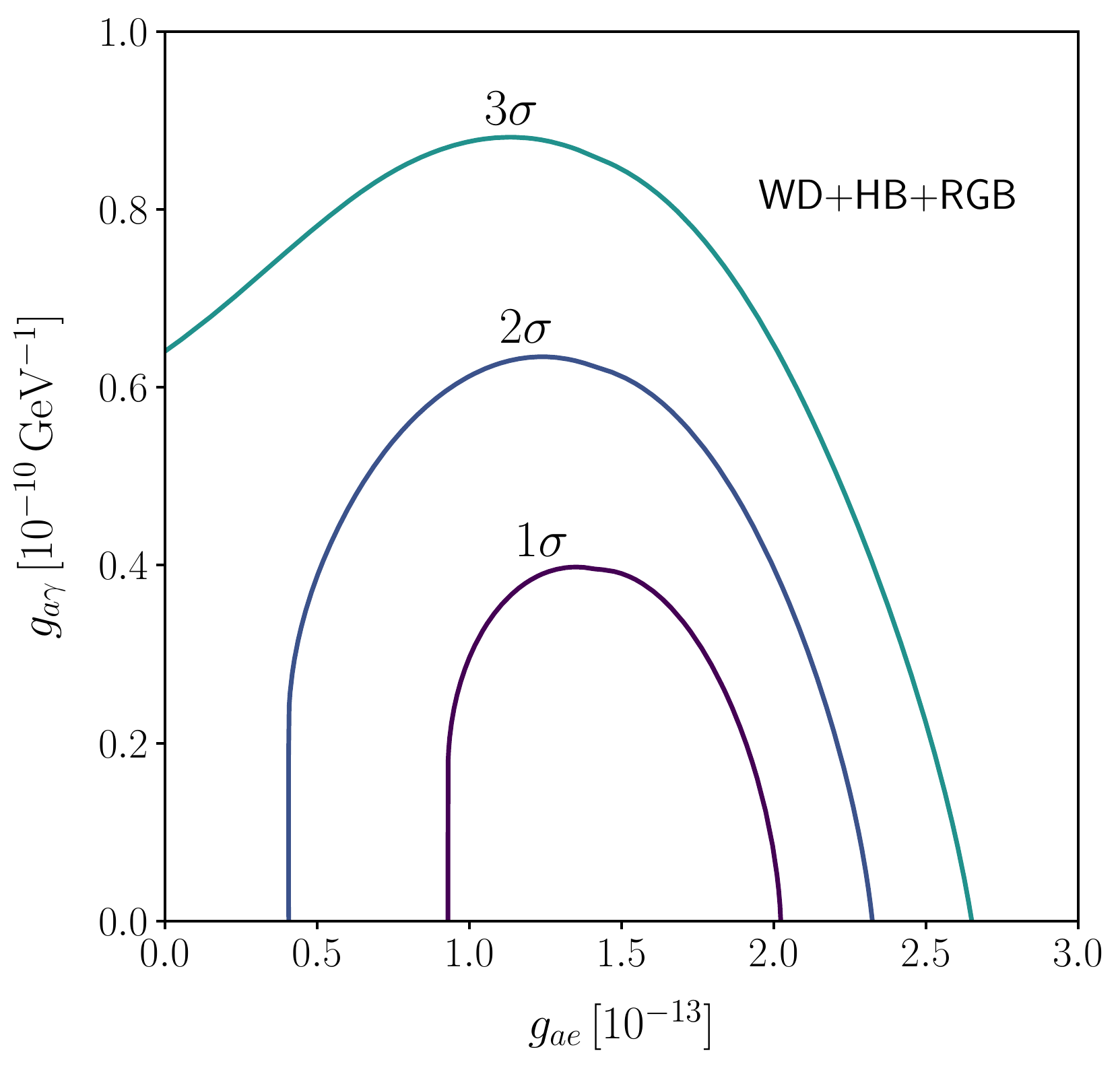}
\caption{
1, 2, and 3\,$\sigma$ hinted regions in the $g_{ae}$-$g_{a\gamma}$ plane found from
the combined analysis of the observational data of WD, HB, and RGB stars.
}
\label{fig:combined_analysis}
\end{figure}

Let us revisit the global analysis of the cooling hints by taking account of several systematic uncertainties that were not included in the analysis of Ref.~\cite{Giannotti:2017hny}.
In Fig.~\ref{fig:combined_analysis}, we show the results for the 1, 2, and 3\,$\sigma$ hinted regions in the $g_{ae}$-$g_{a\gamma}$ plane.
In this analysis we include the systematic uncertainties associated with the effect of stellar rotation in the globular clusters and the model dependencies in the prediction for the
rate of the period change of the WD variable PG 1351+489 (see Secs.~\ref{sec:HB/RGB} and~\ref{sec:WDV}).
Furthermore, we take a different approach to treat the systematic uncertainties in the predicted TRGB brightness.
The details of the analysis method are summarized in Appendix~\ref{app:analysis_method}.

Since we adopt conservative assessments on the systematic uncertainties, the significance of the hints becomes weaker than Ref.~\cite{Giannotti:2017hny}.
In particular, because of the systematic error corresponding to the effect of stellar rotation, 
the hint for a non-zero axion-photon coupling becomes less clear, and any value in the range of $g_{a\gamma}\lesssim 0.07\times 10^{-10}\,\mathrm{GeV}^{-1}$ gives an equally good fit.
A non-zero axion-electron coupling is favored at around 2.4\,$\sigma$, and the best fit value is $g_{ae} = 1.56\times 10^{-13}$ with $\chi^2_{\rm min}/\text{d.o.f.} = 14.7/15$.

Although the anomalous excessive cooling can be explained in a general framework of axion/ALPs where there is no particular relation between the mass and couplings, 
there remains a question of whether typical axion models that provide a solution to the strong CP problem accommodate the hints.
The axion associated with the solution to the strong CP problem acquires a mass from the QCD effects, and the mass $m_a$ is related to the decay constant $f_a$.
On the other hand, the prediction for coupling coefficients $C_{a\gamma}$, $C_{ae}$, $C_{ap}$, and $C_{an}$ is highly model dependent.
By taking account of such model dependencies,
the possibilities to interpret the stellar cooling anomalies in terms of the KSVZ
and DFSZ axion models were explored in Ref.~\cite{Giannotti:2017hny}.

In the KSVZ models, the SM is extended by introducing one singlet complex scalar and exotic heavy quark(s).
According to representations of the exotic quarks under the SM SU(2)$_L \times$ U(1)$_Y$ gauge group,
the low energy effective potential for the axion field can only have a single minimum.
This fact is conventionally described as $N_{\rm DW} = 1$,
where $N_{\rm DW}$ is an integer representing the number of degenerate vacua
and called the domain wall number.
In this case, even though domain walls are formed around the epoch of the QCD phase transition,
they decay immediately after the formation~\cite{Vilenkin:1982ks}.
Hence there is no cosmological problem associated with the domain walls.
However, since the SM leptons do not have PQ charges in these models, the electron coupling 
vanishes at tree level and emerges only at the loop level~\cite{Srednicki:1985xd}, which is too small to explain the anomalous excessive cooling.
This fact excludes the possibility of a pure KSVZ axion interpretation of the stellar cooling anomalies.\footnote{If we extend the KSVZ models
by adding extra three right-handed neutrinos (sometimes called a KSVZ-like axion/majoron model)~\cite{Shin:1987xc,Ballesteros:2016euj,Ballesteros:2016xej},
it becomes possible to obtain good fits to the data since the axion-electron coupling acquires
extra loop contributions from neutrinos that can be adjusted to explain the observed anomalies~\cite{Giannotti:2017hny}.
However, such large loop contributions are generically in tension with the requirement of perturbativity.}

In the DFSZ models, the SM is extended by introducing one singlet complex scalar and two Higgs doublets.
In this case, we can obtain a suitable magnitude of the axion-electron coupling, and there exists a parameter region 
compatible with the observed data of the cooling anomalies~\cite{Giannotti:2017hny}.
However, the DFSZ models lead to multiple degenerate minima in the low energy effective potential, which causes 
the formation of domain walls if the PQ symmetry is broken after inflation~\cite{Sikivie:1982qv}.
In order to avoid the cosmological domain wall problem, 
we have to introduce some extra assumptions such as
the existence of an explicit symmetry breaking term in the effective potential which leads to
the late-time annihilation of domain walls~\cite{Sikivie:1982qv,Chang:1998tb,Hiramatsu:2010yn,Hiramatsu:2012sc,Kawasaki:2014sqa,Ringwald:2015dsf}.

Given the fact that the pure KSVZ models are incompatible with the observations and that
the DFSZ models suffer from the cosmological domain wall problem,
it is reasonable to investigate whether other classes of models could give good fits to the data without causing any problem.
In the following, we study the variant axion models~\cite{Peccei:1986pn,Krauss:1986wx} 
as alternative possibilities to interpret the stellar cooling anomalies.

%%%%%%%%%%%%%%%%%%%%%%%%%%%%%%%%%%%%%%%%%%%%%%%%%%
\section{Variant axion models}
\label{sec:variant_axions}
\setcounter{equation}{0}
%%%%%%%%%%%%%%%%%%%%%%%%%%%%%%%%%%%%%%%%%%%%%%%%%%

We consider the PQ axion models with two Higgs doublets $H_1$, $H_2$ and one singlet scalar $\sigma$,
and assume that the global U(1)$_{\rm PQ}$ symmetry acts on them as
\begin{equation}
H_k \to e^{iX_k\epsilon}H_k, \quad \sigma \to e^{i X_{\sigma}\epsilon}\sigma,
\end{equation}
where $k=1,2$ and $\epsilon$ is a real constant parameter.

In order to provide a solution to the strong CP problem, 
we need to assign the U(1)$_{\rm PQ}$ charges for the quark fields to make the PQ symmetry anomalous for QCD.
In other words, the low energy effective Lagrangian must contain the following term,
\begin{equation}
\mathcal{L} \supset - \frac{\alpha_s}{8\pi} \mathcal{N}\frac{a}{v_{\rm PQ}}G^a_{\mu\nu}\tilde{G}^{a\mu\nu},
\end{equation}
where $\alpha_s$ is the strong coupling constant, $\mathcal{N}$ the QCD anomaly coefficient,
$v_{\rm PQ}$ the PQ symmetry breaking scale, 
$G^a_{\mu\nu}$ the gluon field strength,
and $\tilde{G}^{a\mu\nu}$ its dual.
In the following, we focus on the cases where the QCD anomaly coefficient is given by
\begin{equation}
\mathcal{N} = X_1 - X_2 \quad \text{or} \quad X_2 - X_1, \label{assumption_N}
\end{equation}
which is a crucial element of variant axion models and different from the DFSZ model, for which $\mathcal{N} = 3(X_1-X_2)$ or $3(X_2-X_1)$.

We note that the value of $\mathcal{N}$ can be related to the periodicity of the effective potential for the axion field, 
and hence it determines the vacuum structure of the theory. 
According to the PQ charge assignments, we can consider two possibilities~\cite{Geng:1990dv}: 
If $2X_{\sigma} = X_1 - X_2$, the Lagrangian contains an
interaction term of the form $H_1^{\dagger}H_2\sigma^2$ and its hermitian conjugate, and we have $|\mathcal{N}| = 2|X_{\sigma}|$.
In this case, the periodicity of the axion field becomes twice of that of the QCD $\theta$ term,
and the effective potential for the axion field has two degenerate minima,
leading to the domain wall number $N_{\rm DW} = 2$.
On the other hand, if $X_{\sigma} = X_1 - X_2$, the Lagrangian contains an
interaction term proportional to $H_1^{\dagger}H_2\sigma$ and its hermitian conjugate.
In this case, the periodicity of the axion field becomes the same as that of the QCD $\theta$ term,
which implies $N_{\rm DW} = 1$.
In what follows we consider the latter case, since it avoids the cosmological domain wall problem.
It should be emphasized that the cosmological domain wall problem cannot be solved straightforwardly in the DFSZ model,
since it leads to $|\mathcal{N}| = 6|X_{\sigma}|$ or $3|X_{\sigma}|$, which implies $N_{\rm DW} = 6$ or $3$.
We also note that the physical consequences depend only on the combination $X_1 - X_2$, and that
we can assign $X_{\sigma} = 1$, $X_1 = 0$, and $X_2 = -1$ without loss of generality.

The Yukawa interactions for the quark fields read
\begin{align}
-\mathcal{L}_{\mathrm{Yukawa},q} &= \Gamma^u_i\overline{q}_{iL}\tilde{H}_u u_R + \Gamma^c_i\overline{q}_{iL}\tilde{H}_c c_R + \Gamma^t_i\overline{q}_{iL}\tilde{H}_t t_R \nonumber\\
&\quad + \Gamma^d_i\overline{q}_{iL}H_d d_{R} + \Gamma^s_i\overline{q}_{iL}H_s s_{R} + \Gamma^b_i\overline{q}_{iL}H_b b_{R} + \mathrm{h.c.},
\label{L_yukawa_quarks}
\end{align}
where $H_q$ ($q = u,c,t,d,s,b$) take either $H_1$ or $H_2$, 
$\Gamma^q_i$ are Yukawa couplings with $i = 1,2,3$ being a generation index,
and $\tilde{H}_k = i\sigma_2H_k^*$.
We consider the models where only one quark flavor has a nonzero PQ charge and couples to $H_2$
while the others couple to $H_1$
such that $|\mathcal{N}| = 1$ in Eq.~\eqref{assumption_N} is satisfied.
There are six possibilities (Model U, C, T, D, S, B), 
and the corresponding charge assignments are shown in Table~\ref{tab:charge_quarks}.
For the lepton sector, we consider 
two types of interactions,\footnote{Although we consider flavor-blind PQ charge assignments for the lepton sector for simplicity, 
it is also possible to assign different PQ charges for each lepton flavor.
In particular, the model can be embedded in SU(5) grand unified theory, where only one generation of $\overline{\bf 5}_i$ multiplet has a nonzero PQ charge.
In that case, only one generation of $d_{iR}$ and $\ell_{iL}$ have nonzero PQ charges and couple to $H_2$.}
\begin{equation}
-\mathcal{L}_{\mathrm{Yukawa},\ell} = \left\{
\begin{array}{ll}
\Gamma^{\ell}_{ij}\overline{\ell}_{iL}H_1 e_{jR} + \mathrm{h.c.} & \text{(Type I)},
\vspace{1mm}\\
\Gamma^{\ell}_{ij}\overline{\ell}_{iL}H_2 e_{jR} + \mathrm{h.c.} & \text{(Type II)},
\end{array}
\right.
\label{L_yukawa_leptons}
\end{equation}
where $\Gamma^{\ell}_{ij}$ is the lepton Yukawa matrix with $i,j = 1,2,3$ being generation indices.
These terms are consistent with the charge assignments for the lepton fields shown in Table~\ref{tab:charge_leptons}.

\begin{table}[ht]\centering
\caption{PQ charge assignments for the quark fields.}
\vspace{-\baselineskip}
\begin{equation}{\nonumber
\begin{array}{|l|ccccccc|}
\hline
\text{Model} & q_{iL} & u_R & c_R & t_R & d_R & s_R & b_R \\
\hline
\text{Model U} & 0 & -1 & 0 & 0 & 0 & 0 & 0 \\
\text{Model C} & 0 & 0 & -1 & 0 & 0 & 0 & 0 \\
\text{Model T} & 0 & 0 & 0 & -1 & 0 & 0 & 0 \\
\text{Model D} & 0 & 0 & 0 & 0 & 1 & 0 & 0 \\
\text{Model S} & 0 & 0 & 0 & 0 & 0 & 1 & 0 \\
\text{Model B} & 0 & 0 & 0 & 0 & 0 & 0 & 1 \\
\hline
\end{array}
}
\end{equation}
\label{tab:charge_quarks}
\end{table}

\begin{table}[ht]\centering
\caption{PQ charge assignments for the lepton fields.}
\vspace{-\baselineskip}
\begin{equation}{\nonumber
\begin{array}{|l|cc|}
\hline
\text{Type} & \ell_{iL} & e_{iR} \\
\hline
\text{Type I} & 0 & 0 \\
\text{Type II} & 0 & 1 \\
\hline
\end{array}
}
\end{equation}
\label{tab:charge_leptons}
\end{table}

Let us derive axion couplings to ordinary matter in the variant axion models.
The key ingredient is the PQ current $j^{\rm PQ}_{\mu}$, which is associated with the QCD and EM anomalies,
\begin{equation}
\partial^{\mu}j^{\rm PQ}_{\mu} = \mathcal{N}\frac{\alpha_s}{8\pi} G^a_{\mu\nu}\tilde{G}^{a\mu\nu} + \mathcal{E}\frac{\alpha}{8\pi} F_{\mu\nu}\tilde{F}^{\mu\nu},
\label{PQ_current_anomaly}
\end{equation}
where $\mathcal{E}$ is the EM anomaly coefficient.
Before the spontaneous breaking of the electroweak symmetry, the PQ current is given by
\begin{equation}
j^{\rm PQ}_{\mu} = v_{\sigma}\partial\tilde{a} - \sum_{\psi}\sum_{i}\left[\overline{\psi}_{iL}X_{\psi_{iL}}\gamma^{\mu}\psi_{iL} + \overline{\psi}_{iR}X_{\psi_{iR}}\gamma^{\mu}\psi_{iR}\right],
\end{equation}
where $v_{\sigma}^2 = 2\langle|\sigma|^2\rangle$ is the vacuum expectation value (VEV) of the singlet scalar,
$\tilde{a}$ is its phase direction $\sigma \propto e^{i\tilde{a}/v_{\sigma}}$,
the summation over $\psi$ includes up-type quarks ($u$), down-type quarks ($d$), and leptons ($\ell$),
and $X_{\psi_{iL,R}}$ represent PQ charges for left-handed ($L$) and right-handed ($R$) fermions, which are
specified in Tables~\ref{tab:charge_quarks} and~\ref{tab:charge_leptons}.

After the spontaneous breaking of the electroweak symmetry, the angular field $\tilde{a}$ mixes with a
Nambu-Goldstone boson eaten by the $Z^0$ boson, and we have to redefine the axion field such that
the associated PQ current gives the same QCD and EM anomalies as the original one.
This amounts to the shifts of the PQ charges $X_{\psi_{iL,R}}$ for fermions in the PQ current~\cite{Srednicki:1985xd,Georgi:1986df},
\begin{align}
j^{\rm PQ}_{\mu} &= v_{\rm PQ}\partial_{\mu}a - \sum_{\psi}\sum_{i}\left[\overline{\psi}_{iL}X'_{\psi_{iL}}\gamma^{\mu}\psi_{iL} + \overline{\psi}_{iR}X'_{\psi_{iR}}\gamma^{\mu}\psi_{iR}\right], \label{PQ_current_below_EWSB}\\
X_{\psi_{iL,R}}' &= X_{\psi_{iL,R}} - v^{-2}\left(\sum_k 2X_kY_kv_k^2\right)2Y_{\psi_{iL,R}},
\label{shifted_fermion_PQ_charges}
\end{align}
where $Y_k$ are U(1)$_Y$ charges for two Higgs doublets $H_k$ ($Y_1=Y_2=1/2$), 
$v_k$ represent the VEVs of their neutral components,
$v = \sqrt{v_1^2+v_2^2} \simeq 246\,\mathrm{GeV}$ is the electroweak scale, and
$Y_{\psi_{iL,R}}$ are U(1)$_Y$ charges for the fermions.
The PQ scale $v_{\rm PQ}$ can be related to two symmetry breaking scales $v_{\sigma}$ and $v$ 
as $v_{\rm PQ} = \sqrt{v_{\sigma}^2+v^2\sin^2\beta\cos^2\beta}$, where 
\begin{equation}
\tan\beta \equiv \frac{v_2}{v_1}.
\end{equation}
Based on the PQ current given by Eq.~\eqref{PQ_current_below_EWSB}, 
we can construct the effective Lagrangian at energies below the electroweak scale but above the QCD scale,
\begin{align}
\mathcal{L} &\supset -\frac{1}{2}\partial_{\mu}a\partial^{\mu}a
- \frac{\alpha_s}{8\pi}\frac{a}{f_a}G^a_{\mu\nu}\tilde{G}^{a\mu\nu} - \frac{\alpha}{8\pi}\frac{\mathcal{E}}{\mathcal{N}}\frac{a}{f_a}F_{\mu\nu}\tilde{F}^{\mu\nu} \nonumber\\
&\quad-\frac{\partial_{\mu}a}{f_a} \sum_{\psi}\sum_{i}\left[\overline{\psi}_{iL}\frac{X'_{\psi_{iL}}}{\mathcal{N}}\gamma^{\mu}\psi_{iL} + \overline{\psi}_{iR}\frac{X'_{\psi_{iR}}}{\mathcal{N}}\gamma^{\mu}\psi_{iR}\right],
\label{effective_Lagrangian_1}
\end{align}
where 
\begin{equation}
f_a \equiv \frac{v_{\rm PQ}}{\mathcal{N}}
\label{fa_definition}
\end{equation}
is the axion decay constant.

So far it is implicitly assumed that the fermions $\psi_{iL,R}$ are in the weak interaction basis.
Now we switch to the mass basis by performing unitary transformations $\psi_{L,R} \to U_{\psi_{L,R}}\psi_{L,R}$
that diagonalize the Yukawa matrices. After the transformations, the axion-fermion couplings become
\begin{equation}
\mathcal{L}_{a\psi} = -\frac{\partial_{\mu}a}{2f_a}\sum_{\psi}\sum_{i,j}\overline{\psi}_{i}\gamma^{\mu}\left[\left(C^V_{a\psi}\right)_{ij}-\left(C^A_{a\psi}\right)_{ij}\gamma_5\right]\psi_j,
\label{fermion_couplings}
\end{equation}
where
\begin{align}
C^V_{a\psi} &= \frac{1}{\mathcal{N}}\left(U_{\psi_L}^{\dagger}X'_{\psi_L}U_{\psi_L} + U_{\psi_R}^{\dagger}X'_{\psi_R}U_{\psi_R}\right),\label{C_V_apsi}\\
C^A_{a\psi} &= \frac{1}{\mathcal{N}}\left(U_{\psi_L}^{\dagger}X'_{\psi_L}U_{\psi_L} - U_{\psi_R}^{\dagger}X'_{\psi_R}U_{\psi_R}\right), \label{C_A_apsi}
\end{align}
and $X'_{\psi_{L,R}}$ are understood as diagonal matrices.
If $X'_{\psi_{L,R}}$ are not proportional to the identity matrix, the coupling coefficients $C^V_{a\psi}$ and $C^A_{a\psi}$
depend not only on the effective PQ charges $X'_{\psi_{L,R}}$ but also on the fermion mixing matrices $U_{\psi_{L,R}}$~\cite{Georgi:1986df,Bjorkeroth:2018dzu}.
This is the case for the variant axion models, while in the DFSZ models we can eliminate $U_{\psi_{L,R}}$ as the PQ charges are generation-independent.
In general, both $C^V_{a\psi}$ and $C^A_{a\psi}$ can have off-diagonal components due to the contribution from the fermion mixing matrices,
while diagonal components of $C^V_{a\psi}$ can be eliminated by using the equations of motion.

Applying the PQ charge assignments for the quark fields shown in Table~\ref{tab:charge_quarks} to Eqs.~\eqref{C_V_apsi} and~\eqref{C_A_apsi},
we obtain the explicit forms for the axion-quark couplings in the variant axion models. The results are summarized in Table~\ref{tab:axion_quark_couplings}.
Here we introduced the following matrices,
\begin{align}
V_u &\equiv U^{\dagger}_{u_R}\Lambda U_{u_R}, \label{V_u_definition}\\
V_d &\equiv U^{\dagger}_{d_R}\Lambda U_{d_R}, \label{V_d_definition}\\
\Lambda & \equiv \left\{
\begin{array}{ll}
\mathrm{diag}(0,1,1) & \text{(Model U,\,D)},\\
\mathrm{diag}(1,0,1) & \text{(Model C,\,S)},\\
\mathrm{diag}(1,1,0) & \text{(Model T,\,B)},\\
\end{array}
\right.
\label{Lambda_models}
\end{align}
which represent the corrections arising from the mixings $U_{u_R}, U_{d_R}$ of right-handed up-type and down-type quarks.
The matrices $V_u$ and $V_d$ provide off-diagonal components of the axion-quark couplings, while diagonal components of $V_u$ and $V_d$ are positive.
They satisfy the constraints $\Tr(V_u) = \Tr(V_d) = 2$.
Note that these matrices are different from the Cabibbo-Kobayashi-Maskawa mixing matrix $V_{\rm CKM}=U^{\dagger}_{u_L}U_{d_L}$,
and they should be regarded as new parameters of the theory.
The generation-dependence in the PQ charge assignments is now hidden in the matrices $V_{u,d}$,
and in what follows we group it into two different classes of models,
Model U,\,C,\,T and Model D,\,S,\,B, as shown in Table~\ref{tab:axion_quark_couplings}.

\begin{table}[ht]\centering
\caption{Axion couplings to up-type ($u$) and down-type ($d$) quarks in the variant axion models.}
\vspace{-\baselineskip}
\begin{align}
\begin{array}{|l|cccc|}
\hline
\text{Model} & (C^V_{au})_{i\ne j} & (C^V_{ad})_{i\ne j} & (C^A_{au})_{ij} & (C^A_{ad})_{ij} \\
\hline
\text{Model U,\,C,\,T} & (V_u)_{ij} & 0 & \delta_{ij}\cos^2\beta - (V_u)_{ij} & \delta_{ij}\sin^2\beta \\
\text{Model D,\,S,\,B} & 0 & (V_d)_{ij} & \delta_{ij}\sin^2\beta & \delta_{ij}\cos^2\beta - (V_d)_{ij} \\
\hline
\end{array}
\nonumber
\end{align}
\label{tab:axion_quark_couplings}
\end{table}

Since we consider flavor-blind PQ charge assignments for the lepton fields, there is no correction from mixing matrices to the axion-lepton couplings. 
Hence the coupling matrix $(C^A_{ae})_{ij}$ becomes proportional to the identity matrix, and the coupling strength is solely given by $\tan\beta$.
Now it is straightforward to extract the coefficient for the axion-electron coupling $C_{ae}$ in Eq.~\eqref{g_ae_def}
by using Eq.~\eqref{C_A_apsi} and the PQ charge assignments specified in Table~\ref{tab:charge_leptons}.
In our setup there are four possibilities according to 
different PQ charge assignments for the quark and lepton fields,
which we summarize in Table~\ref{tab:axion_electron_couplings}.\footnote{The difference in the sign of $C_{ae}$ between
Model U,\,C,\,T and Model D,\,S,\,B originates from the sign of $\mathcal{N}$ appearing in Eq.~\eqref{C_A_apsi}.
Note that this model-dependence does not affect the sign of $g_{ae}$, since it is proportional to $C_{ae}/f_a$ and
$f_a$ contains another factor of $1/\mathcal{N}$ [see Eq.~\eqref{fa_definition}].
Hence the electron coupling is solely determined by the Type of the PQ charge assignments for the lepton fields.}

\begin{table}[ht]\centering
\caption{Axion couplings to electrons and photons and the model-dependent coefficient for the axion-photon coupling in the variant axion models.}
\vspace{-\baselineskip}
\begin{align}
\begin{array}{|l|l|ccc|}
\hline
\text{Model} & \text{Type} & C_{ae} & C_{a\gamma} & \mathcal{E}/\mathcal{N} \\
\hline
\text{Model U,\,C,\,T} & \text{Type I} & \sin^2\beta & 0.75(4) & 8/3 \\
 & \text{Type II} & -\cos^2\beta & -5.25(4) & -10/3 \\
 \hline
\text{Model D,\,S,\,B} & \text{Type I} & -\sin^2\beta & -1.25(4) & 2/3 \\
 & \text{Type II} & \cos^2\beta & 4.75(4) & 20/3 \\
\hline
\end{array}
\nonumber
\end{align}
\label{tab:axion_electron_couplings}
\end{table}

Below the QCD scale, the field $a$ mixes with light mesons such as neutral pions, and we have to find a state orthogonal to these mesons
in order to identify the physical axion state at low energies.
The mixing with neutral pions can be eliminated straightforwardly by considering the low energy effective theory emerging from the QCD with
two lightest quarks $q' = (u,d)$.
First, we perform the following chiral transformations of up and down quarks,
\begin{equation}
q' = \left(
\begin{array}{c}
u\\
d
\end{array}
\right)
\to
e^{-i\gamma_5\frac{a}{2f_a}Q_a}
\left(
\begin{array}{c}
u\\
d
\end{array}
\right),
\label{chiral_transf_ud}
\end{equation}
where 
$Q_a$ is a matrix acting on $q' = (u,d)$.
After this redefinition, the coefficient of the $G^a_{\mu\nu}\tilde{G}^{a\mu\nu}$ term in the effective Lagrangian is shifted by 
$\mathrm{Tr}(Q_a)(a/f_a)(\alpha_s/8\pi)$, and we can eliminate this coupling to gluons by taking $\mathrm{Tr}(Q_a) = 1$.
Then, the terms including the lightest two quarks read
\begin{align}
\mathcal{L} \supset \frac{1}{2}\frac{\partial_{\mu}a}{f_a}\bar{q}'\gamma^{\mu}\gamma_5(C_{aq'}-Q_a)q' - (\bar{q}'_L M_a q'_R + \mathrm{h.c.}),
\end{align}
where
\begin{align}
M_a = e^{i\frac{a}{2f_a}Q_a}M_q e^{i\frac{a}{2f_a}Q_a}, \quad 
M_q = \left(
\begin{array}{cc}
m_u & 0\\
0 & m_d
\end{array}
\right), \quad
C_{aq'} = \left(
\begin{array}{cc}
(C_{au}^A)_{11} & 0\\
0 & (C_{ad}^A)_{11}
\end{array}
\right).
\end{align}

The leading order chiral Lagrangian that describes physics below the QCD scale contains the following term,
\begin{align}
\mathcal{L}_{\rm eff} \supset \frac{1}{2}f_{\pi}^2\mu\mathrm{Tr}\left(UM_a^{\dagger} + M_aU^{\dagger}\right), \label{chiral_Lagrangian}
\end{align}
where
\begin{align}
U = e^{i\Pi/f_{\pi}}, \quad 
\Pi = \left(
\begin{array}{cc}
\pi^0 & \sqrt{2}\pi^+ \\
\sqrt{2}\pi^- & - \pi^0
\end{array}
\right),
\end{align}
$f_{\pi} \simeq 92\,\mathrm{MeV}$ is the pion decay constant, and $\mu$ is a parameter that can be related to the pion mass.
Expanding Eq.~\eqref{chiral_Lagrangian} to quadratic order in the fields, we find the axion-pion mixing, which can be eliminated by choosing
$Q_a = M_q^{-1}/\mathrm{Tr}(M_q^{-1})$~\cite{Georgi:1986df}.
After eliminating the axion-pion mixing and fixing the parameter $\mu$ in terms of the pion mass $m_{\pi}$, we can also extract the mass of the physical axion,
\begin{equation}
m_a^2 = \frac{m_um_d}{(m_u+m_d)^2}\frac{m_{\pi}^2f_{\pi}^2}{f_a^2}.
\end{equation}

The above formula for the axion mass is derived based on the tree-level axion-pion mixing in
the effective theory with two quark flavors.
More precise computation including the extra contributions from the strange quark was performed
based on the next to leading order (NLO)~\cite{diCortona:2015ldu} and next-to-next to leading order (NNLO)~\cite{Gorghetto:2018ocs} chiral perturbation theory.
Here we quote the NNLO result obtained in Ref.~\cite{Gorghetto:2018ocs},
\begin{equation}
m_a = 5.691(51)\,\mu\mathrm{eV}\left(\frac{10^{12}\,\mathrm{GeV}}{f_a}\right).
\end{equation}
This result is consistent with that of the direct calculation of the topological susceptibility in lattice QCD performed in Ref.~\cite{Borsanyi:2016ksw}.

The axion-photon coupling [Eq.~\eqref{g_agamma_def}] consists of the model-dependent contribution 
given by the term proportional to $\mathcal{E}/\mathcal{N}$ in Eq.~\eqref{effective_Lagrangian_1}
and the model-independent contribution arising from the coupling to QCD.
The tree level contribution to the latter can be extracted from a shift of the $F_{\mu\nu}\tilde{F}^{\mu\nu}$ term
due to the rotation of two light quarks~\eqref{chiral_transf_ud}.
A more precise result including the NLO corrections~\cite{diCortona:2015ldu} reads
\begin{align}
C_{a\gamma} = \frac{\mathcal{E}}{\mathcal{N}} - 1.92(4).
\end{align}
In Table~\ref{tab:axion_electron_couplings}, 
we summarize the values of the model-dependent coefficient $\mathcal{E}/\mathcal{N}$ 
as well as the values of $C_{a\gamma}$ in the variant axion models.

The axion-nucleon couplings [Eq.~\eqref{g_aN_def}] can be derived by matching the low energy effective Lagrangian 
involving nucleons to the ultraviolet (UV) Lagrangian involving quarks.
Similarly to the axion-photon coupling, the axion-nucleon couplings can be written in terms of 
the model-independent contributions from the mixing with mesons and
the model-dependent contributions arising from the couplings to quarks in the UV Lagrangian~\eqref{fermion_couplings}.
Here we adopt the results obtained in Ref.~\cite{diCortona:2015ldu},
\begin{align}
C_{ap} &= -0.47(3) + 0.88(3)(C^A_{au})_{11} - 0.012(5)(C^A_{au})_{22} - 0.0035(4)(C^A_{au})_{33} \nonumber\\
&\qquad\qquad\quad -0.39(2)(C^A_{ad})_{11} - 0.038(5)(C^A_{ad})_{22} - 0.009(2)(C^A_{ad})_{33},\\
C_{an} &= -0.02(3) - 0.39(2)(C^A_{au})_{11} - 0.012(5)(C^A_{au})_{22} - 0.0035(4)(C^A_{au})_{33} \nonumber\\
&\qquad\qquad\quad +0.88(3)(C^A_{ad})_{11} - 0.038(5)(C^A_{ad})_{22} - 0.009(2)(C^A_{ad})_{33}.
\end{align}
Substituting the diagonal components of $(C^A_{au})_{ij}$ and $(C^A_{ad})_{ij}$ shown in Table~\ref{tab:axion_quark_couplings} to
the above equations, we obtain
\begin{align}
C_{ap} &\approx 
\left\{
\begin{array}{ll}
0.39 -1.30\sin^2\beta - 0.88(V_u)_{11} + \delta_N & \text{(Model U,\,C,\,T)},\\
-0.91 +1.30\sin^2\beta + 0.39(V_d)_{11} + \delta_N  & \text{(Model D,\,S,\,B)},
\end{array}
\right. \label{C_ap_VA}\\
C_{an} &\approx 
\left\{
\begin{array}{ll}
-0.43 +1.24\sin^2\beta + 0.39(V_u)_{11} + \delta_N & \text{(Model U,\,C,\,T)},\\
0.81 -1.24\sin^2\beta - 0.88(V_d)_{11} + \delta_N  & \text{(Model D,\,S,\,B)},
\end{array}
\right. \label{C_an_VA}
\end{align}
where $\delta_N$ is a correction term containing $(V_{u,d})_{22}$ and $(V_{u,d})_{33}$.
This correction term is small compared to the leading terms as it originates from contributions of heavy quarks.\footnote{
Changing the value of $(V_{u,d})_{22}$ from $0$ to $2$ with imposing the constraint $(V_{u,d})_{22}+(V_{u,d})_{33} = 2$,
we obtain $\delta_N \lesssim 0.024$ for Model U,\,C,\,T and $\delta_N \lesssim 0.076$ for Model D,\,S,\,B.
Note that the constant terms in $C_{ap}$ and $C_{an}$ have uncertainty comparable to these upper limits on $\delta_N$.}
Hereafter we simply ignore this term and marginalize over the parameters $(V_{u,d})_{22}$ and $(V_{u,d})_{33}$.
After this simplification, the nucleon couplings can be described by two parameters $\sin\beta$ and $(V_{u,d})_{11}$,
where $(V_{u,d})_{11}$ take some values within the range of $0$ to $2$.

\begin{figure}[htbp]
\centering
\includegraphics[width=160mm]{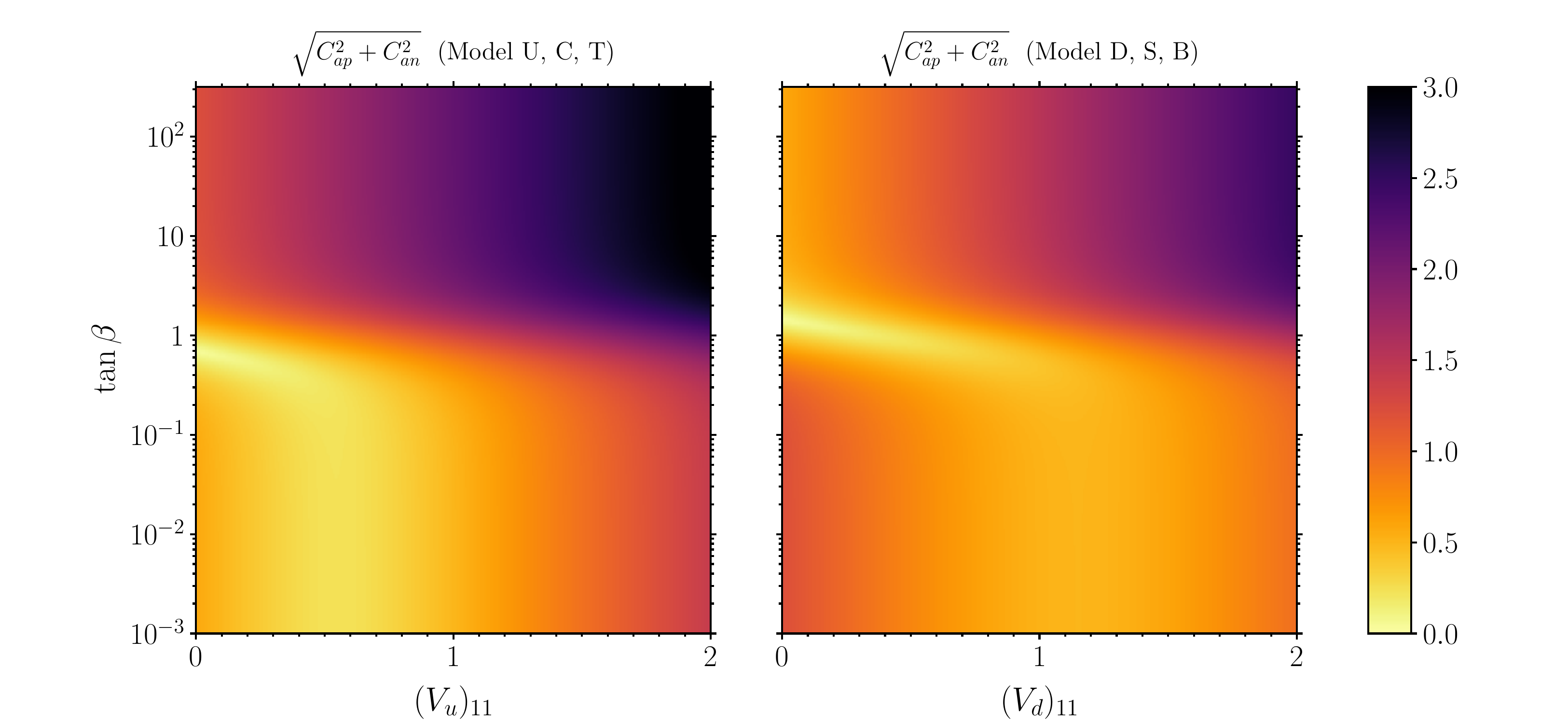}
\caption{
The value of $\sqrt{C_{ap}^2+C_{an}^2}$ based on Eqs.~\eqref{C_ap_VA} and~\eqref{C_an_VA} 
in the $(V_u)_{11}$-$\tan\beta$ plane for Model U,\,C,\,T (left panel)
and in the $(V_d)_{11}$-$\tan\beta$ plane for Model D,\,S,\,B (right panel).
}
\label{fig:nucleon_couplings}
\end{figure}

In Fig.~\ref{fig:nucleon_couplings}, we show the value of the combined quantity $\sqrt{C_{ap}^2 + C_{an}^2}$,
which is relevant to the bound from SN 1987A observations [Eq.~\eqref{SN_bound}].
Since this quantity depends on the new parameter $(V_{u,d})_{11}$ in addition to $\tan\beta$,
it shows a non-trivial structure in the two dimensional parameter space.
In particular, the nucleon couplings become much smaller than unity around $\tan\beta \sim \mathcal{O}(1)$ and $(V_{u,d,})_{11}\sim 0$.
To be more precise, $\sqrt{C_{ap}^2 + C_{an}^2}$ takes a minimum value of 0.043 at $\tan\beta \simeq 0.70$ and $(V_u)_{11} \simeq 0$
for Model U,\,C,\,T, and 0.042 at $\tan\beta \simeq 1.5$ and $(V_d)_{11} \simeq 0$ for Model D,\,S,\,B.
The fact that there exists a parameter region where the nucleon couplings get suppressed will play a role
in the interpretation of observational results described in the next section.\footnote{It is possible to arrange 
the model such that the nucleon couplings are further suppressed, $C_{ap} \approx C_{an} \approx 0$~\cite{DiLuzio:2017ogq}.
Indeed, this is the case for Model D,\,S,\,B if we take account of the small correction $\delta_N$ in addition to the dependence on $\tan\beta$ and $(V_d)_{11}$.
Note that, however, such a cancellation requires some tuning of the additional parameters $(V_d)_{22}$ and $(V_d)_{33}$.
In this work, we do not consider such a tuned case and focus on the general consequences following from non-vanishing nucleon couplings.} 

Finally, it might be worthwhile to compare the axion couplings in the variant axion models derived above
with those in the DFSZ axion models~\cite{Zhitnitsky:1980tq,Dine:1981rt}.
The DFSZ models are built based on the generation-independent PQ charge assignments for the SM quarks and leptons,
and we can consider two possibilities according to the following Yukawa interaction terms,
\begin{align}
\mathcal{L}_{\mathrm{Yukawa}} &= \mathcal{L}_{\mathrm{Yukawa},q} + \mathcal{L}_{\mathrm{Yukawa},\ell}, \nonumber\\
-\mathcal{L}_{\mathrm{Yukawa},q} &= \Gamma^d_{ij}\overline{q}_{iL}H_1 d_{jR} + \Gamma^u_{ij}\overline{q}_{iL}\tilde{H}_2 u_{jR}, \nonumber\\
-\mathcal{L}_{\mathrm{Yukawa},\ell} &= \left\{
\begin{array}{ll}
\Gamma^{\ell}_{ij}\overline{\ell}_{iL}H_1 e_{jR} + \mathrm{h.c.} &
\text{(DFSZ I)}, 
\vspace{1mm}\\
\Gamma^{\ell}_{ij}\overline{\ell}_{iL}H_2 e_{jR} + \mathrm{h.c.} &
\text{(DFSZ II)}.
\end{array}
\right.
\end{align}
These models lead to the following photon and electron couplings at low energies~\cite{Srednicki:1985xd,diCortona:2015ldu}
\begin{align}
&C_{a\gamma}^{\mathrm{DFSZ\ I}} = \frac{8}{3} - 1.92(4), \quad C_{a\gamma}^{\mathrm{DFSZ\ II}} = \frac{2}{3} - 1.92(4), \\
&C_{ae}^{\mathrm{DFSZ\ I}} = \frac{1}{3}\sin^2\beta, \quad C_{ae}^{\mathrm{DFSZ\ II}} = -\frac{1}{3}\cos^2\beta.
\end{align}
Comparing with Table~\ref{tab:axion_electron_couplings}, we see that $|C_{ae}|$ in Type I (Type II) of the variant axion models
is a factor three larger than that of DFSZ I (DFSZ II).
Furthermore, the nucleon couplings in the DFSZ models are given by~\cite{diCortona:2015ldu}
\begin{align}
C_{ap}^{\mathrm{DFSZ}} &= -0.182 - 0.435\sin^2\beta \pm 0.025,\\
C_{an}^{\mathrm{DFSZ}} &= -0.160 + 0.414\sin^2\beta \pm 0.025.
\end{align}
In contrast with the variant axion models, where the value of $\sqrt{C_{ap}^2 + C_{an}^2}$ becomes as small as $\sim 0.04$ up to the choice of parameters,
in the DFSZ models the suppression is milder, $\sqrt{C_{ap}^2 + C_{an}^2} \gtrsim 0.24$.
These differences lead to different consequences on the parameter regions hinted by the stellar cooling observations, 
as we discuss in the following section.

%%%%%%%%%%%%%%%%%%%%%%%%%%%%%%%%%%%%%%%%%%%%%%%%%%
\section{Interpretation of stellar cooling anomalies}
\label{sec:interpretation_of_anomalies}
\setcounter{equation}{0}
%%%%%%%%%%%%%%%%%%%%%%%%%%%%%%%%%%%%%%%%%%%%%%%%%%

Based on the axion couplings to ordinary matter derived in the previous section, 
we now investigate the parameter region where the variant axion models provide good fits for the stellar cooling hints.
See Appendix~\ref{app:analysis_method} for the details of the analysis method.

In addition to the stellar cooling hints, we take account of the constraint from SN 1987A.
In order to see how the SN 1987A bound affects the hinted parameter space,
we first perform the global fits without including it, and add it afterwards.
Following the procedure in Ref.~\cite{Giannotti:2017hny}, we include the SN 1987A bound by taking it as
a $1\sigma$ hint of $g_{ap}^2 + g_{an}^2 = 0$ with the error specified by the right-hand side of Eq.~\eqref{SN_bound}.
On the other hand, we do not include the bounds or hints obtained by the observation of NSs, 
since there remain several controversies on the interpretation of observational results
as mentioned in Sec.~\ref{sec:stellar_cooling}.

In the variant axion models, the axion-nucleon couplings depend on the new parameters $(V_{u,d})_{11}$
[see Eqs.~\eqref{C_ap_VA} and~\eqref{C_an_VA}],
and their values affect the constraint from SN 1987A.
These parameters arise from the mixings of right-handed up-type and down-type quarks.
In principle they can take any values within the range of 0 to 2.
On the other hand, it is also possible to guess their values if we make some assumption about the UV-completion of the models.
In Sec.~\ref{sec:general_mixings} we first consider the case of general quark mixings where $(V_{u,d})_{11}$
are treated as extra free parameters.
After that, in Sec.~\ref{sec:small_mixings} we consider more specific cases where the parameters $(V_{u,d})_{11}$
are fixed to some well-motivated values.

%%%%%%%%%%%%%%%%%%%%%%%%%%%%%%%%%%%%%%%%%%%%%%%%%%
\subsection{Models with general quark mixings}
\label{sec:general_mixings}
%%%%%%%%%%%%%%%%%%%%%%%%%%%%%%%%%%%%%%%%%%%%%%%%%%

\begin{figure}[htbp]
\centering
\includegraphics[width=110mm]{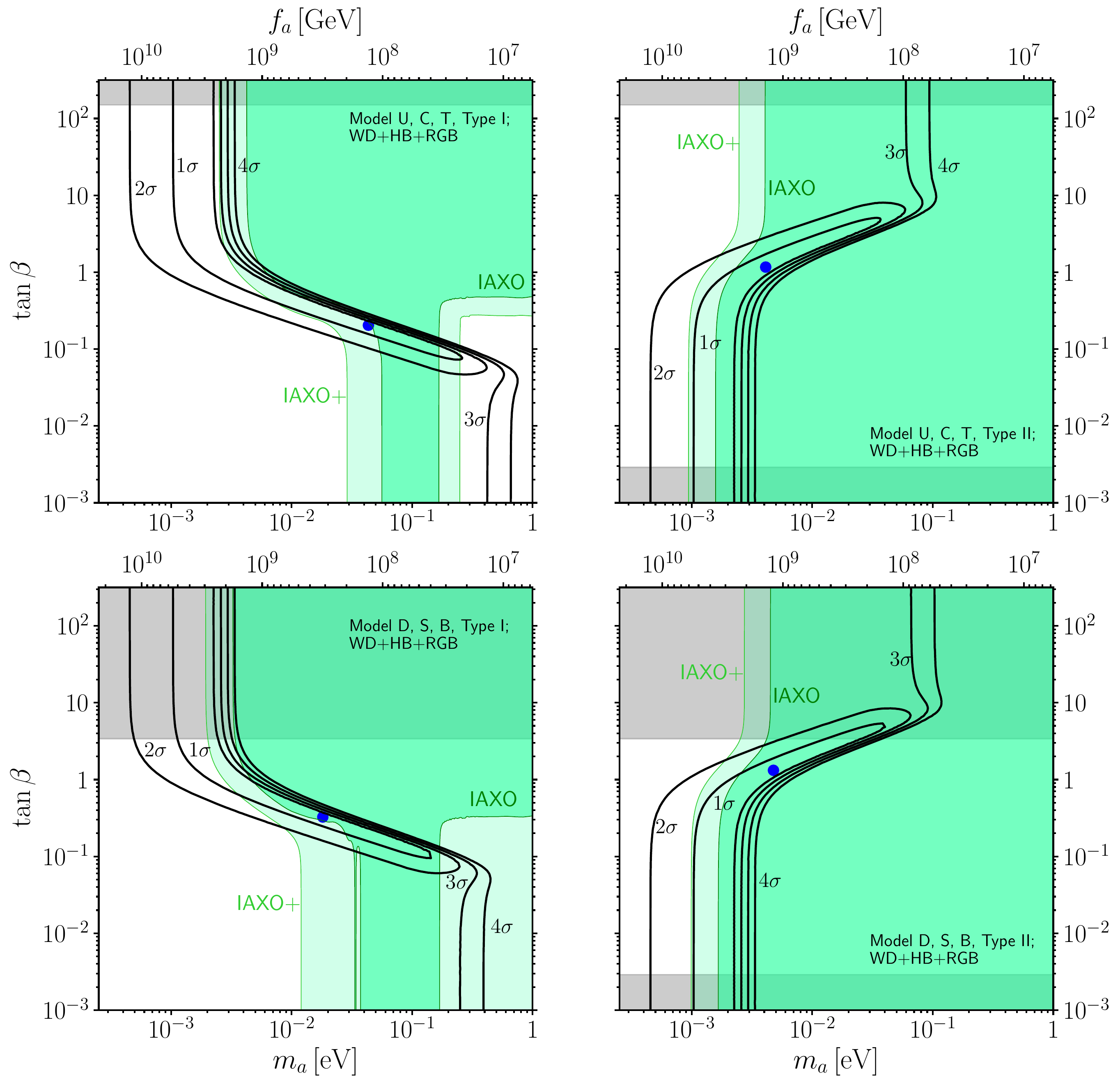}
\caption{
1, 2, 3, 4\,$\sigma$ contours in the $m_a$-$\tan\beta$ plane
from a fit to the data of WD, HB, and RGB cooling observations
for Model U,\,C,\,T, Type I (top left), Model U,\,C,\,T, Type II (top right), 
Model D,\,S,\,B, Type I (bottom left), and Model D,\,S,\,B, Type II (bottom right).
Blue dots represent the best fit parameters shown in Table~\ref{tab:best_fit_parameters}.
Gray regions correspond to the parameter space 
outside a typical range compatible with perturbativity of Yukawa interactions [Eq.~\eqref{range_tanbeta}].
Projected sensitivities of IAXO (green) and IAXO+ (light green) are also shown.
}
\label{fig:contour_noSN}
\end{figure}

If we do not consider the SN 1987A bound, the observational results can be explained in terms of
the axion-photon coupling $g_{a\gamma}$ and axion-electron coupling $g_{ae}$.
In the variant axion models these couplings are specified by two parameters $f_a$ (or $m_a$) and $\tan\beta$,
in a similar manner to the DFSZ models.
Figure~\ref{fig:contour_noSN} shows 1, 2, 3, 4\,$\sigma$ hinted regions in the parameter space of $m_a$ and $\tan\beta$
obtained based on the data of WD, HB, and RGB cooling anomalies.
Here we also show the projected sensitivities of IAXO and its upgrades (IAXO+)~\cite{Giannotti:2017hny,Armengaud:2019uso}.
Furthermore, the best fit parameter values are shown in blue dots, and they are also summarized in Table~\ref{tab:best_fit_parameters}.
Note that there exist upper and/or lower limits on the value of $\tan\beta$ due to the requirement of perturbativity of Yukawa interactions
(see Appendix~\ref{app:yukawa_interactions}), and they are shown as gray shaded regions.

\begin{table}[ht]\centering
\caption{
Best fit parameters and $\chi^2_{\rm min}/\text{d.o.f.}$ for the interpretation of the stellar cooling anomalies in the variant axion models
with general quark mixing.
The column of $(V_{u,d})_{11}$ shows the value of $(V_u)_{11}$ for Model U,\,C,\,T and that of $(V_d)_{11}$ for Model D,\,S,\,B.
}
\small
\vspace{-\baselineskip}
\begin{align}
\begin{array}{|l|l|l|c|c|c|c|c|}
\hline
\text{Model} & \text{Type} & \text{Global fit includes} & f_a\,[10^8\,\mathrm{GeV}] & m_a\,[\mathrm{meV}] & \tan\beta & (V_{u,d})_{11} & \chi^2_{\rm min}/\text{d.o.f.}\\
\hline
\text{Model U,\,C,\,T} & \text{Type I} & \text{WD,HB,RGB} & 1.3 & 43 & 0.20 & & 14.7/15 \\
 & & \text{WD,HB,RGB,SN} & 11 & 5.4 & 0.69 & 0 & 14.7/15 \\
 & \text{Type II} & \text{WD,HB,RGB} & 14 & 4.1 & 1.2 & & 14.7/15 \\
 & & \text{WD,HB,RGB,SN} & 22 & 2.5 & 0.68 & 0 & 14.7/15 \\
\hline
\text{Model D,\,S,\,B} & \text{Type I} & \text{WD,HB,RGB} & 3.1 & 18 & 0.33 & & 14.7/15 \\
 & & \text{WD,HB,RGB,SN} & 22 & 2.6 & 1.4 & 0 & 14.7/15 \\
 & \text{Type II} & \text{WD,HB,RGB} & 12 & 4.8 & 1.3 & & 14.7/15 \\
 & & \text{WD,HB,RGB,SN} & 11 & 5.2 & 1.4 & 0 & 14.7/15 \\
 \hline
\end{array}
\nonumber
\end{align}
\label{tab:best_fit_parameters}
\end{table}

Since most of the data prefer the electron coupling of $g_{ae} \sim 1.6 \times 10^{-13}$,
the overall shape of 1, 2, 3, 4\,$\sigma$ contours in Fig.~\ref{fig:contour_noSN} can be understood by considering the parameter dependence of $g_{ae}$.
For Type I (Type II) of variant axion models the electron coupling is given by $|g_{ae}| = m_e\sin^2\beta/f_a$ ($m_e\cos^2\beta/f_a$),
and the contours become vertical at large (small) $\tan\beta$ since the coefficient becomes constant, $|C_{ae}| \simeq 1$ in that region.
On the other hand, at small (large) $\tan\beta$ in Type I (Type II) models the coefficient $|C_{ae}|$ becomes suppressed and smaller values of $f_a$
are preferred in order to maintain the desired value of $g_{ae}$.
This behavior ends when the value of $f_a$ becomes too small to make the photon coupling $g_{a\gamma}$ compatible with
the constraint from the $R$-parameter.
If we fix the type of the electron coupling and do not include the bound from SN 1987A, 
the difference between Model U,\,C,\,T and Model D,\,S,\,B is minor and only affects the behavior at higher $m_a$,
which essentially comes from a difference in the model-dependent coefficient $\mathcal{E}/\mathcal{N}$
for the axion-photon coupling (see Table~\ref{tab:axion_electron_couplings}).

We note that the parameter range accessible to future helioscope experiments is also model-dependent.
The sensitivity of IAXO for a generic axion model is given by~\cite{Giannotti:2017hny}
\begin{equation}
g_{a\gamma} \ge g_{a\gamma}^{\rm min} = d_1 \left(1+\left(\frac{d_1}{d_2}\right)^4\zeta^2\right)^{-1/4},
\end{equation}
where $g_{a\gamma}^{\rm min}$ is the minimal value of $g_{a\gamma}$ accessible to IAXO,
$d_1$ is the data of the minimal value of $g_{a\gamma}$ accessible to IAXO if solar axions are produced 
solely through the processes involving $g_{a\gamma}$, and
$d_2$ is the data of the minimal value of $\sqrt{g_{a\gamma}g_{ae}}$ accessible to IAXO if solar axions are produced 
solely through the processes involving $g_{ae}$. $\zeta$ is a model-dependent dimensionless factor defined as
\begin{equation}
\zeta = \left|\frac{g_{ae}}{g_{a\gamma}}\right|
\end{equation}
with $g_{a\gamma}$ specified in units of $\mathrm{GeV}^{-1}$.
We summarize the values of $\zeta$ in the variant axion models in Table~\ref{tab:IAXO_potential}.

If $\zeta$ is sufficiently small, the production of solar axions through the interaction with electrons becomes irrelevant.
This case corresponds to the limit of $\tan\beta \ll 1$ ($\tan\beta \gg 1$) in Type I (Type II) variant axion models.
In this case, the sensitivity is simply given by $g_{a\gamma}^{\rm min} \simeq d_1$, from which we can define minimal mass ranges
accessible to IAXO regardless of the value of $\tan\beta$. Such mass ranges are shown in Table~\ref{tab:IAXO_potential}.
On the other hand, if $\zeta$ becomes sufficiently large, which corresponds to the limit of $\tan\beta \gg 1$ ($\tan\beta \ll 1$) in Type I (Type II) variant axion models,
the sensitivity is improved as $g_{a\gamma}^{\rm min} \simeq d_2/\sqrt{\zeta}$.
In any case, the experimental sensitivity becomes improved if a model predicts a higher value of the photon coupling $|C_{a\gamma}|$.
Since the value of $|C_{a\gamma}|$ in Type II models is larger than that in Type I models (see Table~\ref{tab:axion_electron_couplings}),
a wider mass range can be probed for the former models compared to the latter ones.

\begin{table}[ht]\centering
\caption{The coefficient $\zeta$ for the estimation of experimental sensitivities and minimal axion mass ranges accessible to IAXO
and IAXO+ in the variant axion models.}
\vspace{-\baselineskip}
\begin{align}
\begin{array}{|l|l|c|c|c|}
\hline
\text{Model} & \text{Type} & \zeta & \text{Mass range (IAXO)} &  \text{Mass range (IAXO+)} \\
\hline
\text{Model U,\,C,\,T} & \text{Type I} & 0.59\sin^2\beta & 0.056\,\mathrm{eV} \lesssim m_a \lesssim 0.17\,\mathrm{eV} & 0.029\,\mathrm{eV} \lesssim m_a \lesssim 0.25\,\mathrm{eV} \\
 & \text{Type II} & 0.08\cos^2\beta & 0.0041\,\mathrm{eV} \lesssim m_a & 0.0025\,\mathrm{eV} \lesssim m_a \\
 \hline
\text{Model D,\,S,\,B} & \text{Type I} & 0.35\sin^2\beta & 0.033\,\mathrm{eV} \lesssim m_a \lesssim 0.17\,\mathrm{eV} & 0.012\,\mathrm{eV} \lesssim m_a \\
 & \text{Type II} & 0.09\cos^2\beta & 0.0045\,\mathrm{eV} \lesssim m_a & 0.0027\,\mathrm{eV} \lesssim m_a \\
\hline
\end{array}
\nonumber
\end{align}
\label{tab:IAXO_potential}
\end{table}

\begin{figure}[htbp]
\centering
\vspace{-5mm}
$\begin{array}{c}
\subfigure{
\includegraphics[width=100mm]{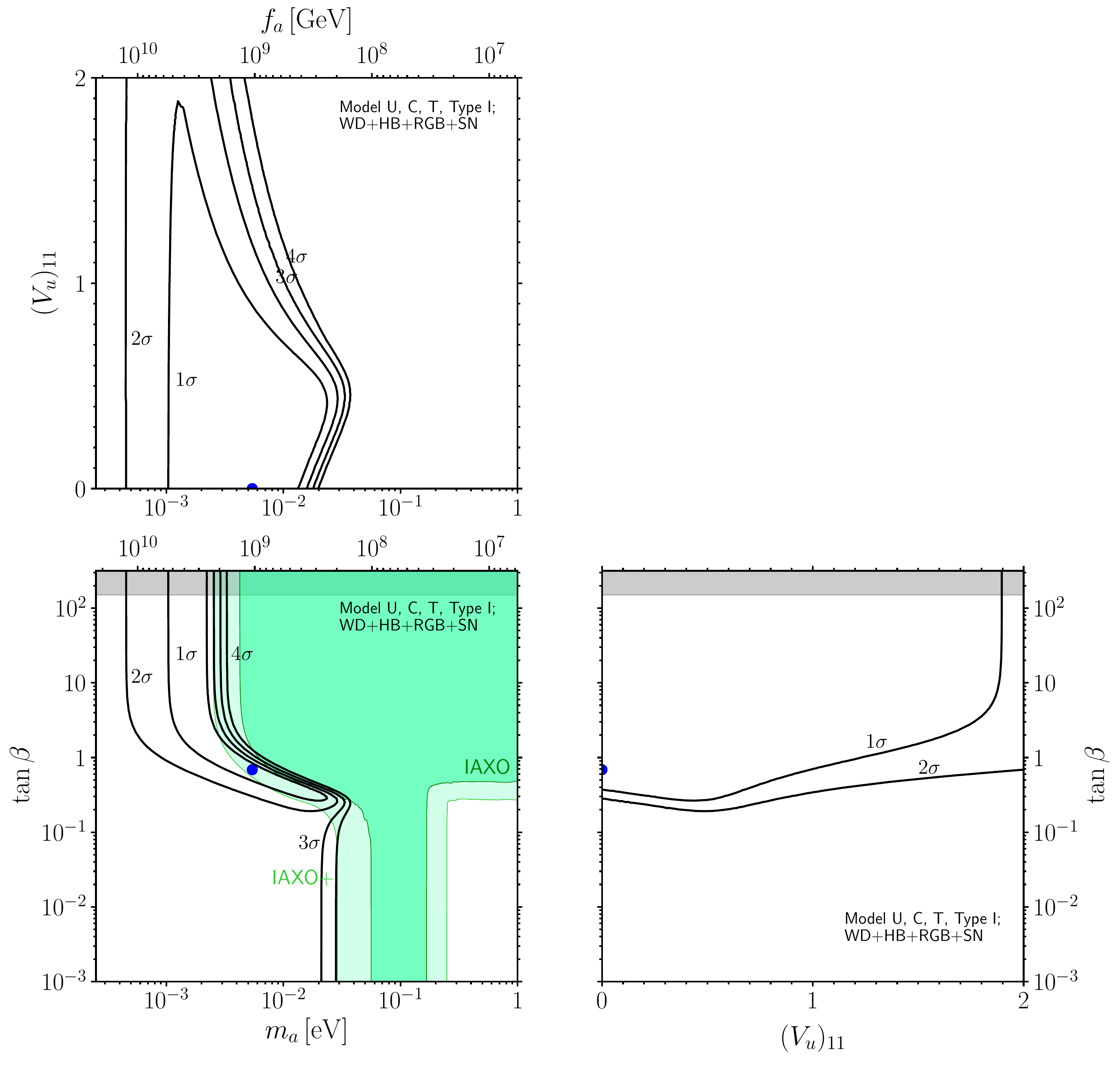}}
\vspace{2mm}
\\
\subfigure{
\includegraphics[width=100mm]{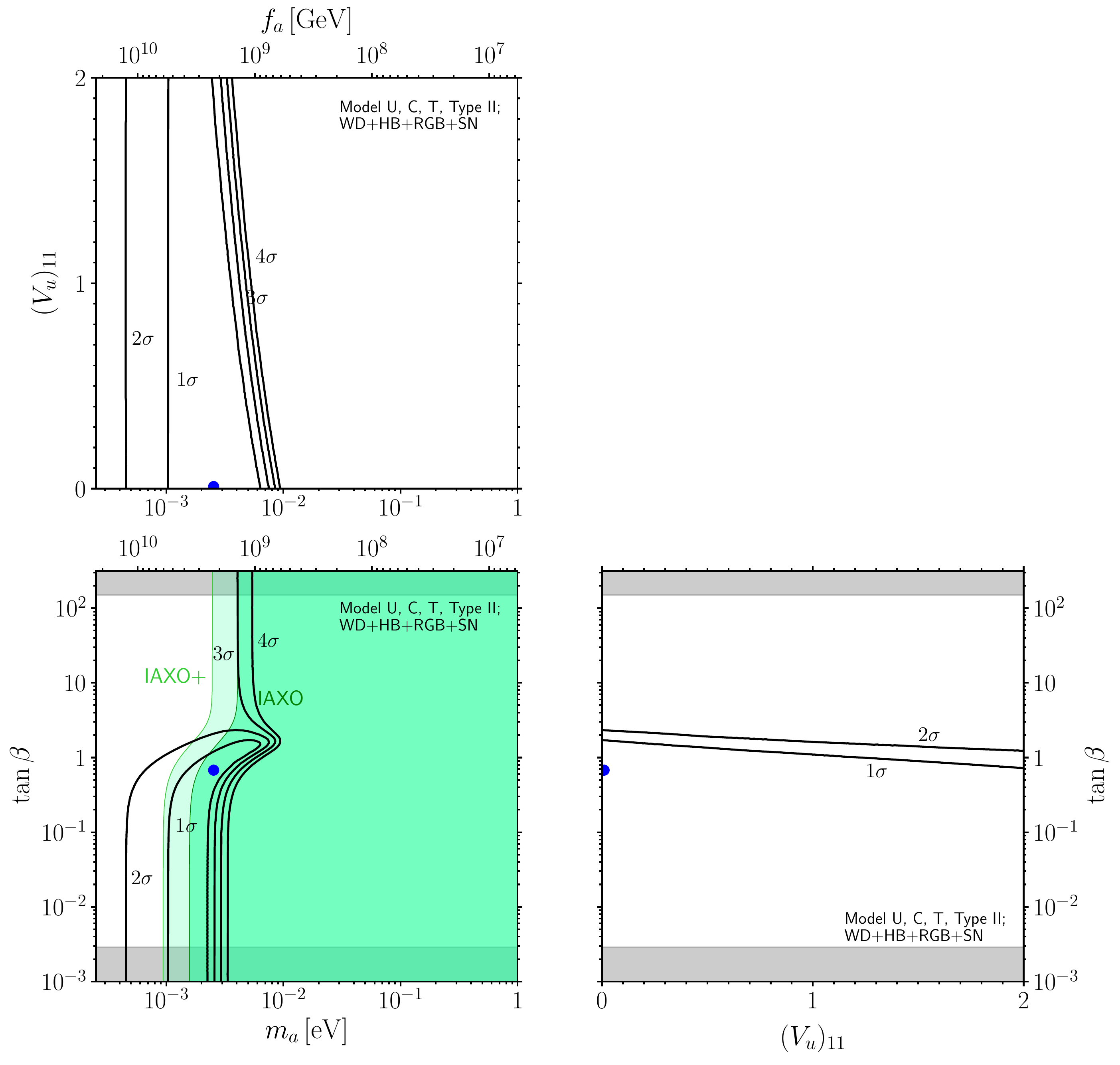}}
\end{array}$
\caption{
1, 2, 3, 4\,$\sigma$ contours 
from a fit to the data including SN 1987A in addition to WD, HB, and RGB cooling observations
for Model U,\,C,\,T, Type I (top panels) and Type II (bottom panels).
Each figure shows a projection to two dimensional parameter space from three dimensional parameter space
of $m_a$, $\tan\beta$, and $(V_u)_{11}$, where the contours are obtained by minimizing over the third parameter.
Blue dots represent the best fit parameters shown in Table~\ref{tab:best_fit_parameters}.
Gray regions correspond to the parameter space 
outside a typical range compatible with perturbativity of Yukawa interactions [Eq.~\eqref{range_tanbeta}].
Projected sensitivities of IAXO (green) and IAXO+ (light green) are also shown on the plot in the $m_a$-$\tan\beta$ plane.
}
\label{fig:contour_UCT_SN}
\end{figure}

\begin{figure}[htbp]
\centering
\vspace{-5mm}
$\begin{array}{c}
\subfigure{
\includegraphics[width=100mm]{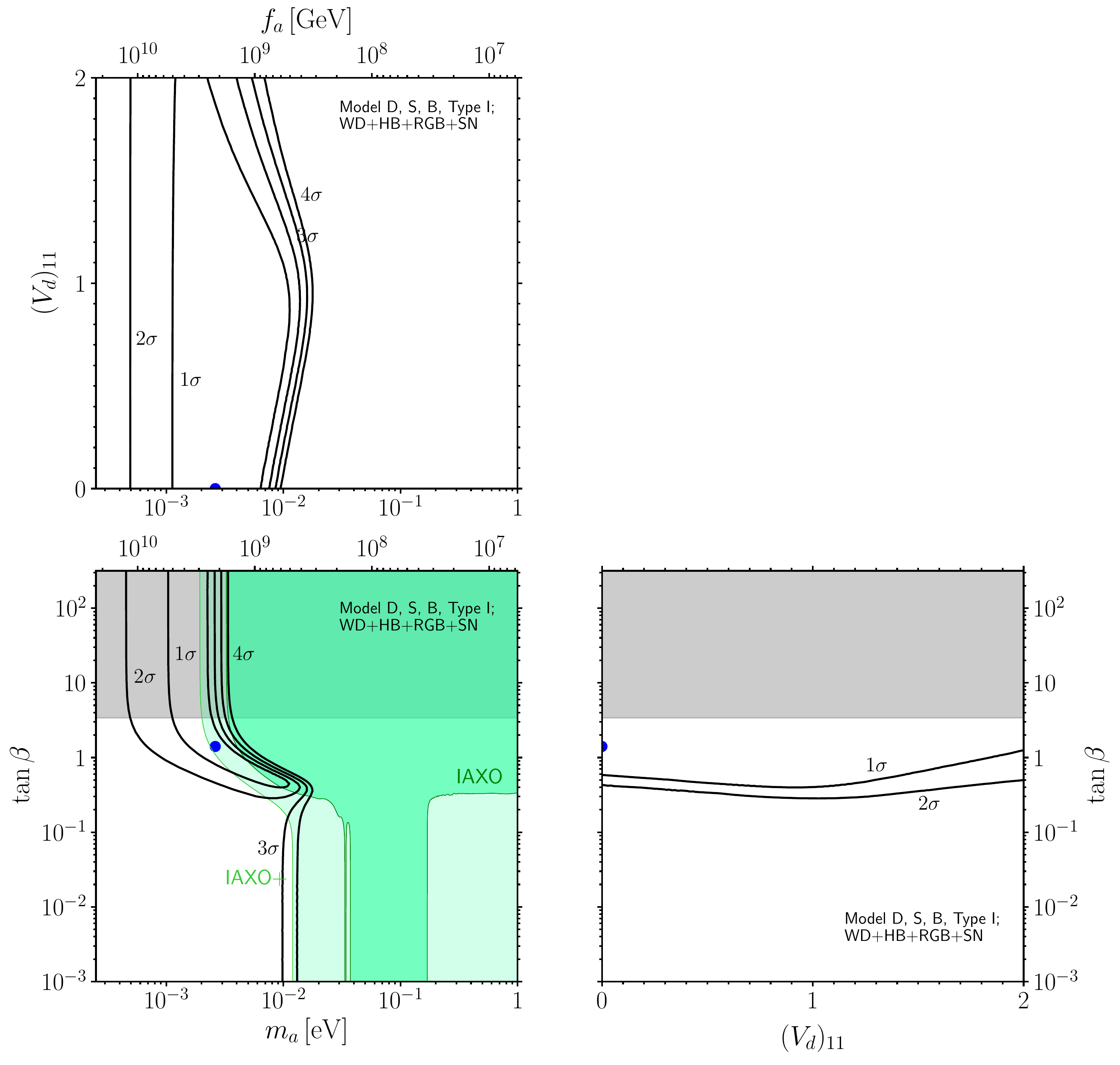}}
\vspace{5mm}
\\
\subfigure{
\includegraphics[width=100mm]{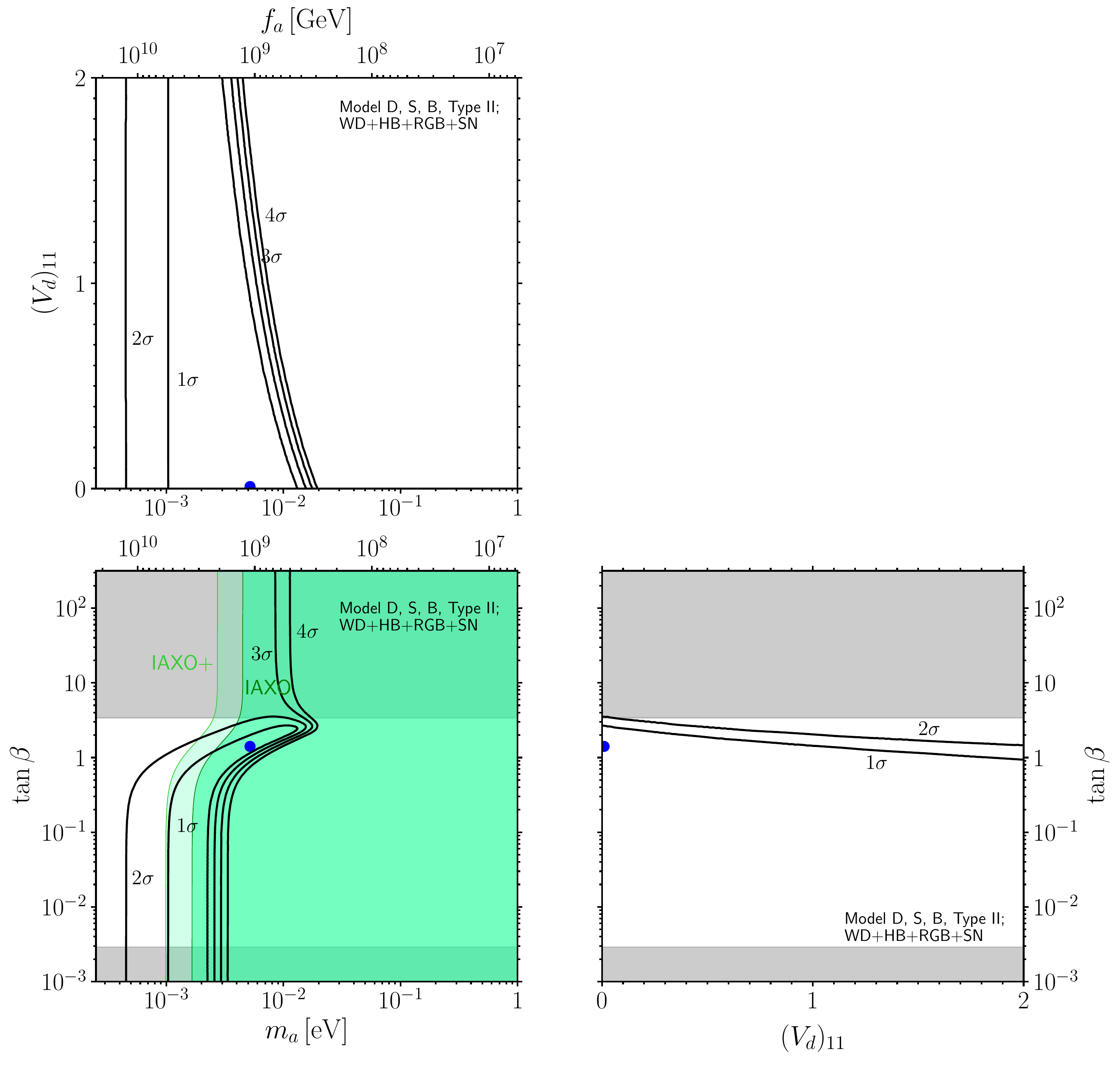}}
\end{array}$
\caption{
The same figure as Fig.~\ref{fig:contour_UCT_SN} but 1, 2, 3, 4\,$\sigma$ contours, constraints from the requirement of 
perturbativity of Yukawa interactions, and projected sensitivities of IAXO and IAXO+ are
plotted for Model D,\,S,\,B, Type I (top panels) and Type II (bottom panels).
Each figure shows a projection to two dimensional parameter space from three dimensional parameter space
of $m_a$, $\tan\beta$, and $(V_d)_{11}$, where the contours are obtained by minimizing over the third parameter.
}
\label{fig:contour_DSB_SN}
\end{figure}

Now, let us include the bound from SN 1987A.
As nucleon couplings depend on new parameters $(V_{u,d})_{11}$, here we scan over three dimensional parameter space of
$m_a$, $\tan\beta$, and $(V_{u,d})_{11}$.
Since we add an extra data point for the SN constraint, the number of d.o.f. remains unchanged.
The best fit parameters and the values of $\chi^2_{\rm min}/\text{d.o.f.}$ for the three dimensional scans
are shown in Table~\ref{tab:best_fit_parameters}.
In Figs.~\ref{fig:contour_UCT_SN} and~\ref{fig:contour_DSB_SN}, we also show the projection of hinted regions to two dimensional parameter space 
$(a,b) = (m_a,\tan\beta)$, $(m_a,(V_{u,d})_{11})$, or $((V_{u,d})_{11},\tan\beta)$ 
by computing $\Delta\chi^2(a,b)$, which is obtained by minimizing $\Delta\chi^2(a,b,c) = \chi^2(a,b,c)-\chi^2_{\rm min}$ over the third parameter $c$.
Comparing the plots on the $m_a$-$\tan\beta$ plane
with those for the results without including the SN data (Fig.~\ref{fig:contour_noSN}), 
we see that every model still shows a good fit to the observed data in the lower mass ranges,
while the higher mass ranges become incompatible with the SN bound.

The plots for the $m_a$-$(V_{u,d})_{11}$ plane in Figs.~\ref{fig:contour_UCT_SN} and~\ref{fig:contour_DSB_SN}
show that the value of $(V_{u,d})_{11}$ does not have significant effects on the fits at lower mass ranges,
where the bound from SN 1987A becomes less important.
On the other hand, smaller values of $(V_{u,d})_{11}$ are preferred at higher mass ranges,
since the SN bound becomes important but gets relaxed due to the reduction of the nucleon couplings 
for smaller values of $(V_{u,d})_{11}$ (see Fig.~\ref{fig:nucleon_couplings}).
To be more precise, in order to suppress the nucleon couplings the value of $\tan\beta$ should also be adjusted,
which results in some model dependencies.
In Type I models, smaller values of $\tan\beta$ are required at higher mass ranges in order to keep $g_{ae}$ close to the value hinted by the cooling anomalies,
and for such smaller $\tan\beta$ values a non-vanishing value of $(V_{u,d})_{11}$ is slightly preferred over $(V_{u,d})_{11}=0$ as
it leads to smaller values of the nucleon couplings in the small $\tan\beta$ region.
This can be contrasted with the case of Type II models, where larger values of $\tan\beta$ are required at higher mass ranges 
and $(V_{u,d})_{11}=0$ is preferred such that the nucleon couplings become the smallest in the large $\tan\beta$ region.

The shape of the contours in the $(V_{u,d})_{11}$-$\tan\beta$ plane in Figs.~\ref{fig:contour_UCT_SN} and~\ref{fig:contour_DSB_SN}
can also be understood in terms of the hinted value of $g_{ae}$ and the parameter dependence of the nucleon couplings.
In Type I models, the large $\tan\beta$ region is preferred by WD, HB, and RGB cooling observations,
and in the lower end of such a region the SN bound becomes relevant, which exhibits a slight preference for a non-vanishing value of $(V_{u,d})_{11}$.
On the other hand, in Type II models, the small $\tan\beta$ region is preferred, and in the upper end of such a region
$(V_{u,d})_{11}$ is forced to be zero in order to alleviate the SN bound by reducing the nucleon couplings.

\begin{figure}[htbp]
\centering
\includegraphics[width=110mm]{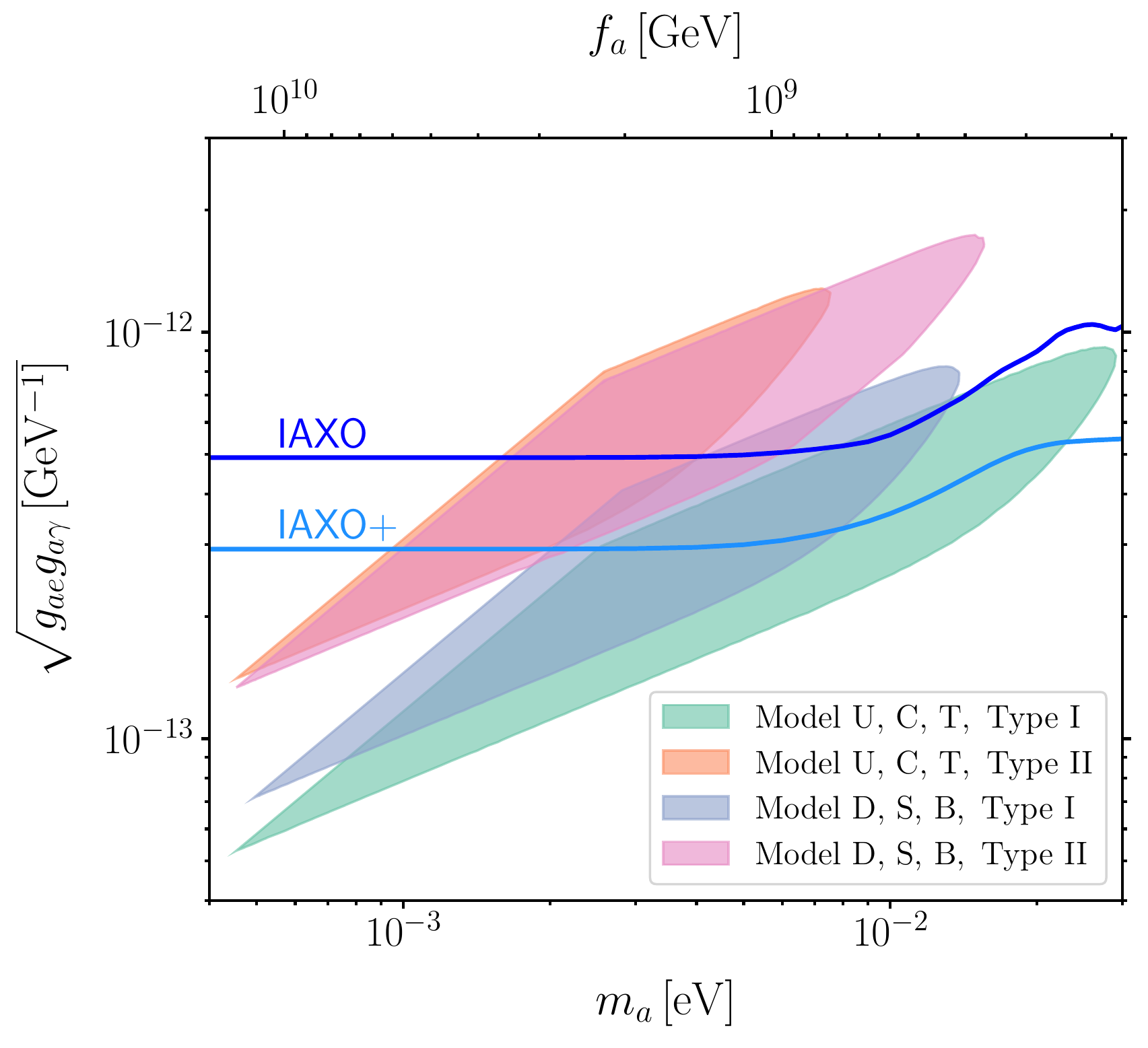}
\caption{
Parameter regions in the plane of the axion mass and the coupling product $\sqrt{g_{ae}g_{a\gamma}}$,
corresponding to the 2\,$\sigma$ ranges of WD, HB, and RGB cooling anomalies 
compatible with the bound from SN 1987A and the requirement of perturbativity of Yukawa interactions
for four different cases of the variant axion models.
Projected sensitivities of IAXO and IAXO+ are also shown.
}
\label{fig:model_predictions}
\end{figure}

In Fig.~\ref{fig:model_predictions}, we show the 2\,$\sigma$ predicted regions 
in the parameter space of $m_a$ and $\sqrt{g_{ae}g_{a\gamma}}$ together with the projected sensitivities of
IAXO and IAXO+. Numerical values of the axion mass and couplings
for the corresponding ranges are also summarized in Table~\ref{tab:predicted_ranges}.
We see that in any cases the lower end of the predicted mass ranges lie around $m_a \sim 0.45\,\mathrm{meV}$.
This region corresponds to the limit of $\tan\beta \gg 1$ ($\tan\beta \ll 1$) in Type I (Type II) models,
where the electron coupling becomes $|g_{ae}| \simeq m_e/f_a \simeq 4\times 10^{-14}\,(m_a/0.45\,\mathrm{meV})$
regardless of the value of $\tan\beta$, and hence the hinted value of $|g_{ae}|$ fixes the value of $m_a$.

\begin{table}[ht]\centering
\caption{2\,$\sigma$ ranges in the axion mass and couplings to electrons, photons, neutrons, and protons,
hinted by WD, HB, and RGB cooling anomalies
and compatible with the bound from SN 1987A and the requirement of perturbativity of Yukawa interactions
for different cases of the variant axion models.}
\small
\vspace{-\baselineskip}
\begin{align}
\begin{array}{|l|l|c|c|c|c|c|}
\hline
\text{Model} & \text{Type} & m_a\,[\mathrm{meV}] & g_{ae}\,[10^{-13}] & g_{a\gamma}\,[10^{-12}\,\mathrm{GeV}^{-1}] & g_{an}\,[10^{-10}] & g_{ap}\,[10^{-10}] \\
\hline
\text{Model U,\,C,\,T} & \text{Type I} & (0.46,\,29) & (0.40,\,2.3) & (0.069,\,4.4) & (-8.6,\,3.4) & (-3.8,\,-0.049) \\
 & \text{Type II} & (0.45,\,7.5) & (-2.3,\,-0.41) & (-8.1,\,-0.49) & (-0.91,\,6.1) & (-7.2,\,-0.021) \\
 \hline
\text{Model D,\,S,\,B} & \text{Type I} & (0.49,\,14) & (-2.3,\,-0.41) & (-3.5,\,-0.13) & (-3.8,\,-0.022) & (-8.6,\,1.3) \\
 & \text{Type II} & (0.45,\,16) & (0.40,\,2.3) & (0.44,\,15) & (-7.2,\,-0.043) & (-2.0,\,6.0) \\
\hline
\text{Model U} & \text{Type I} & (0.45,\,16) & (0.41,\,2.3) & (0.069,\,2.4) & (-7.3,\,3.4) & (-3.8,\,6.1) \\
\text{(small mixings)} & \text{Type II} & (0.45,\,7.6) & (-2.3,\,-0.41) & (-8.1,\,-0.49) & (-1.8,\,6.2) & (-7.2,\,1.7) \\
 \hline
\text{Model D} & \text{Type I} & (0.45,\,7.6) & (-2.3,\,-0.41) & (-1.9,\,-0.12) & (-1.8,\,6.2) & (-7.2,\,1.7) \\
\text{(small mixings)} & \text{Type II} & (0.45,\,16) & (0.40,\,2.3) & (0.44,\,15) & (-7.2,\,3.4) & (-3.8,\,6.0) \\
\hline
\end{array}
\nonumber
\end{align}
\label{tab:predicted_ranges}
\end{table}

From Fig.~\ref{fig:model_predictions}, we also see that the higher end of the predicted mass ranges becomes different according to the models.
This difference can be understood as follows:
First, we note that the contours in the $m_a$-$\tan\beta$ plane for Type II models (like those shown in Fig.~\ref{fig:contour_noSN})
should be obtained by flipping those for Type I models vertically with respect to the line of $\tan\beta = 1$ if
we only take care of the structure of the electron coupling.
This symmetric property along $\tan\beta$ direction is broken when we add the SN bound and consider the structure of the nucleon couplings.
From left panel of Fig.~\ref{fig:nucleon_couplings}, we see that in Model U,\,C,\,T the nucleon couplings become larger
at large $\tan\beta$ region, while they become smaller at small $\tan\beta$ region.
Therefore, the SN bound becomes weaker (stronger) for Type I (Type II) of Model U,\,C,\,T, where smaller (larger) values of $\tan\beta$
are favored at higher masses, which results in a less (more) stringent upper limit on the axion mass.
On the other hand, in Model D,\,S,\,B the asymmetry of the nucleon couplings along $\tan\beta$ direction is less significant at $(V_d)_{11} \simeq 0$
(see right panel of Fig.~\ref{fig:nucleon_couplings}), and the strength of the SN bound remains almost the same for Type I and Type II models.
This explains the fact that there is little difference in the upper limit of the predicted mass ranges between Type I and Type II of Model D,\,S,\,B.

In addition to the difference in the predicted mass ranges described above,
Fig.~\ref{fig:model_predictions} shows that the overall magnitude of the coupling product $g_{ae}g_{a\gamma}$ differs according to the models.
Such a difference can be understood in terms of the value of the coefficient $|C_{a\gamma}|$ of the axion-photon coupling (see Table~\ref{tab:axion_electron_couplings}).
For Type I models, some part of the predicted region is inaccessible to IAXO since the values of $|C_{a\gamma}|$ in these models are relatively small.
On the other hand, most part of the predicted parameter region is covered by the projected sensitivity of IAXO for Type II models 
as they predict larger values of $|C_{a\gamma}|$,
and hence we expect that these models can be tested by the forthcoming experiments.

Before closing this subsection, we note that the models 
can potentially be constrained by the observation of heavy meson decays~\cite{Gelmini:1982zz},
since they predict flavor-changing couplings with quarks given by the off-diagonal elements of the matrices $(V_{u,d})_{ij}$.
Currently the strongest constraint is obtained from a search for a decay $K^+ \to \pi^+ + a$~\cite{Adler:2008zza}, 
which leads to a bound~\cite{Bjorkeroth:2018dzu}
\begin{equation}
f_a/|(C^A_{ad})_{21}| > 3.5 \times 10^{11}\,\mathrm{GeV}. \label{Kaon_bound}
\end{equation}
This bound is trivially satisfied in Model U,\,C,\,T, since they predict $(C^A_{ad})_{21} = 0$ (see Table~\ref{tab:axion_quark_couplings}),
while Model D,\,S,\,B can have a non-vanishing value of $|(C^A_{ad})_{21}| = |(V_d)_{21}|$ and in such cases the bound could become relevant.
However, at this point we should treat $(V_d)_{21}$ as an extra free parameter in the same way as $(V_{u,d})_{11}$,
and cannot deduce a definite bound from Eq.~\eqref{Kaon_bound} unless we fix the pattern of quark mixings to obtain a specific value of $(V_d)_{21}$.

%%%%%%%%%%%%%%%%%%%%%%%%%%%%%%%%%%%%%%%%%%%%%%%%%%
\subsection{Models with small quark mixings}
\label{sec:small_mixings}
%%%%%%%%%%%%%%%%%%%%%%%%%%%%%%%%%%%%%%%%%%%%%%%%%%

So far we have considered the general cases of the variant axion models
where the mixings of the right-handed quarks are arbitrary.
Although such models already show good fits to the observational data,
it is enlightening to explore a more specific scenario of the quark mixings.
As a well-motivated possibility, 
here we consider a pattern of quark mixings 
inspired by the Frogatt-Nielsen mechanism~\cite{Froggatt:1978nt,Buchmuller:1998zf},
which was introduced to provide a natural explanation to 
the SM quark and lepton mass hierarchy.

In the Frogatt-Nielsen scenario, the flavor structure is related to 
a spontaneously broken global U(1)$_{\rm FN}$ symmetry.\footnote{Note that this global symmetry
is different from the PQ symmetry in our setup.
Some attempts to relate the U(1)$_{\rm FN}$ symmetry to the PQ symmetry
are found in Refs.~\cite{Ema:2016ops,Calibbi:2016hwq}.}
The Yukawa interactions of the SM quarks and leptons
originate from non-renormalizable interactions with a singlet field $\Phi$
suppressed by a cutoff scale $M$,
and the mass hierarchy of quarks and leptons is generated as
a power of $\varepsilon = \langle\Phi\rangle/M$
after the U(1)$_{\rm FN}$ symmetry is spontaneously broken and the singlet field acquires a VEV $\langle\Phi\rangle$.
The magnitude of fermion mixings is also specified as
$(U_{\psi_L})_{ij} \sim \varepsilon^{|Q_{\psi_{iL}}-Q_{\psi_{jL}}|}$ and 
$(U_{\psi_R})_{ij} \sim \varepsilon^{|Q_{\psi_{iR}}-Q_{\psi_{jR}}|}$~\cite{Froggatt:1978nt},
where $Q_{\psi_{iL,R}}$ are U(1)$_{\rm FN}$ charges of the fermions.
The observed quark and lepton masses imply $\varepsilon^2 \sim 1/300$.

Assuming that the flavor structure follows a pattern predicted by the Frogatt-Nielsen mechanism,
we can guess the magnitude of the parameters $(V_{u,d})_{11}$ appearing in the axion-nucleon couplings.
In particular, adopting the U(1)$_{\rm FN}$ charges specified in Ref.~\cite{Buchmuller:1998zf},
from Eqs.~\eqref{V_u_definition},~\eqref{V_d_definition}, and \eqref{Lambda_models} 
we obtain $(V_{u,d})_{11} \sim \mathcal{O}(\varepsilon^2)$ for Model U and D.
This fact motivates us to study the consequences of Model U and D
with a negligibly small value of $(V_{u,d})_{11}$.

\begin{figure}[htbp]
\centering
\includegraphics[width=110mm]{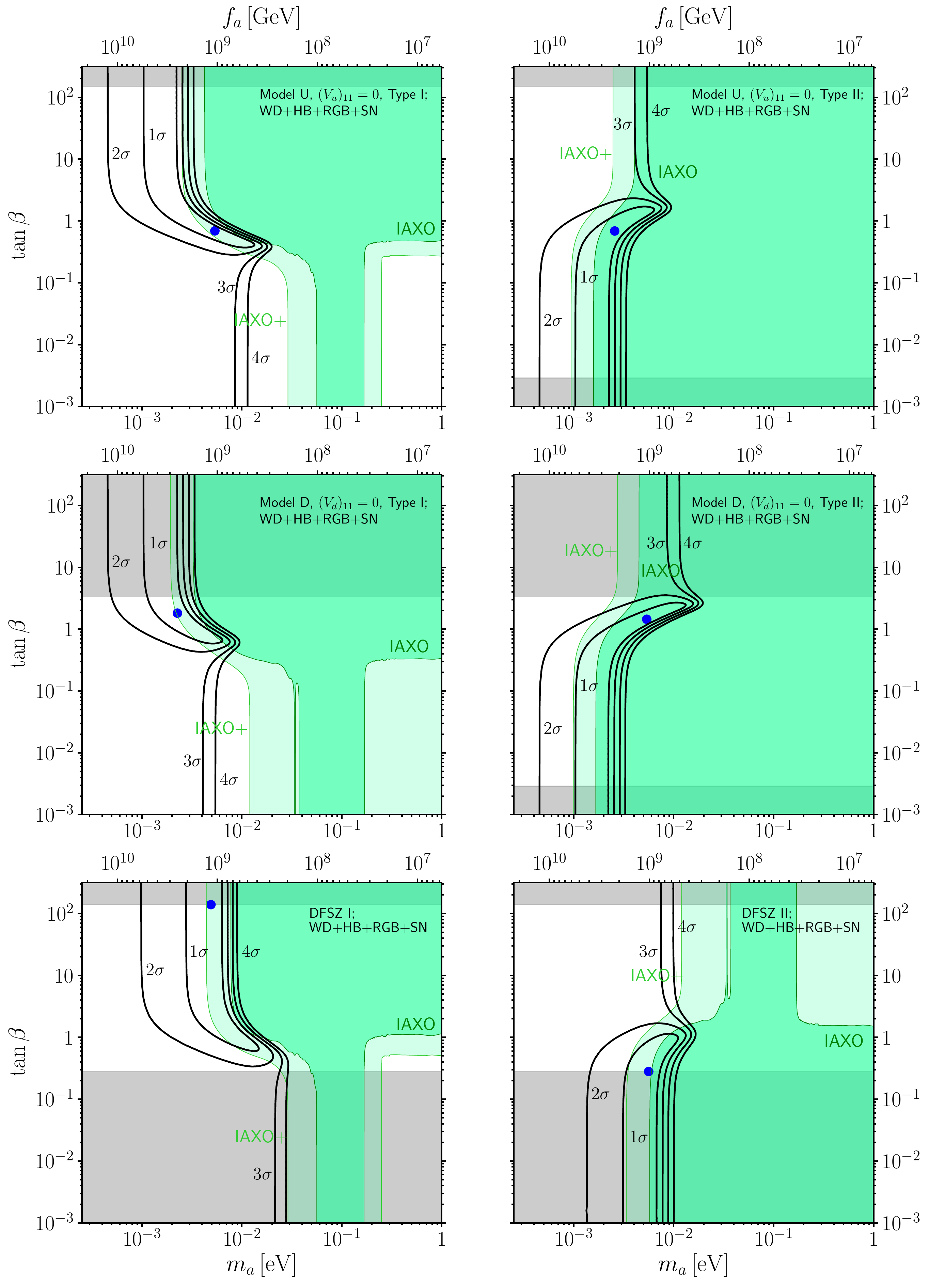}
\caption{
1, 2, 3, 4\,$\sigma$ contours in the $m_a$-$\tan\beta$ plane
from a fit to the data including SN 1987A in addition to WD, HB, and RGB cooling observations
for Model U with $(V_u)_{11}=0$, Type I (top left),
Model U with $(V_u)_{11}=0$, Type II (top right),
Model D with $(V_d)_{11}=0$, Type I (middle left),
Model D with $(V_d)_{11}=0$, Type II (middle right),
DFSZ I (bottom left), and DFSZ II (bottom right).
Blue dots represent the best fit parameters, and gray regions correspond to the parameter space 
outside a typical range compatible with perturbativity of Yukawa interactions.
Projected sensitivities of IAXO (green) and IAXO+ (light green) are also shown.
}
\label{fig:contour_UD_DFSZ_SN}
\end{figure}

In Fig.~\ref{fig:contour_UD_DFSZ_SN}, 
we show the 1, 2, 3, 4\,$\sigma$ hinted regions from the stellar cooling observations including the bound from
SN 1987A for Model U and D with $(V_{u,d})_{11}=0$.
The values of the axion mass and couplings corresponding to the 2$\,\sigma$ regions are also summarized in Table~\ref{tab:predicted_ranges}.
In these models $(V_{u,d})_{11}$ are no longer free parameters, and we just scan over
two dimensional parameter space of $m_a$ and $\tan\beta$.
Overall, the shapes of the contours for Model U and D with $(V_{u,d})_{11}=0$ are almost the same as
those for models with general quark mixings shown in Figs.~\ref{fig:contour_UCT_SN} and~\ref{fig:contour_DSB_SN}.
The difference between general and specific cases only appear in the region with higher values of $m_a$ 
where the effect of the axion-nucleon couplings becomes relevant because of the constraint from SN 1987A.
Comparing Fig.~\ref{fig:contour_UD_DFSZ_SN} with Figs.~\ref{fig:contour_UCT_SN} and~\ref{fig:contour_DSB_SN},
we see that the contours extend to $m_a \lesssim 10^{-2}\,\mathrm{eV}$ in Type I of Model U and D with small quark mixings,
while they reach slightly higher mass regions in Type I models with general quark mixings.
This difference arises from the fact that $(V_{u,d})_{11}$ are fixed to be zero in the former cases while
in the latter cases they can be adjusted to compensate the increase in the nucleon couplings due to the lower values of $f_a$.
Such a difference is not clearly seen in the contours for Type II models,
since for these models the general setup already shows a preference for $(V_{u,d})_{11}=0$
even at higher masses.

The best fit parameter values and the values of $\chi^2_{\rm min}/\text{d.o.f.}$ for the interpretation of the stellar cooling anomalies in the framework of 
the variant axion models with a Frogatt-Nielsen like flavor structure are summarized in Table~\ref{tab:best_fit_parameters_UD}.
From Table~\ref{tab:best_fit_parameters_UD} we see that the fits including WD, RGB, HB, and SN data
give $\chi^2_{\rm min}/\mathrm{d.o.f.} = 14.7/16$, 
which is contrasted with $\chi^2_{\rm min}/\mathrm{d.o.f.} = 15.4/16$ for DFSZ I~\cite{Giannotti:2017hny}.
This improvement originates from a specific structure of the axion-nucleon and axion-electron couplings in the variant axion models.
In Sec.~\ref{sec:variant_axions}, we see that the combined nucleon coupling coefficient $\sqrt{C_{ap}^2+C_{an}^2}$ can become as small as 
$\sim 0.04$ in the variant axion models with $(V_{u,d})_{11}\simeq 0$, while it becomes larger than $\gtrsim 0.24$ in the DFSZ models.
Even if we take account of the variation due to the change of $\tan\beta$, the magnitude of the nucleon couplings
remains comparable between the variant and DFSZ models as long as the value of $\tan\beta$ is not far from $\mathcal{O}(1)$.
On the other hand, the coefficient of the axion-electron coupling $|C_{ae}|$ of the variant axion models
is a factor three larger than that of the DFSZ models.
Combining these facts, we see that the ratio $\sqrt{C_{ap}^2+C_{an}^2}/|C_{ae}|$ in the variant axion models
can be smaller than that in the DFSZ models for the relevant range of $\tan\beta$.
This fact slightly relaxes the bound from SN 1987A, providing better fits to the stellar cooling hints.

\begin{table}[ht]\centering
\caption{
Best fit parameters and $\chi^2_{\rm min}/\text{d.o.f.}$ for the interpretation of the stellar cooling anomalies in
Model U and D with the assumption of $(V_{u,d})_{11} = 0$
and in DFSZ models.
}
%\small
\vspace{-\baselineskip}
\begin{align}
\begin{array}{|l|l|l|c|c|c|c|}
\hline
\text{Model} & \text{Type} & \text{Global fit includes} & f_a\,[10^8\,\mathrm{GeV}] & m_a\,[\mathrm{meV}] & \tan\beta & \chi^2_{\rm min}/\text{d.o.f.}\\
\hline
\text{Model U} & \text{Type I} & \text{WD,HB,RGB,SN} & 11 & 5.3 & 0.69 & 14.7/16 \\
\text{(small mixings)} & \text{Type II} & \text{WD,HB,RGB,SN} & 22 & 2.6 & 0.69 & 14.7/16 \\
\hline
\text{Model D} & \text{Type I} & \text{WD,HB,RGB,SN} & 25 & 2.3 & 1.8 & 14.7/16 \\
\text{(small mixings)} & \text{Type II} & \text{WD,HB,RGB,SN} & 11 & 5.4 & 1.4 & 14.7/16 \\
 \hline
 \text{DFSZ} & \text{Type I} & \text{WD,HB,RGB,SN} & 12 & 4.9 & 140 & 15.4/16 \\
 & \text{Type II} & \text{WD,HB,RGB,SN} & 10 & 5.6 & 0.28 & 14.7/16 \\
 \hline
\end{array}
\nonumber
\end{align}
\label{tab:best_fit_parameters_UD}
\end{table}

For the sake of comparison, in Fig.~\ref{fig:contour_UD_DFSZ_SN} we also show the results for the DFSZ models.
Here we see that the difference in the coefficient of the axion-electron coupling 
between the variant and DFSZ axion models
affects the axion mass range favored by the stellar cooling hints.
The factor three difference in $|C_{ae}|$ is compensated by shifting the value of $f_a$ by the same amount.
As a result, the variant axion models provide good fits even in the mass ranges
lower than those predicted in the DFSZ models.

Finally, we note that the bound from the kaon decay [Eq.~\eqref{Kaon_bound}] is trivially satisfied
in Model U, since it predicts $(C^A_{ad})_{21} = 0$, which is exactly the same as 
in the general cases of Model U,\,C,\,T discussed in the previous subsection.
On the other hand, this bound might give rise to a tension with the interpretation of the stellar cooling anomalies in
Model D with Froggatt-Nielsen like quark mixings, since we have $(V_d)_{21} \sim \mathcal{O}(\varepsilon)$ in this case.

%%%%%%%%%%%%%%%%%%%%%%%%%%%%%%%%%%%%%%%%%%%%%%%%%%
\section{Discussion and conclusions}
\label{sec:conclusion}
%%%%%%%%%%%%%%%%%%%%%%%%%%%%%%%%%%%%%%%%%%%%%%%%%%

In this paper, we have revisited the global analysis of the stellar cooling anomalies and their axion/ALP interpretation.
We have adopted a conservative approach to include possible systematic uncertainties associated with
the effect of stellar rotation in the globular clusters, prediction for the TRGB brightness, and different models 
for the pulsating WD PG 1351+489.
As a result of this procedure, the significance of the cooling hints becomes weaker, pointing to a non-vanishing
axion-electron coupling of $g_{ae} \sim 1.56\times 10^{-13}$ at around 2.4\,$\sigma$.
Furthermore, we emphasize that there could be potentially large systematic uncertainties associated with survey incompleteness
in the observed WDLF (see Sec.~\ref{sec:WDLF}), which we have not included in the analysis presented in this paper.
These situations should be reviewed once further observational data become available.

With the revised results of the global fits, we have considered the variant axion models
as possible explanation for the anomalous cooling observed in WD, HB, and RGB stars.
The models are constructed by considering six different flavor-dependent PQ charge assignments for the SM quarks
and two different flavor-blind PQ charge assignments for the SM leptons,
which result in four different possibilities for axion couplings to ordinary matter.
For each case, coupling coefficients are derived systematically, and they are summarized in
Tables~\ref{tab:axion_quark_couplings} and~\ref{tab:axion_electron_couplings}, 
and Eqs.~\eqref{C_ap_VA} and~\eqref{C_an_VA}.
By using the derived couplings, we have performed the global fits with the WD, HB, RGB data
and the bound from SN 1987A by considering both the general cases for quark mixings 
and specific cases motivated by the Froggatt-Nielsen mechanism.
Every model shows a quite good fit to the data within mass ranges around $0.45\,\mathrm{meV}\lesssim m_a \lesssim 30\,\mathrm{meV}$
as shown in Figs.~\ref{fig:contour_noSN},~\ref{fig:contour_UCT_SN},~\ref{fig:contour_DSB_SN}, and~\ref{fig:contour_UD_DFSZ_SN}, 
and Tables~\ref{tab:best_fit_parameters} and~\ref{tab:best_fit_parameters_UD}.
The 2\,$\sigma$ parameter regions preferred by the cooling hints and compatible with the SN bound and perturbativity requirements
are shown in Fig.~\ref{fig:model_predictions} and Table~\ref{tab:predicted_ranges}.

We emphasize that the models presented in this paper can resolve
two fundamental issues that prevent the canonical KSVZ and DFSZ axion models
from interpreting the stellar cooling anomalies straightforwardly:
\begin{enumerate}
\item The KSVZ models do not account for the large axion-electron coupling preferred by the stellar cooling hints.
A KSVZ-like axion/majoron model might account for it due to extra loop contributions from neutrinos, 
but there is a tension with perturbativity requirements~\cite{Giannotti:2017hny}.
This drawback is absent in the variant axion models, since a sizable electron coupling arises naturally at tree level.
\item The DFSZ models predict a sufficiently large axion-electron coupling and hence provide good fits to the observational data,
but they suffer from the cosmological domain wall problem.
In the variant axion models, it is possible to avoid the domain wall problem (i.e.~$N_{\rm DW} = 1$)
with keeping the structure of the axion-electron coupling similar to that in the DFSZ models.
\end{enumerate}

In addition to the above two important points, we have also found that
the models where a nonzero PQ charge is assigned to first generation quarks 
give a slightly better fit than the DFSZ models 
if we make an assumption of small mixings of right-handed quarks
as suggested by the Froggatt-Nielsen mechanism.
This is because the constraint from SN 1987A is slightly relaxed
due to the fact that
the ratio of the axion-nucleon coupling to the electron coupling 
becomes smaller than that in the DFSZ models
in the relevant parameter regions.

Another important feature of the variant axion models is that they predict flavor-changing couplings
with quarks parameterized by the matrices $(V_{u,d})_{ij}$ (see Table~\ref{tab:axion_quark_couplings}).
One consequence of such flavor-changing couplings is the possibility of rare decays of heavy mesons into axions.
In particular, the search for the decay process $K^+ \to \pi^+ + a$ could lead to a severe constraint on this class of models.
We have found that such a process is absent and the constraint is trivially satisfied
in the models with a PQ charge assigned for an up-type quark,
while it may give rise to some tension in those with a PQ charge assigned for a down-type quark.

In addition to the above constraint from the decay of kaons, 
processes involving other off-diagonal elements of $(V_{u,d})_{ij}$ can be probed by various other decay channels. 
Present and expected future limits on the flavor-changing couplings 
from heavy meson decays are summarized in Ref.~\cite{Bjorkeroth:2018dzu}.
Furthermore, if the extra heavier states coming from two Higgs doublets (see Appendix~\ref{app:yukawa_interactions}) have masses comparable to the electroweak scale,
the models predict flavor-changing neutral-current processes that can be searched by collider experiments~\cite{Chiang:2015cba,Chiang:2017fjr,Chiang:2018bnu}.
Such laboratory searches can complement the results of astrophysical observations discussed in this paper.
Indeed, exploring flavor-changing interactions would be crucial to distinguish between the variant axion models
and other flavor-blind models such as the DFSZ models.

The mass range $m_a \gtrsim 0.45\,\mathrm{meV}$ predicted by the variant axion interpretation of
the stellar cooling anomalies might have intriguing implications for cosmology.
In this mass range, the axion decay constant is of order $f_a \sim 10^9\,\mathrm{GeV}$,
and it is preferable to assume that the PQ symmetry is broken after inflation 
since otherwise there is a severe constraint from isocurvature fluctuations as mentioned in Sec.~\ref{sec:Introduction}.
In such a post-inflationary PQ symmetry breaking scenario, axions can be produced by the decay of strings and domain walls,
and they behave as cold dark matter in the present universe~\cite{Davis:1986xc,Lyth:1991bb}.
Although the previous estimates of the relic axion abundance from the decay of string-wall systems for models with $N_{\rm DW} = 1$
showed that such axions would comprise only a small fraction of the cold dark matter 
in the mass range $m_a \gtrsim 1\,\mathrm{meV}$ [see e.g.~Refs.~\cite{Hiramatsu:2012gg,Kawasaki:2014sqa,Fleury:2015aca,Klaer:2017ond}], 
it has been pointed out recently that there could be a much broader uncertainty in those estimates~\cite{Gorghetto:2018myk}.
Within the broad uncertainty suggested in Ref.~\cite{Gorghetto:2018myk}, there remains a possibility that axions become the main constituent of
dark matter up to the mass of $m_a \lesssim 4.4\,\mathrm{meV}$~\cite{Armengaud:2019uso}. 
Indeed, preliminary results of the state of the art simulations~\cite{Gorghetto:2019patras,Saikawa:2019patras} 
show a preference for higher dark matter mass ranges close to this upper limit on $m_a$.
Therefore, there is a possibility that the $\mathrm{meV}$ mass variant axions could account for dark matter 
as well as providing the explanation of the stellar cooling anomalies.
Direct detection of axion dark matter in the $\mathrm{meV}$ mass range is quite challenging,
but several techniques are proposed in the literature, such as a dish antenna~\cite{Horns:2012jf},
absorption in superconductors~\cite{Hochberg:2016ajh}, and topological insulators~\cite{Marsh:2018dlj}.

The mass range preferred by the stellar cooling anomalies will also be probed by 
the proposed experiment ARIADNE~\cite{Arvanitaki:2014dfa},
which is the search for long range forces mediated by axions~\cite{Moody:1984ba}.
This experiment does not rely on any assumption 
on astrophysics (e.g.~axion emissivity from the Sun)
or cosmology (e.g.~local dark matter density),
and hence it would play a complementary role with respect to other axion searches.
However, in order to achieve a reasonable sensitivity it is necessary to assume that there exists
a sizable CP-violating interaction between axions and nuclei saturating the limit on the neutron electric dipole moment.
Such a large CP-violating coupling is not necessarily guaranteed in the models considered in this paper.

Crucially, the most part of the parameter regions predicated by the stellar cooling hints
will be covered by the next generation helioscope IAXO.
Once axions are detected in such experiments, it would be even possible
to measure the axion mass and couplings to photons and electrons 
if the detectors have enough energy resolution~\cite{Jaeckel:2018mbn,Dafni:2018tvj}.
The predictions of Type II of the variant axion models exhibit relatively high values for
the coupling product $g_{ae}g_{a\gamma}$ (see Fig.~\ref{fig:model_predictions}), 
and hence this class of models would potentially be distinguishable from other models 
through such measurements in the future helioscope experiments.

%%%%%%%%%%%%%%%%%%%%%%%%%%%%%%%%%%%%%%%%%%%%%%%%%%
\section*{Acknowledgments}
%%%%%%%%%%%%%%%%%%%%%%%%%%%%%%%%%%%%%%%%%%%%%%%%%%
K.~S. would like to thank Maurizio Giannotti and Javier Redondo for a lot of important comments.
We also wish to thank the anonymous referee for his/her constructive comments and suggestions 
that lead to great improvements in the quality of the paper.
K.~S. acknowledges partial support by the Deutsche Forschungsgemeinschaft through Grant No.\ EXC 153 (Excellence
Cluster ``Universe'') and Grant No.\ SFB 1258 (Collaborative Research
Center ``Neutrinos, Dark Matter, Messengers'') as well as by the
European Union through Grant No.\ H2020-MSCA-ITN-2015/674896 (Innovative Training Network ``Elusives'').
T.~T.~Y. was supported in part by the China Grant for Talent 
Scientific Start-Up Project and 
the JSPS Grant-in-Aid for 
Scientific Research No. 16H02176, 
and No. 17H02878 and by World Premier International Research Center Initiative (WPI Initiative), MEXT, Japan. T. T. Y. thanks to Hamamatsu Photonics.
%%%%%%%%%%%%%%%%%%%%%%%%%%%%%%%%%%%%%%%%%%%%%%%%%%

%%%%%%%%%%%%%%%%%%%%%%%%%%%%%%%%%%%%%%%%%%%%%%%%%%
\appendix
%%%%%%%%%%%%%%%%%%%%%%%%%%%%%%%%%%%%%%%%%%%%%%%%%% 

%%%%%%%%%%%%%%%%%%%%%%%%%%%%%%%%%%%%%%%%%%%%%%%%%% 
\section{Analysis method}
\setcounter{equation}{0}
\label{app:analysis_method}
%%%%%%%%%%%%%%%%%%%%%%%%%%%%%%%%%%%%%%%%%%%%%%%%%%

For the global fits we take a slightly different procedure than Ref.~\cite{Giannotti:2017hny}.
In Sec.~\ref{sec:stellar_cooling}, we see that there are several sources of systematic uncertainties, 
which might not be regarded as simple Gaussian-like fluctuations.
One possible way to deal with such systematic errors is to introduce additional parameters (nuisance parameters)
that are marginalized to construct a profile likelihood function [e.g. Ref.~\cite{Cowan}].

Our general procedure to marginalize the systematic variables is as follows.
Suppose that an outcome of an observation is measured as a value $M$ with an error $\sigma$.
Let $T(\bm{\theta})$ denotes a theoretical prediction for this quantity, which depends on a set of parameters $\bm{\theta} = (\theta_1,\dots,\theta_N)$.
We assume that $T(\bm{\theta})$ has some systematic uncertainty, whose typical magnitude is given by $\Delta T(\bm{\theta})$.
The effect of the systematic uncertainty can be embedded as a shift of $T(\bm{\theta})-M$ by an amount $\xi\Delta T(\bm{\theta})$,
where $\xi$ is an additional parameter spanning a finite interval.
The values of $\bm{\theta}$ compatible with the observational data are found by evaluating the following quantity,
\begin{equation}
\chi^2 = \frac{(M-T(\bm{\theta})-\hat{\xi}(\bm{\theta})\Delta T(\bm{\theta}))^2}{\sigma^2},
\end{equation}
where $\hat{\xi}(\bm{\theta})$ is a value that minimizes $\chi^2$ for specified values of $\bm{\theta}$.
The above formula resembles the pull $\chi^2$ function~\cite{Stump:2001gu,Fogli:2002pt},
but here we assume that the variable $\xi$ obeys a flat probability distribution rather than the Gaussian distribution.

One can evaluate $\hat{\xi}(\bm{\theta})$ almost trivially: 
It takes the boundary value $\xi_{\rm min}$ or $\xi_{\rm max}$ of the specified domain $\xi \in [\xi_{\rm min},\xi_{\rm max}]$,
otherwise $\hat{\xi}(\bm{\theta}) = (M-T(\bm{\theta}))/\Delta T(\bm{\theta})$, for which $\chi^2=0$.
The appearance of the flat direction $\xi = (M-T(\bm{\theta}))/\Delta T(\bm{\theta})$ just implies that
there is a degeneracy between the systematic effect and the effect of theoretical parameters,
and that the values of the parameters $\bm{\theta}$ cannot be determined unless the corresponding systematic uncertainty is resolved.

We apply the above procedure to the analysis of the astrophysical data.
For the $R$-parameter, the observed value is $R^{\rm obs} = 1.39\pm0.03$, and the theoretical model is given by Eq.~\eqref{R_theory}.
We use the value $Y = 0.2535 \pm 0.0036$ as representative of the mass fraction in low metallicity environments~\cite{Aver:2013wba,Ayala:2014pea}, 
and the corresponding error $\sigma_Y = 0.026$ is added to the observational error $\sigma_R = 0.03$.
Furthermore, we introduce a parameter $\xi_{\rm rot}$ corresponding to the systematic uncertainty due to the stellar rotation.
$\xi_{\rm rot}$ is allowed to vary in the range $[-1,0]$: It is negative since the effect of rotation 
just shortens the helium-burning lifetime~\cite{Piersanti:2013rya}.
The magnitude of this uncertainty is estimated as $3\,\%$ of the observed value, $\Delta R = 0.042$.

For the TRGB of the globular cluster M5, the observed value is $M_{I,\mathrm{TRGB}}^{\rm obs} = -4.17 \pm 0.13\,\mathrm{mag}$.
We use the theoretical model of Eq.~\eqref{MI_TRGB_theory}, but here we drop the last term
$\delta M_{I,\mathrm{TRGB}}^{\rm th}$, since it was introduced in order to correct the asymmetric ranges of
systematic errors and convert them to a Gaussian error in Ref.~\cite{Viaux:2013hca,Viaux:2013lha}, which we do not follow in our analysis.
Instead, we introduce a parameter $\xi_{\rm BC}$, which is defined in the range $[-1,1]$, to represent the systematic uncertainties.
We use the error estimate for bolometric corrections as the magnitude of the uncertainty, $\Delta M_{I,\mathrm{TRGB}} = (0.08 + 0.02 g_{13})\,\mathrm{mag}$,
since it gives the largest range among the possible sources of systematic uncertainties enumerated in Refs.~\cite{Viaux:2013hca,Viaux:2013lha}.

For the WDLF, we follow the same procedure as in Ref.~\cite{Giannotti:2017hny}.
We take 11 binned data points in the luminosity range $7<M_{\rm bol} < 12.25$ from Refs.~\cite{Bertolami:2014wua,Bertolami:2014noa}.
For the theoretical model, we allow the normalization factor to vary as a function of $g_{ae}$ and marginalize it such that it leads to the smallest $\chi^2$.

For the period change of the WD variables, we adopt the observational data of R548 [$\dot{\Pi} = (3.3\pm1.1)\times 10^{-15}\,\mathrm{s/s}$]~\cite{Corsico:2012sh},
PG 1351+489 [$\dot{\Pi} = (2.0\pm0.9)\times 10^{-13}\,\mathrm{s/s}$]~\cite{Corsico:2014mpa,Battich:2016htm},
and two pulsation modes of L19-2 [$\dot{\Pi} = (3.0\pm0.6)\times 10^{-15}\,\mathrm{s/s}$ for both modes]~\cite{Corsico:2016okh}.
Following the suggestion in Ref.~\cite{Giannotti:2017hny}, we exclude the data of G117-B15A~\cite{Corsico:2012ki} and
add the theoretical uncertainties for R548 [$\sigma_{\dot{\Pi}_{\rm th}} = 0.09\times 10^{-15}\,\mathrm{s/s}$]
and for L19-2 [$\sigma_{\dot{\Pi}_{\rm th}} = 0.85\times 10^{-15}\,\mathrm{s/s}$ and $\sigma_{\dot{\Pi}_{\rm th}} = 1.45\times 10^{-15}\,\mathrm{s/s}$] in quadrature.
For PG 1351+489, instead of adding the $\sigma_{\dot{\Pi}_{\rm th}}^2$ error, we introduce a parameter $\xi_{\rm model}$ to take account of the systematic uncertainty due to different models.
We extract 7 different model curves from Figs.~6 and~7 of Ref.~\cite{Battich:2016htm} and estimate the ``theoretical" prediction $\dot{\Pi}(g_{ae})$ from the median between
the largest and smallest value among the 7 curves.
Then the magnitude of the systematic uncertainty $\Delta\dot{\Pi}$ is identified by using half of the difference between the largest and smallest value among those curves.
We marginalize it by allowing $\xi_{\rm model}$ to vary in the range $[-1,1]$.

In summary, the $\chi^2$ function is
\begin{align}
\chi^2 &= \frac{(R^{\rm obs}-R^{\rm th}(g_{ae},g_{a\gamma})-\hat{\xi}_{\rm rot}(g_{ae},g_{a\gamma})\Delta R)^2}{\sigma_R^2 + \sigma_Y^2} \nonumber\\
&\quad+ \frac{(M_{I,\mathrm{TRGB}}^{\rm obs}-M_{I,\mathrm{TRGB}}^{0}(g_{ae})-\hat{\xi}_{\rm BC}(g_{ae})\Delta M_{I,\mathrm{TRGB}}(g_{ae}))^2}{\sigma_{M_{I,\mathrm{TRGB}}}^2} \nonumber\\
&\quad+ \sum^{\rm WDLF}_{i=1,\dots,11}\frac{(M^{\rm obs}_i-N(g_{ae})M_i(g_{ae}))}{\sigma^2_{M_i}}
+ \sum_{s=1,2,3}^{\rm R548,L19\mathchar`-2}\frac{(\dot{\Pi}^{\rm obs}_s-\dot{\Pi}_s(g_{ae}))^2}{\sigma_{\dot{\Pi}_s}^2+\sigma_{\dot{\Pi}_s^{\rm th}}^2} \nonumber\\
&\quad+ \frac{(\dot{\Pi}^{\rm obs}_{\rm PG 1351+489}-\dot{\Pi}(g_{ae})-\hat{\xi}_{\rm model}(g_{ae})\Delta\dot{\Pi}(g_{ae}))^2}{\sigma_{\dot{\Pi}_{\rm PG 1351+489}}^2},
\end{align}
where $M_{I,\mathrm{TRGB}}^{0}(g_{ae}) = M_{I,\mathrm{TRGB}}^{\rm th}(g_{ae})-\delta M_{I,\mathrm{TRGB}}^{\rm th}$.
When we include the SN 1987A bound, we add
\begin{equation}
\left(\frac{g_{ap}^2+g_{an}^2}{3.6\times 10^{-19}}\right)^2.
\end{equation}

%%%%%%%%%%%%%%%%%%%%%%%%%%%%%%%%%%%%%%%%%%%%%%%%%% 
\section{Higgs sector and Yukawa interactions}
\setcounter{equation}{0}
\label{app:yukawa_interactions}
%%%%%%%%%%%%%%%%%%%%%%%%%%%%%%%%%%%%%%%%%%%%%%%%%%

In this appendix, we investigate the structure of the Higgs sector and Yukawa interactions in the variant axion models
for the purpose of obtaining typical ranges of the parameter $\tan\beta$ compatible with perturbativity of the Yukawa interactions.
The most general renormalizable Higgs potential at energies below the PQ scale but above the electroweak scale reads~\cite{Chiang:2015cba}
\begin{align}
V(H_1,H_2) &= m_{11}^2H^{\dagger}_1 H_1 + m_{22}^2H^{\dagger}_2H_2 - \left(m_{12}^2H^{\dagger}_1H_2 + \mathrm{h.c.}\right)
+ \frac{\lambda_1}{2}\left(H^{\dagger}_1 H_1\right)^2 + \frac{\lambda_2}{2}\left(H^{\dagger}_2 H_2\right)^2\nonumber\\
&\quad+\lambda_3\left(H^{\dagger}_1 H_1\right)\left(H^{\dagger}_2 H_2\right) + \lambda_4\left(H^{\dagger}_1 H_2\right)\left(H^{\dagger}_2 H_1\right),
\end{align}
where the $m_{12}^2$ terms originate from the interaction terms with the singlet scaler (i.e.~terms proportional to $H_1^{\dagger}H_2\sigma$ 
and its hermitian conjugate) in the UV-complete theory (see Sec.~\ref{sec:variant_axions}), 
and we can make the parameter $m_{12}^2$ real and positive by the PQ symmetry transformation. 
All the other parameters $m_{11}^2$, $m_{22}^2$, and $\lambda_{1,2,3,4}$ are real. 

After the electroweak symmetry breaking, two Higgs doublet fields can be decomposed into their VEVs $v_k$ and component fields, 
$H_k = (H_k^+,(v_k + h_k + iA_k)/\sqrt{2})$.
These fields can be related to the SM Higgs field $H^{\rm SM}$ and the orthogonal field $H'$,
\begin{equation}
\left(
\begin{array}{c}
H_1 \\
H_2
\end{array}
\right)
=
\left(
\begin{array}{cc}
\cos\beta & -\sin\beta \\
\sin\beta & \cos\beta
\end{array}
\right)
\left(
\begin{array}{c}
H^{\rm SM} \\
H'
\end{array}
\right),
\end{equation}
where
\begin{equation}
H^{\rm SM} = 
\left(
\begin{array}{c}
G^+ \\
(v + h^{\rm SM} + iG^0)/\sqrt{2}
\end{array}
\right)
\quad \text{and} \quad
H' = 
\left(
\begin{array}{c}
H^+ \\
(h' + iA^0)/\sqrt{2}
\end{array}
\right).
\end{equation}
Among eight field degrees of freedom, $G^{\pm}$ and $G^0$ correspond to the Nambu-Goldstone bosons which get eaten by
$W^{\pm}$ and $Z^0$ bosons. The charged scalar $H^{\pm}$ and neutral pseudoscalar $A^0$ are mass eigenstates, 
while neutral scalars $h^{\rm SM}$ and $h'$ are in general not mass eigenstates.
We define the rotation angle $\alpha$ which relates these neutral scalars to the lighter ($h$) and heavier ($H$) mass eigenstates,
\begin{equation}
\left(
\begin{array}{c}
H \\
h
\end{array}
\right)
=
\left(
\begin{array}{cc}
\cos\alpha & \sin\alpha \\
-\sin\alpha & \cos\alpha
\end{array}
\right)
\left(
\begin{array}{c}
h_1 \\
h_2
\end{array}
\right)
=
\left(
\begin{array}{cc}
\cos(\beta-\alpha) & -\sin(\beta-\alpha) \\
\sin(\beta-\alpha) & \cos(\beta-\alpha)
\end{array}
\right)
\left(
\begin{array}{c}
h^{\rm SM} \\
h'
\end{array}
\right).
\end{equation}

Note that all the heavier states $H^{\pm}$, $A^0$ and $H$ decouple in the limit $m_{12}^2 \to \infty$, and
these states become irrelevant to low energy physics, in which the lightest eigenstate $h$ is identified as the SM Higgs boson.
However, even in such a case, it is possible to restrict parameters of the models 
by considering perturbativity of Yukawa interactions.

Yukawa interactions of neutral Higgs bosons can be parameterized as~\cite{Aoki:2009ha}
\begin{equation}
\mathcal{L}_{\rm Yukawa} = -\sum_{ij}\frac{m_{\psi_i}}{v}\left[\xi^{h\psi}_{ij}h\overline{\psi}_{iL}\psi_{jR} + \xi^{H\psi}_{ij}H\overline{\psi}_{iL}\psi_{jR} - i\xi^{A\psi}_{ij}A^0\overline{\psi}_{iL}\psi_{jR}\right] + \mathrm{h.c.},
\end{equation}
where the fermions $\psi_i$ are taken to be mass eigenstates. 
We recall that the Yukawa interactions are different according to the PQ charge assignments 
for the quark and lepton fields [see Eqs.~\eqref{L_yukawa_quarks} and~\eqref{L_yukawa_leptons}].
The coupling coefficients read
\begin{align}
\xi^{hu}_{ij} &= \left\{
\begin{array}{ll}
\displaystyle
\frac{\cos\alpha}{\sin\beta}\delta_{ij}-\left(\frac{\cos\alpha}{\sin\beta}+\frac{\sin\alpha}{\cos\beta}\right)(V_u)_{ij} & \text{(Model U,\,C,\,T)},
\vspace{1mm}\\
\displaystyle
-\frac{\sin\alpha}{\cos\beta}\delta_{ij}  & \text{(Model D,\,S,\,B)},
\end{array}
\right.\\
\xi^{Hu}_{ij} &= \left\{
\begin{array}{ll}
\displaystyle
\frac{\sin\alpha}{\sin\beta}\delta_{ij}+\left(\frac{\cos\alpha}{\cos\beta}-\frac{\sin\alpha}{\sin\beta}\right)(V_u)_{ij} & \text{(Model U,\,C,\,T)},
\vspace{1mm}\\
\displaystyle
\frac{\cos\alpha}{\cos\beta}\delta_{ij}  & \text{(Model D,\,S,\,B)},
\end{array}
\right.\\
\xi^{Au}_{ij} &= \left\{
\begin{array}{ll}
\displaystyle
\cot\beta\delta_{ij}-(\cot\beta + \tan\beta)(V_u)_{ij} & \text{(Model U,\,C,\,T)},
\vspace{1mm}\\
\displaystyle
-\tan\beta\delta_{ij}  & \text{(Model D,\,S,\,B)},
\end{array}
\right.
\end{align}
for the up-type quarks, where $V_u$ is defined in Eq.~\eqref{V_u_definition},
\begin{align}
\xi^{hd}_{ij} &= \left\{
\begin{array}{ll}
\displaystyle
-\frac{\sin\alpha}{\cos\beta}\delta_{ij} & \text{(Model U,\,C,\,T)},
\vspace{1mm}\\
\displaystyle
\frac{\cos\alpha}{\sin\beta}\delta_{ij}-\left(\frac{\cos\alpha}{\sin\beta}+\frac{\sin\alpha}{\cos\beta}\right)(V_d)_{ij}  & \text{(Model D,\,S,\,B)},
\end{array}
\right.\\
\xi^{Hd}_{ij} &= \left\{
\begin{array}{ll}
\displaystyle
\frac{\cos\alpha}{\cos\beta}\delta_{ij} & \text{(Model U,\,C,\,T)},
\vspace{1mm}\\
\displaystyle
\frac{\sin\alpha}{\sin\beta}\delta_{ij}+\left(\frac{\cos\alpha}{\cos\beta}-\frac{\sin\alpha}{\sin\beta}\right)(V_d)_{ij} & \text{(Model D,\,S,\,B)},
\end{array}
\right.\\
\xi^{Ad}_{ij} &= \left\{
\begin{array}{ll}
\displaystyle
\tan\beta\delta_{ij} & \text{(Model U,\,C,\,T)},
\vspace{1mm}\\
\displaystyle
-\cot\beta\delta_{ij}+(\cot\beta + \tan\beta)(V_d)_{ij} & \text{(Model D,\,S,\,B)},
\end{array}
\right.
\end{align}
for the down-type quarks, where $V_d$ is defined in Eq.~\eqref{V_d_definition}, and
\begin{align}
\xi^{h\ell}_{ij} &= \left\{
\begin{array}{ll}
\displaystyle
-\frac{\sin\alpha}{\cos\beta}\delta_{ij} & \text{(Type I)},
\vspace{1mm}\\
\displaystyle
\frac{\cos\alpha}{\sin\beta}\delta_{ij} & \text{(Type II)},
\end{array}
\right.\\
\xi^{H\ell}_{ij} &= \left\{
\begin{array}{ll}
\displaystyle
\frac{\cos\alpha}{\cos\beta}\delta_{ij} & \text{(Type I)},
\vspace{1mm}\\
\displaystyle
\frac{\sin\alpha}{\sin\beta}\delta_{ij} & \text{(Type II)},
\end{array}
\right.\\
\xi^{A\ell}_{ij} &= \left\{
\begin{array}{ll}
\displaystyle
\tan\beta\delta_{ij} & \text{(Type I)},
\vspace{1mm}\\
\displaystyle
-\cot\beta\delta_{ij} & \text{(Type II)},
\end{array}
\right.
\end{align}
for leptons.

In order to guarantee perturbativity of the Yukawa interactions, we require
\begin{equation}
|y^{\phi\psi}_{ij}|^2 < 4\pi,
\end{equation}
where
\begin{equation}
y^{\phi\psi}_{ij} = \frac{\sqrt{2}m_{\psi_i}}{v}\xi^{\phi\psi}_{ij}
\end{equation}
for $\phi = h, H, A^0$.
Requiring that the above constraints are satisfied for any value of $\alpha$,
we obtain
\begin{align}
&\, \tan\beta < 150 \quad \text{(Model U,\,C,\,T, Type I)},\nonumber\\
0.0029 <&\, \tan\beta < 150 \quad \text{(Model U,\,C,\,T, Type II)},\nonumber\\
&\, \tan\beta < 3.4 \quad \text{(Model D,\,S,\,B, Type I)},\nonumber\\
0.0029 <&\, \tan\beta < 3.4 \quad \text{(Model D,\,S,\,B, Type II)},
\label{range_tanbeta}
\end{align}
where the upper limit for Model U, C, T comes from bottom Yukawa interactions, 
that for Model D, S, B comes from top Yukawa interactions,
and the lower limit for Type II comes from tau Yukawa interactions.
Here we have omitted the bounds obtained from couplings that depend on 
$(V_u)_{ij}$ and $(V_d)_{ij}$, whose values are unknown.\footnote{
Although we have specified the values of $(V_u)_{11}$ and $(V_d)_{11}$ in the plots shown in Sec.~\ref{sec:interpretation_of_anomalies},
the bounds obtained from couplings that depend on these parameters are quite weak because of the smallness of 
up and down quark masses, and they are irrelevant in the range of $\tan\beta$ shown in those figures.}
Note that, however, more stringent limits could be obtained according to their values.
In the main text, we have used Eq.~\eqref{range_tanbeta} as a typical range compatible with perturbativity requirements.

%%%%%%%%%%%%%%%%%%%%%%%%%%%%%%%%%%%%%%%%%%%%%%%%%%%%
%%%%%%%%%%%%%%%%%%%%%%%%%%%%%%%%%%%%%%%%%%%%%%%%%%%%


\begin{thebibliography}{200}

%\cite{Raffelt:1996wa}
\bibitem{Raffelt:1996wa} 
  G.~G.~Raffelt,
  ``Stars as laboratories for fundamental physics : The astrophysics of neutrinos, axions, and other weakly interacting particles,''
  Chicago Univ. Pr., Chicago U.S.A., (1996).

%\cite{Weinberg:1977ma}
\bibitem{Weinberg:1977ma} 
  S.~Weinberg,
  %``A New Light Boson?,''
  Phys.\ Rev.\ Lett.\  {\bf 40}, 223 (1978).
  %doi:10.1103/PhysRevLett.40.223
  %%CITATION = doi:10.1103/PhysRevLett.40.223;%%

%\cite{Wilczek:1977pj}
\bibitem{Wilczek:1977pj} 
  F.~Wilczek,
  %``Problem of Strong  $P$  and  $T$  Invariance in the Presence of Instantons,''
  Phys.\ Rev.\ Lett.\  {\bf 40}, 279 (1978).
  %doi:10.1103/PhysRevLett.40.279
  %%CITATION = doi:10.1103/PhysRevLett.40.279;%%

%\cite{Peccei:1977hh}
\bibitem{Peccei:1977hh} 
  R.~D.~Peccei and H.~R.~Quinn,
  %``CP Conservation in the Presence of Instantons,''
  Phys.\ Rev.\ Lett.\  {\bf 38}, 1440 (1977).
  %doi:10.1103/PhysRevLett.38.1440
  %%CITATION = doi:10.1103/PhysRevLett.38.1440;%%

%\cite{Raffelt:2011ft}
\bibitem{Raffelt:2011ft} 
  G.~G.~Raffelt, J.~Redondo and N.~Viaux Maira,
  %``The meV mass frontier of axion physics,''
  Phys.\ Rev.\ D {\bf 84}, 103008 (2011)
  %doi:10.1103/PhysRevD.84.103008
  [arXiv:1110.6397 [hep-ph]].
  %%CITATION = doi:10.1103/PhysRevD.84.103008;%%

%\cite{Ringwald:2015lqa}
\bibitem{Ringwald:2015lqa} 
  A.~Ringwald,
  %``The hunt for axions,''
  PoS NEUTEL {\bf 2015}, 021 (2015)
  %doi:10.22323/1.244.0021
  [arXiv:1506.04259 [hep-ph]].
  %%CITATION = doi:10.22323/1.244.0021;%%

%\cite{Giannotti:2015dwa}
\bibitem{Giannotti:2015dwa} 
  M.~Giannotti,
  %``ALP hints from cooling anomalies,''
  %doi:10.3204/DESY-PROC-2015-02/giannotti_maurizio
  arXiv:1508.07576 [astro-ph.HE].
  %%CITATION = doi:10.3204/DESY-PROC-2015-02/giannotti_maurizio;%%

%\cite{Giannotti:2015kwo}
\bibitem{Giannotti:2015kwo} 
  M.~Giannotti, I.~Irastorza, J.~Redondo and A.~Ringwald,
  %``Cool WISPs for stellar cooling excesses,''
  JCAP {\bf 1605}, no. 05, 057 (2016)
  %doi:10.1088/1475-7516/2016/05/057
  [arXiv:1512.08108 [astro-ph.HE]].
  %%CITATION = doi:10.1088/1475-7516/2016/05/057;%%

%\cite{Giannotti:2016hnk}
\bibitem{Giannotti:2016hnk} 
  M.~Giannotti,
  %``Hints of new physics from stars,''
  PoS ICHEP {\bf 2016}, 076 (2016)
  %doi:10.22323/1.282.0076
  [arXiv:1611.04651 [astro-ph.HE]].
  %%CITATION = doi:10.22323/1.282.0076;%%

%\cite{Giannotti:2017hny}
\bibitem{Giannotti:2017hny} 
  M.~Giannotti, I.~G.~Irastorza, J.~Redondo, A.~Ringwald and K.~Saikawa,
  %``Stellar Recipes for Axion Hunters,''
  JCAP {\bf 1710}, no. 10, 010 (2017)
  %doi:10.1088/1475-7516/2017/10/010
  [arXiv:1708.02111 [hep-ph]].
  %%CITATION = doi:10.1088/1475-7516/2017/10/010;%%

%\cite{Irastorza:2018dyq}
\bibitem{Irastorza:2018dyq} 
  I.~G.~Irastorza and J.~Redondo,
  %``New experimental approaches in the search for axion-like particles,''
  Prog.\ Part.\ Nucl.\ Phys.\  {\bf 102}, 89 (2018)
  %doi:10.1016/j.ppnp.2018.05.003
  [arXiv:1801.08127 [hep-ph]].
  %%CITATION = doi:10.1016/j.ppnp.2018.05.003;%%

\bibitem{IAXOloe} 
  I.~G.~Irastorza {\it et al.} [IAXO Collaboration],
  The International Axion Observatory IAXO. Letter of Intent to the CERN SPS committee,
  CERN-SPSC-2013-022 / SPSC-I-242 (8 August 2013).

%\cite{Armengaud:2014gea}
\bibitem{Armengaud:2014gea} 
  E.~Armengaud {\it et al.},
  %``Conceptual Design of the International Axion Observatory (IAXO),''
  JINST {\bf 9}, T05002 (2014)
  %doi:10.1088/1748-0221/9/05/T05002
  [arXiv:1401.3233 [physics.ins-det]].
  %%CITATION = doi:10.1088/1748-0221/9/05/T05002;%%

%\cite{Armengaud:2019uso}
\bibitem{Armengaud:2019uso} 
  E.~Armengaud {\it et al.} [IAXO Collaboration],
  %``Physics potential of the International Axion Observatory (IAXO),''
  JCAP {\bf 1906}, no. 06, 047 (2019)
  %doi:10.1088/1475-7516/2019/06/047
  [arXiv:1904.09155 [hep-ph]].
  %%CITATION = doi:10.1088/1475-7516/2019/06/047;%%

%\cite{Linde:1985yf}
\bibitem{Linde:1985yf} 
  A.~D.~Linde,
  %``Generation of Isothermal Density Perturbations in the Inflationary Universe,''
  Phys.\ Lett.\  {\bf 158B}, 375 (1985).
  %doi:10.1016/0370-2693(85)90436-8
  %%CITATION = doi:10.1016/0370-2693(85)90436-8;%%
  
%\cite{Seckel:1985tj}
\bibitem{Seckel:1985tj} 
  D.~Seckel and M.~S.~Turner,
  %``Isothermal Density Perturbations in an Axion Dominated Inflationary Universe,''
  Phys.\ Rev.\ D {\bf 32}, 3178 (1985).
  %doi:10.1103/PhysRevD.32.3178
  %%CITATION = doi:10.1103/PhysRevD.32.3178;%%
  
%\cite{Kobayashi:2013nva}
\bibitem{Kobayashi:2013nva} 
  T.~Kobayashi, R.~Kurematsu and F.~Takahashi,
  %``Isocurvature Constraints and Anharmonic Effects on QCD Axion Dark Matter,''
  JCAP {\bf 1309}, 032 (2013)
  %doi:10.1088/1475-7516/2013/09/032
  [arXiv:1304.0922 [hep-ph]].
  %%CITATION = doi:10.1088/1475-7516/2013/09/032;%%

%\cite{Preskill:1982cy}
\bibitem{Preskill:1982cy} 
  J.~Preskill, M.~B.~Wise and F.~Wilczek,
  %``Cosmology of the Invisible Axion,''
  Phys.\ Lett.\ B {\bf 120}, 127 (1983)
  [Phys.\ Lett.\  {\bf 120B}, 127 (1983)].
  %doi:10.1016/0370-2693(83)90637-8
  %%CITATION = doi:10.1016/0370-2693(83)90637-8;%%
  
%\cite{Abbott:1982af}
\bibitem{Abbott:1982af} 
  L.~F.~Abbott and P.~Sikivie,
  %``A Cosmological Bound on the Invisible Axion,''
  Phys.\ Lett.\ B {\bf 120}, 133 (1983)
  [Phys.\ Lett.\  {\bf 120B}, 133 (1983)].
  %doi:10.1016/0370-2693(83)90638-X
  %%CITATION = doi:10.1016/0370-2693(83)90638-X;%%

%\cite{Dine:1982ah}
\bibitem{Dine:1982ah} 
  M.~Dine and W.~Fischler,
  %``The Not So Harmless Axion,''
  Phys.\ Lett.\ B {\bf 120}, 137 (1983)
  [Phys.\ Lett.\  {\bf 120B}, 137 (1983)].
  %doi:10.1016/0370-2693(83)90639-1
  %%CITATION = doi:10.1016/0370-2693(83)90639-1;%%

%\cite{Sikivie:1982qv}
\bibitem{Sikivie:1982qv} 
  P.~Sikivie,
  %``Of Axions, Domain Walls and the Early Universe,''
  Phys.\ Rev.\ Lett.\  {\bf 48}, 1156 (1982).
  %doi:10.1103/PhysRevLett.48.1156
  %%CITATION = doi:10.1103/PhysRevLett.48.1156;%%

%\cite{Zeldovich:1974uw}
\bibitem{Zeldovich:1974uw} 
  Y.~B.~Zeldovich, I.~Y.~Kobzarev and L.~B.~Okun,
  %``Cosmological Consequences of the Spontaneous Breakdown of Discrete Symmetry,''
  Zh.\ Eksp.\ Teor.\ Fiz.\  {\bf 67}, 3 (1974)
  [Sov.\ Phys.\ JETP {\bf 40}, 1 (1974)].
  %%CITATION = ZETFA,67,3;%%

%\cite{Zhitnitsky:1980tq}
\bibitem{Zhitnitsky:1980tq} 
  A.~R.~Zhitnitsky,
  %``On Possible Suppression of the Axion Hadron Interactions. (In Russian),''
  Sov.\ J.\ Nucl.\ Phys.\  {\bf 31}, 260 (1980)
  [Yad.\ Fiz.\  {\bf 31}, 497 (1980)].
  %%CITATION = SJNCA,31,260;%%
  
%\cite{Dine:1981rt}
\bibitem{Dine:1981rt} 
  M.~Dine, W.~Fischler and M.~Srednicki,
  %``A Simple Solution to the Strong CP Problem with a Harmless Axion,''
  Phys.\ Lett.\  {\bf 104B}, 199 (1981).
  %doi:10.1016/0370-2693(81)90590-6
  %%CITATION = doi:10.1016/0370-2693(81)90590-6;%%

%\cite{Kim:1979if}
\bibitem{Kim:1979if} 
  J.~E.~Kim,
  %``Weak Interaction Singlet and Strong CP Invariance,''
  Phys.\ Rev.\ Lett.\  {\bf 43}, 103 (1979).
  %doi:10.1103/PhysRevLett.43.103
  %%CITATION = doi:10.1103/PhysRevLett.43.103;%%

%\cite{Shifman:1979if}
\bibitem{Shifman:1979if} 
  M.~A.~Shifman, A.~I.~Vainshtein and V.~I.~Zakharov,
  %``Can Confinement Ensure Natural CP Invariance of Strong Interactions?,''
  Nucl.\ Phys.\ B {\bf 166}, 493 (1980).
  %doi:10.1016/0550-3213(80)90209-6
  %%CITATION = doi:10.1016/0550-3213(80)90209-6;%%

%\cite{Shin:1987xc}
\bibitem{Shin:1987xc} 
  M.~Shin,
  %``Light Neutrino Masses and Strong {CP} Problem,''
  Phys.\ Rev.\ Lett.\  {\bf 59}, 2515 (1987)
  Erratum: [Phys.\ Rev.\ Lett.\  {\bf 60}, 383 (1988)].
  %doi:10.1103/PhysRevLett.60.383, 10.1103/PhysRevLett.59.2515
  %%CITATION = doi:10.1103/PhysRevLett.60.383, 10.1103/PhysRevLett.59.2515;%%

%\cite{Ballesteros:2016euj}
\bibitem{Ballesteros:2016euj} 
  G.~Ballesteros, J.~Redondo, A.~Ringwald and C.~Tamarit,
  %``Unifying inflation with the axion, dark matter, baryogenesis and the seesaw mechanism,''
  Phys.\ Rev.\ Lett.\  {\bf 118}, no. 7, 071802 (2017)
  %doi:10.1103/PhysRevLett.118.071802
  [arXiv:1608.05414 [hep-ph]].
  %%CITATION = doi:10.1103/PhysRevLett.118.071802;%%

%\cite{Ballesteros:2016xej}
\bibitem{Ballesteros:2016xej} 
  G.~Ballesteros, J.~Redondo, A.~Ringwald and C.~Tamarit,
  %``Standard Model—axion—seesaw—Higgs portal inflation. Five problems of particle physics and cosmology solved in one stroke,''
  JCAP {\bf 1708}, no. 08, 001 (2017)
  %doi:10.1088/1475-7516/2017/08/001
  [arXiv:1610.01639 [hep-ph]].
  %%CITATION = doi:10.1088/1475-7516/2017/08/001;%%

%\cite{Peccei:1986pn}
\bibitem{Peccei:1986pn} 
  R.~D.~Peccei, T.~T.~Wu and T.~Yanagida,
  %``A Viable Axion Model,''
  Phys.\ Lett.\ B {\bf 172}, 435 (1986).
 % doi:10.1016/0370-2693(86)90284-4
  %%CITATION = doi:10.1016/0370-2693(86)90284-4;%%

%\cite{Krauss:1986wx}
\bibitem{Krauss:1986wx} 
  L.~M.~Krauss and F.~Wilczek,
  %``A Shortlived Axion Variant,''
  Phys.\ Lett.\ B {\bf 173}, 189 (1986).
  %doi:10.1016/0370-2693(86)90244-3
  %%CITATION = doi:10.1016/0370-2693(86)90244-3;%%
  
%\cite{Geng:1988nc}
\bibitem{Geng:1988nc} 
  C.~Q.~Geng and J.~N.~Ng,
  %``Flavor Connections and Neutrino Mass Hierarchy Invariant Invisible Axion Models Without Domain Wall Problem,''
  Phys.\ Rev.\ D {\bf 39}, 1449 (1989).
  %doi:10.1103/PhysRevD.39.1449
  %%CITATION = doi:10.1103/PhysRevD.39.1449;%%

%\cite{Hindmarsh:1997ac}
\bibitem{Hindmarsh:1997ac} 
  M.~Hindmarsh and P.~Moulatsiotis,
  %``Constraints on variant axion models,''
  Phys.\ Rev.\ D {\bf 56}, 8074 (1997)
  %doi:10.1103/PhysRevD.56.8074
  [hep-ph/9708281].
  %%CITATION = doi:10.1103/PhysRevD.56.8074;%%

%\cite{Ema:2016ops}
\bibitem{Ema:2016ops} 
  Y.~Ema, K.~Hamaguchi, T.~Moroi and K.~Nakayama,
  %``Flaxion: a minimal extension to solve puzzles in the standard model,''
  JHEP {\bf 1701}, 096 (2017)
  %doi:10.1007/JHEP01(2017)096
  [arXiv:1612.05492 [hep-ph]].
  %%CITATION = doi:10.1007/JHEP01(2017)096;%%

%\cite{Calibbi:2016hwq}
\bibitem{Calibbi:2016hwq} 
  L.~Calibbi, F.~Goertz, D.~Redigolo, R.~Ziegler and J.~Zupan,
  %``Minimal axion model from flavor,''
  Phys.\ Rev.\ D {\bf 95}, no. 9, 095009 (2017)
  %doi:10.1103/PhysRevD.95.095009
  [arXiv:1612.08040 [hep-ph]].
  %%CITATION = doi:10.1103/PhysRevD.95.095009;%%
  
  %\cite{Bjorkeroth:2018dzu}
\bibitem{Bjorkeroth:2018dzu} 
  F.~Bj\"orkeroth, E.~J.~Chun and S.~F.~King,
  %``Flavourful Axion Phenomenology,''
  JHEP {\bf 1808}, 117 (2018)
  %doi:10.1007/JHEP08(2018)117
  [arXiv:1806.00660 [hep-ph]].
  %%CITATION = doi:10.1007/JHEP08(2018)117;%%
  
  %\cite{Bjorkeroth:2018ipq}
\bibitem{Bjorkeroth:2018ipq} 
  F.~Bj\"orkeroth, L.~Di Luzio, F.~Mescia and E.~Nardi,
  %``$U(1)$ flavour symmetries as Peccei-Quinn symmetries,''
  JHEP {\bf 1902}, 133 (2019)
  %doi:10.1007/JHEP02(2019)133
  [arXiv:1811.09637 [hep-ph]].
  %%CITATION = doi:10.1007/JHEP02(2019)133;%%

%\cite{Viaux:2013hca}
\bibitem{Viaux:2013hca} 
  N.~Viaux, M.~Catelan, P.~B.~Stetson, G.~G.~Raffelt, J.~Redondo, A.~A.~R.~Valcarce and A.~Weiss,
  %``Particle-physics constraints from the globular cluster M5: Neutrino Dipole Moments,''
  Astron.\ Astrophys.\  {\bf 558}, A12 (2013)
  %doi:10.1051/0004-6361/201322004
  [arXiv:1308.4627 [astro-ph.SR]].
  %%CITATION = doi:10.1051/0004-6361/201322004;%%

%\cite{Viaux:2013lha}
\bibitem{Viaux:2013lha} 
  N.~Viaux, M.~Catelan, P.~B.~Stetson, G.~G.~Raffelt, J.~Redondo, A.~A.~R.~Valcarce and A.~Weiss,
  %``Neutrino and axion bounds from the globular cluster M5 (NGC 5904),''
  Phys.\ Rev.\ Lett.\  {\bf 111}, 231301 (2013)
  %doi:10.1103/PhysRevLett.111.231301
  [arXiv:1311.1669 [astro-ph.SR]].
  %%CITATION = doi:10.1103/PhysRevLett.111.231301;%%

%\cite{Ayala:2014pea}
\bibitem{Ayala:2014pea} 
  A.~Ayala, I.~Dom\'inguez, M.~Giannotti, A.~Mirizzi and O.~Straniero,
  %``Revisiting the bound on axion-photon coupling from Globular Clusters,''
  Phys.\ Rev.\ Lett.\  {\bf 113}, no. 19, 191302 (2014)
  %doi:10.1103/PhysRevLett.113.191302
  [arXiv:1406.6053 [astro-ph.SR]].
  %%CITATION = doi:10.1103/PhysRevLett.113.191302;%%

%\cite{Aver:2013wba}
\bibitem{Aver:2013wba}
  E.~Aver, K.~A.~Olive, R.~L.~Porter and E.~D.~Skillman,
  %``The primordial helium abundance from updated emissivities,''
  JCAP {\bf 1311} (2013) 017
  %doi:10.1088/1475-7516/2013/11/017
  [arXiv:1309.0047 [astro-ph.CO]].
  %%CITATION = doi:10.1088/1475-7516/2013/11/017;%%
  %77 citations counted in INSPIRE as of 27 Jan 2020

%\cite{Straniero:2015nvc}
\bibitem{Straniero:2015nvc}
  O.~Straniero, A.~Ayala, M.~Giannotti, A.~Mirizzi and I.~Dominguez,
  ``Axion-Photon Coupling: Astrophysical Constraints,''
  in 11th Patras Workshop on Axions, WIMPs and WISPs,
  Zaragoza, Spain, 22-26 June 2015,
  DESY-PROC-2015-02.
  %doi:10.3204/DESY-PROC-2015-02/straniero_oscar
  %%CITATION = doi:10.3204/DESY-PROC-2015-02/straniero_oscar;%%

%\cite{Straniero:2014}
\bibitem{Straniero:2014}
  O.~Straniero, S.~Cristallo and L.~Piersanti,
  %``Heavy elements in globular clusters: The role of asymptotic giant branch stars,""
  Astrophys.\ J.\  {\bf 785} (2014) 77
  %doi:10.1088/0004-637X/785/1/77

%\cite{Straniero:2005hc}
\bibitem{Straniero:2005hc}
  O.~Straniero, R.~Gallino and S.~Cristallo,
  %``s-Process in low-mass asymptotic giant branch stars,''
  Nucl.\ Phys.\ A {\bf 777} (2006) 311
  %doi:10.1016/j.nuclphysa.2005.01.011
  [astro-ph/0501405].
  %%CITATION = doi:10.1016/j.nuclphysa.2005.01.011;%%

%\cite{Dominguez:1999qr}
\bibitem{Dominguez:1999qr}
  I.~Dominguez, A.~Chieffi, M.~Limongi and O.~Straniero,
  %``Intermediate mass stars: updated models,''
  Astrophys.\ J.\  {\bf 524} (1999) 226
  %doi:10.1086/307787
  [astro-ph/9906030].
  %%CITATION = doi:10.1086/307787;%%

%\cite{Piersanti:2013rya}
\bibitem{Piersanti:2013rya}
  L.~Piersanti, S.~Cristallo and O.~Straniero,
  %``The effects of rotation on the s-process nucleosynthesys in Asymptotic Giant Branch stars,''
  Astrophys.\ J.\  {\bf 774} (2013) 98
  %doi:10.1088/0004-637X/774/2/98
  [arXiv:1307.2017 [astro-ph.SR]].
  %%CITATION = doi:10.1088/0004-637X/774/2/98;%%

%\cite{Bertolami:2014wua}
\bibitem{Bertolami:2014wua} 
  M.~M.~Miller Bertolami, B.~E.~Melendez, L.~G.~Althaus and J.~Isern,
  %``Revisiting the axion bounds from the Galactic white dwarf luminosity function,''
  JCAP {\bf 1410}, no. 10, 069 (2014)
  %doi:10.1088/1475-7516/2014/10/069
  [arXiv:1406.7712 [hep-ph]].
  %%CITATION = doi:10.1088/1475-7516/2014/10/069;%%

%\cite{Isern:2008nt}
\bibitem{Isern:2008nt}
  J.~Isern, E.~Garcia-Berro, S.~Torres and S.~Catalan,
  %``Axions and the cooling of white dwarf stars,''
  Astrophys.\ J.\  {\bf 682} (2008) L109
  %doi:10.1086/591042
  [arXiv:0806.2807 [astro-ph]].
  %%CITATION = doi:10.1086/591042;%%

%\cite{Rowell:2013vsa}
\bibitem{Rowell:2013vsa}
  N.~Rowell,
  %``The Star Formation History of the Solar Neighbourhood from the White Dwarf Luminosity Function,''
  Mon.\ Not.\ Roy.\ Astron.\ Soc.\  {\bf 434} (2013) 1549
  %doi:10.1093/mnras/stt1110
  [arXiv:1306.4195 [astro-ph.GA]].
  %%CITATION = doi:10.1093/mnras/stt1110;%%

%\cite{Bertolami:2014noa}
\bibitem{Bertolami:2014noa} 
  M.~M.~Miller Bertolami,
  %``Limits on the neutrino magnetic dipole moment from the luminosity function of hot white dwarfs,''
  Astron.\ Astrophys.\ {\bf 562}, A123 (2014)
  %doi:10.1051/0004-6361/201322641
  [arXiv:1407.1404 [hep-ph]].
  %%CITATION = doi:10.1051/0004-6361/201322641;%%

%\cite{Harris:2005gd}
\bibitem{Harris:2005gd}
  H.~C.~Harris {\it et al.},
  %``The white dwarf luminosity function from sdss imaging data,''
  Astron.\ J.\  {\bf 131} (2006) 571
  %doi:10.1086/497966
  [astro-ph/0510820].
  %%CITATION = doi:10.1086/497966;%%

%\cite{Krzesinski:2009}
\bibitem{Krzesinski:2009} 
  J.~Krzesinski, S.~J.~Kleinman, A.~Nitta, S.~H\"{u}gelmeyer, S.~Dreizler, J.~Liebert and H.~Harris,
  %``A hot white dwarf luminosity function from the Sloan Digital Sky Survey,''
  Astron.\ Astrophys.\ {\bf 508}, 339 (2009).

%\cite{Rowell:2011}
\bibitem{Rowell:2011}
  N.~Rowell and N.~C.~Hambly,
  %``White dwarfs in the SuperCOSMOS Sky Survey: the thin disc, thick disc and spheroid luminosity functions,''
  Mon.\ Not.\ Roy.\ Astron.\ Soc.\  {\bf 417} (2011) 93.

%\cite{Isern:1992gia}
\bibitem{Isern:1992gia} 
  J.~Isern, M.~Hernanz and E.~Garcia-Berro,
  %``Axion cooling of white dwarfs,''
  Astrophys.\ J.\  {\bf 392}, L23 (1992).
  %doi:10.1086/186416
  %%CITATION = doi:10.1086/186416;%%

%\cite{Corsico:2012ki}
\bibitem{Corsico:2012ki} 
  A.~H.~C\'{o}rsico, L.~G.~Althaus, M.~M.~M.~Bertolami, A.~D.~Romero, E.~Garc\'{i}a-Berro, J.~Isern and S.~O.~Kepler,
  %``The rate of cooling of the pulsating white dwarf star G117$-$B15A: a new asteroseismological inference of the axion mass,''
  Mon.\ Not.\ Roy.\ Astron.\ Soc.\  {\bf 424}, 2792 (2012)
  %doi:10.1111/j.1365-2966.2012.21401.x
  [arXiv:1205.6180 [astro-ph.SR]].
  %%CITATION = doi:10.1111/j.1365-2966.2012.21401.x;%%

%\cite{Corsico:2012sh}
\bibitem{Corsico:2012sh} 
  A.~H.~C\'{o}rsico, L.~G.~Althaus, A.~D.~Romero, A.~S.~Mukadam, E.~Garc\'{i}a-Berro, J.~Isern, S.~O.~Kepler and M.~A.~Corti,
  %``An independent limit on the axion mass from the variable white dwarf star R548,''
  JCAP {\bf 1212}, 010 (2012)
  %doi:10.1088/1475-7516/2012/12/010
  [arXiv:1211.3389 [astro-ph.SR]].
  %%CITATION = doi:10.1088/1475-7516/2012/12/010;%%

%\cite{Corsico:2014mpa}
\bibitem{Corsico:2014mpa} 
  A.~H.~C\'{o}rsico, L.~G.~Althaus, M.~M.~Miller Bertolami, S.~O.~Kepler and E.~Garc\'{i}a-Berro,
  %``Constraining the neutrino magnetic dipole moment from white dwarf pulsations,''
  JCAP {\bf 1408}, 054 (2014)
  %doi:10.1088/1475-7516/2014/08/054
  [arXiv:1406.6034 [astro-ph.SR]].
  %%CITATION = doi:10.1088/1475-7516/2014/08/054;%%

%\cite{Battich:2016htm}
\bibitem{Battich:2016htm} 
  T.~Battich, A.~H.~C\'{o}rsico, L.~G.~Althaus, M.~M.~Miller Bertolami and M.~M.~M.~Bertolami,
  %``First axion bounds from a pulsating helium-rich white dwarf star,''
  JCAP {\bf 1608}, no. 08, 062 (2016)
  %doi:10.1088/1475-7516/2016/08/062
  [arXiv:1605.07668 [astro-ph.SR]].
  %%CITATION = doi:10.1088/1475-7516/2016/08/062;%%

%\cite{Corsico:2016okh}
\bibitem{Corsico:2016okh} 
  A.~H.~C\'{o}rsico {\it et al.},
  %``An asteroseismic constraint on the mass of the axion from the period drift of the pulsating DA white dwarf star L19-2,''
  JCAP {\bf 1607}, no. 07, 036 (2016)
  %doi:10.1088/1475-7516/2016/07/036
  [arXiv:1605.06458 [astro-ph.SR]].
  %%CITATION = doi:10.1088/1475-7516/2016/07/036;%%

%\cite{Raffelt:1987yt}
\bibitem{Raffelt:1987yt} 
  G.~G.~Raffelt and D.~Seckel,
  %``Bounds on Exotic Particle Interactions from SN 1987a,''
  Phys.\ Rev.\ Lett.\  {\bf 60}, 1793 (1988).
  %doi:10.1103/PhysRevLett.60.1793
  %%CITATION = doi:10.1103/PhysRevLett.60.1793;%%

%\cite{Turner:1987by}
\bibitem{Turner:1987by} 
  M.~S.~Turner,
  %``Axions from SN 1987a,''
  Phys.\ Rev.\ Lett.\  {\bf 60}, 1797 (1988).
  %doi:10.1103/PhysRevLett.60.1797
  %%CITATION = doi:10.1103/PhysRevLett.60.1797;%%

%\cite{Mayle:1987as}
\bibitem{Mayle:1987as} 
  R.~Mayle, J.~R.~Wilson, J.~R.~Ellis, K.~A.~Olive, D.~N.~Schramm and G.~Steigman,
  %``Constraints on Axions from SN 1987a,''
  Phys.\ Lett.\ B {\bf 203}, 188 (1988).
  %doi:10.1016/0370-2693(88)91595-X
  %%CITATION = doi:10.1016/0370-2693(88)91595-X;%%

%\cite{Raffelt:1991pw}
\bibitem{Raffelt:1991pw} 
  G.~G.~Raffelt and D.~Seckel,
  %``Multiple scattering suppression of the bremsstrahlung emission of neutrinos and axions in supernovae,''
  Phys.\ Rev.\ Lett.\  {\bf 67}, 2605 (1991).
  %doi:10.1103/PhysRevLett.67.2605
  %%CITATION = doi:10.1103/PhysRevLett.67.2605;%%

%\cite{Raffelt:1993ix}
\bibitem{Raffelt:1993ix} 
  G.~G.~Raffelt and D.~Seckel,
  %``A selfconsistent approach to neutral current processes in supernova cores,''
  Phys.\ Rev.\ D {\bf 52}, 1780 (1995)
  %doi:10.1103/PhysRevD.52.1780
  [astro-ph/9312019].
  %%CITATION = doi:10.1103/PhysRevD.52.1780;%%

%\cite{Keil:1996ju}
\bibitem{Keil:1996ju} 
  W.~Keil, H.~T.~Janka, D.~N.~Schramm, G.~Sigl, M.~S.~Turner and J.~R.~Ellis,
  %``A Fresh look at axions and SN-1987A,''
  Phys.\ Rev.\ D {\bf 56}, 2419 (1997)
  %doi:10.1103/PhysRevD.56.2419
  [astro-ph/9612222].
  %%CITATION = doi:10.1103/PhysRevD.56.2419;%%

%\cite{Fischer:2016cyd}
\bibitem{Fischer:2016cyd} 
  T.~Fischer, S.~Chakraborty, M.~Giannotti, A.~Mirizzi, A.~Payez and A.~Ringwald,
  %``Probing axions with the neutrino signal from the next galactic supernova,''
  Phys.\ Rev.\ D {\bf 94}, no. 8, 085012 (2016)
  %doi:10.1103/PhysRevD.94.085012
  [arXiv:1605.08780 [astro-ph.HE]].
  %%CITATION = doi:10.1103/PhysRevD.94.085012;%%
  
%\cite{Janka:1995ir}
\bibitem{Janka:1995ir} 
  H.~T.~Janka, W.~Keil, G.~G.~Raffelt and D.~Seckel,
  %``Nucleon spin fluctuations and the supernova emission of neutrinos and axions,''
  Phys.\ Rev.\ Lett.\  {\bf 76}, 2621 (1996)
  %doi:10.1103/PhysRevLett.76.2621
  [astro-ph/9507023].
  %%CITATION = doi:10.1103/PhysRevLett.76.2621;%%
  
%\cite{Sigl:1995ac}
\bibitem{Sigl:1995ac} 
  G.~Sigl,
  %``Weak interactions in supernova cores and saturation of nucleon spin fluctuations,''
  Phys.\ Rev.\ Lett.\  {\bf 76}, 2625 (1996)
  %doi:10.1103/PhysRevLett.76.2625
  [astro-ph/9508046].
  %%CITATION = doi:10.1103/PhysRevLett.76.2625;%%

%\cite{Chang:2018rso}
\bibitem{Chang:2018rso} 
  J.~H.~Chang, R.~Essig and S.~D.~McDermott,
  %``Supernova 1987A Constraints on Sub-GeV Dark Sectors, Millicharged Particles, the QCD Axion, and an Axion-like Particle,''
  JHEP {\bf 1809}, 051 (2018)
  %doi:10.1007/JHEP09(2018)051
  [arXiv:1803.00993 [hep-ph]].
  %%CITATION = doi:10.1007/JHEP09(2018)051;%%

%\cite{Raffelt:2006cw}
\bibitem{Raffelt:2006cw} 
  G.~G.~Raffelt,
  %``Astrophysical axion bounds,''
  Lect.\ Notes Phys.\  {\bf 741}, 51 (2008)
  %doi:10.1007/978-3-540-73518-2_3
  [hep-ph/0611350].
  %%CITATION = doi:10.1007/978-3-540-73518-2_3;%%

%\cite{Carenza:2019pxu}
\bibitem{Carenza:2019pxu}
  P.~Carenza, T.~Fischer, M.~Giannotti, G.~Guo, G.~Mart\'{i}nez-Pinedo and A.~Mirizzi,
  %``Improved axion emissivity from a supernova via nucleon-nucleon bremsstrahlung,''
  JCAP {\bf 1910} (2019) no.10,  016
  %doi:10.1088/1475-7516/2019/10/016
  [arXiv:1906.11844 [hep-ph]].
  %%CITATION = doi:10.1088/1475-7516/2019/10/016;%%

%\cite{Keller:2012yr}
\bibitem{Keller:2012yr} 
  J.~Keller and A.~Sedrakian,
  %``Axions from cooling compact stars,''
  Nucl.\ Phys.\ A {\bf 897}, 62 (2013)
  %doi:10.1016/j.nuclphysa.2012.11.004
  [arXiv:1205.6940 [astro-ph.CO]].
  %%CITATION = doi:10.1016/j.nuclphysa.2012.11.004;%%

%\cite{Sedrakian:2015krq}
\bibitem{Sedrakian:2015krq} 
  A.~Sedrakian,
  %``Axion cooling of neutron stars,''
  Phys.\ Rev.\ D {\bf 93}, no. 6, 065044 (2016)
  %doi:10.1103/PhysRevD.93.065044
  [arXiv:1512.07828 [astro-ph.HE]].
  %%CITATION = doi:10.1103/PhysRevD.93.065044;%%

%\cite{Sedrakian:2018kdm}
\bibitem{Sedrakian:2018kdm} 
  A.~Sedrakian,
  %``Axion cooling of neutron stars. II. Beyond hadronic axions,''
  Phys.\ Rev.\ D {\bf 99}, no. 4, 043011 (2019)
  %doi:10.1103/PhysRevD.99.043011
  [arXiv:1810.00190 [astro-ph.HE]].

%\cite{Leinson:2014ioa}
\bibitem{Leinson:2014ioa} 
  L.~B.~Leinson,
  %``Axion mass limit from observations of the neutron star in Cassiopeia A,''
  JCAP {\bf 1408}, 031 (2014)
  %doi:10.1088/1475-7516/2014/08/031
  [arXiv:1405.6873 [hep-ph]].
  %%CITATION = doi:10.1088/1475-7516/2014/08/031;%%

%\cite{Hamaguchi:2018oqw}
\bibitem{Hamaguchi:2018oqw} 
  K.~Hamaguchi, N.~Nagata, K.~Yanagi and J.~Zheng,
  %``Limit on the Axion Decay Constant from the Cooling Neutron Star in Cassiopeia A,''
  Phys.\ Rev.\ D {\bf 98}, no. 10, 103015 (2018)
  %doi:10.1103/PhysRevD.98.103015
  [arXiv:1806.07151 [hep-ph]].
  %%CITATION = doi:10.1103/PhysRevD.98.103015;%%

%\cite{Beznogov:2018fda}
\bibitem{Beznogov:2018fda} 
  M.~V.~Beznogov, E.~Rrapaj, D.~Page and S.~Reddy,
  %``Constraints on Axion-like Particles and Nucleon Pairing in Dense Matter from the Hot Neutron Star in HESS J1731-347,''
  Phys.\ Rev.\ C {\bf 98}, no. 3, 035802 (2018)
  %doi:10.1103/PhysRevC.98.035802
  [arXiv:1806.07991 [astro-ph.HE]].
  %%CITATION = doi:10.1103/PhysRevC.98.035802;%%

%\cite{Elshamouty:2013nfa}
\bibitem{Elshamouty:2013nfa}
  K.~G.~Elshamouty, C.~O.~Heinke, G.~R.~Sivakoff, W.~C.~G.~Ho, P.~S.~Shternin, D.~G.~Yakovlev, D.~J.~Patnaude and L.~David,
  %``Measuring the Cooling of the Neutron Star in Cassiopeia A with all Chandra X-ray Observatory Detectors,''
  Astrophys.\ J.\  {\bf 777} (2013) 22
  %doi:10.1088/0004-637X/777/1/22
  [arXiv:1306.3387 [astro-ph.HE]].
  %%CITATION = doi:10.1088/0004-637X/777/1/22;%%

%\cite{Posselt:2013xva}
\bibitem{Posselt:2013xva}
  B.~Posselt, G.~G.~Pavlov, V.~Suleimanov and O.~Kargaltsev,
  %``New constraints on the cooling of the Central Compact Object in Cas A,''
  Astrophys.\ J.\  {\bf 779} (2013) 186
  %doi:10.1088/0004-637X/779/2/186
  [arXiv:1311.0888 [astro-ph.HE]].
  %%CITATION = doi:10.1088/0004-637X/779/2/186;%%

%\cite{Posselt:2018xaf}
\bibitem{Posselt:2018xaf}
  B.~Posselt and G.~G.~Pavlov,
  %``Upper limits on the rapid cooling of the Central Compact Object in Cas A,''
  Astrophys.\ J.\  {\bf 864} (2018) no.2,  135
  %doi:10.3847/1538-4357/aad7fc
  [arXiv:1808.00531 [astro-ph.HE]].
  %%CITATION = doi:10.3847/1538-4357/aad7fc;%%

%\cite{Vilenkin:1982ks}
\bibitem{Vilenkin:1982ks} 
  A.~Vilenkin and A.~E.~Everett,
  %``Cosmic Strings and Domain Walls in Models with Goldstone and PseudoGoldstone Bosons,''
  Phys.\ Rev.\ Lett.\  {\bf 48}, 1867 (1982).
  %doi:10.1103/PhysRevLett.48.1867
  %%CITATION = doi:10.1103/PhysRevLett.48.1867;%%

%\cite{Srednicki:1985xd}
\bibitem{Srednicki:1985xd} 
  M.~Srednicki,
  %``Axion Couplings to Matter. 1. CP Conserving Parts,''
  Nucl.\ Phys.\ B {\bf 260}, 689 (1985).
  %doi:10.1016/0550-3213(85)90054-9
  %%CITATION = doi:10.1016/0550-3213(85)90054-9;%%

%\cite{Chang:1998tb}
\bibitem{Chang:1998tb} 
  S.~Chang, C.~Hagmann and P.~Sikivie,
  %``Studies of the motion and decay of axion walls bounded by strings,''
  Phys.\ Rev.\ D {\bf 59}, 023505 (1999)
  %doi:10.1103/PhysRevD.59.023505
  [hep-ph/9807374].
  %%CITATION = doi:10.1103/PhysRevD.59.023505;%%

%\cite{Hiramatsu:2010yn}
\bibitem{Hiramatsu:2010yn} 
  T.~Hiramatsu, M.~Kawasaki and K.~Saikawa,
  %``Evolution of String-Wall Networks and Axionic Domain Wall Problem,''
  JCAP {\bf 1108}, 030 (2011)
  %doi:10.1088/1475-7516/2011/08/030
  [arXiv:1012.4558 [astro-ph.CO]].
  %%CITATION = doi:10.1088/1475-7516/2011/08/030;%%

%\cite{Hiramatsu:2012sc}
\bibitem{Hiramatsu:2012sc} 
  T.~Hiramatsu, M.~Kawasaki, K.~Saikawa and T.~Sekiguchi,
  %``Axion cosmology with long-lived domain walls,''
  JCAP {\bf 1301}, 001 (2013)
  %doi:10.1088/1475-7516/2013/01/001
  [arXiv:1207.3166 [hep-ph]].
  %%CITATION = doi:10.1088/1475-7516/2013/01/001;%%
  
  %\cite{Kawasaki:2014sqa}
\bibitem{Kawasaki:2014sqa} 
  M.~Kawasaki, K.~Saikawa and T.~Sekiguchi,
  %``Axion dark matter from topological defects,''
  Phys.\ Rev.\ D {\bf 91}, no. 6, 065014 (2015)
  %doi:10.1103/PhysRevD.91.065014
  [arXiv:1412.0789 [hep-ph]].
  %%CITATION = doi:10.1103/PhysRevD.91.065014;%%
  
  %\cite{Ringwald:2015dsf}
\bibitem{Ringwald:2015dsf} 
  A.~Ringwald and K.~Saikawa,
  %``Axion dark matter in the post-inflationary Peccei-Quinn symmetry breaking scenario,''
  Phys.\ Rev.\ D {\bf 93}, no. 8, 085031 (2016)
  Addendum: [Phys.\ Rev.\ D {\bf 94}, no. 4, 049908 (2016)]
  %doi:10.1103/PhysRevD.93.085031, 10.1103/PhysRevD.94.049908
  [arXiv:1512.06436 [hep-ph]].
  %%CITATION = doi:10.1103/PhysRevD.93.085031, 10.1103/PhysRevD.94.049908;%%

%\cite{Geng:1990dv}
\bibitem{Geng:1990dv} 
  C.~Q.~Geng and J.~N.~Ng,
  %``The Domain Wall Number In Various Invisible Axion Models,''
  Phys.\ Rev.\ D {\bf 41}, 3848 (1990).
  %doi:10.1103/PhysRevD.41.3848
  %%CITATION = doi:10.1103/PhysRevD.41.3848;%%

%\cite{Georgi:1986df}
\bibitem{Georgi:1986df} 
  H.~Georgi, D.~B.~Kaplan and L.~Randall,
  %``Manifesting the Invisible Axion at Low-energies,''
  Phys.\ Lett.\  {\bf 169B}, 73 (1986).
  %doi:10.1016/0370-2693(86)90688-X
  %%CITATION = doi:10.1016/0370-2693(86)90688-X;%%
  
%\cite{diCortona:2015ldu}
\bibitem{diCortona:2015ldu} 
  G.~Grilli di Cortona, E.~Hardy, J.~Pardo Vega and G.~Villadoro,
  %``The QCD axion, precisely,''
  JHEP {\bf 1601}, 034 (2016)
  %doi:10.1007/JHEP01(2016)034
  [arXiv:1511.02867 [hep-ph]].
  %%CITATION = doi:10.1007/JHEP01(2016)034;%%

%\cite{Gorghetto:2018ocs}
\bibitem{Gorghetto:2018ocs} 
  M.~Gorghetto and G.~Villadoro,
  %``Topological Susceptibility and QCD Axion Mass: QED and NNLO corrections,''
  JHEP {\bf 1903}, 033 (2019)
  %doi:10.1007/JHEP03(2019)033
  [arXiv:1812.01008 [hep-ph]].
  %%CITATION = doi:10.1007/JHEP03(2019)033;%%

%\cite{Borsanyi:2016ksw}
\bibitem{Borsanyi:2016ksw} 
  S.~Borsanyi {\it et al.},
  %``Calculation of the axion mass based on high-temperature lattice quantum chromodynamics,''
  Nature {\bf 539}, no. 7627, 69 (2016)
  %doi:10.1038/nature20115
  [arXiv:1606.07494 [hep-lat]].
  %%CITATION = doi:10.1038/nature20115;%%

%\cite{DiLuzio:2017ogq}
\bibitem{DiLuzio:2017ogq} 
  L.~Di Luzio, F.~Mescia, E.~Nardi, P.~Panci and R.~Ziegler,
  %``Astrophobic Axions,''
  Phys.\ Rev.\ Lett.\  {\bf 120}, no. 26, 261803 (2018)
  %doi:10.1103/PhysRevLett.120.261803
  [arXiv:1712.04940 [hep-ph]].
  %%CITATION = doi:10.1103/PhysRevLett.120.261803;%%

%\cite{Gelmini:1982zz}
\bibitem{Gelmini:1982zz} 
  G.~B.~Gelmini, S.~Nussinov and T.~Yanagida,
  %``Does Nature Like Nambu-Goldstone Bosons?,''
  Nucl.\ Phys.\ B {\bf 219}, 31 (1983).
  %doi:10.1016/0550-3213(83)90426-1
  %%CITATION = doi:10.1016/0550-3213(83)90426-1;%%

%\cite{Adler:2008zza}
\bibitem{Adler:2008zza} 
  S.~Adler {\it et al.} [E949 and E787 Collaborations],
  %``Measurement of the K+ --> pi+ nu nu branching ratio,''
  Phys.\ Rev.\ D {\bf 77}, 052003 (2008)
  %doi:10.1103/PhysRevD.77.052003
  [arXiv:0709.1000 [hep-ex]].
  %%CITATION = doi:10.1103/PhysRevD.77.052003;%%

%\cite{Froggatt:1978nt}
\bibitem{Froggatt:1978nt} 
  C.~D.~Froggatt and H.~B.~Nielsen,
  %``Hierarchy of Quark Masses, Cabibbo Angles and CP Violation,''
  Nucl.\ Phys.\ B {\bf 147}, 277 (1979).
  %doi:10.1016/0550-3213(79)90316-X
  %%CITATION = doi:10.1016/0550-3213(79)90316-X;%%

%\cite{Buchmuller:1998zf}
\bibitem{Buchmuller:1998zf} 
  W.~Buchmuller and T.~Yanagida,
  %``Quark lepton mass hierarchies and the baryon asymmetry,''
  Phys.\ Lett.\ B {\bf 445}, 399 (1999)
  %doi:10.1016/S0370-2693(98)01480-4
  [hep-ph/9810308].
  %%CITATION = doi:10.1016/S0370-2693(98)01480-4;%%

%\cite{Chiang:2015cba}
\bibitem{Chiang:2015cba} 
  C.~W.~Chiang, H.~Fukuda, M.~Takeuchi and T.~T.~Yanagida,
  %``Flavor-Changing Neutral-Current Decays in Top-Specific Variant Axion Model,''
  JHEP {\bf 1511}, 057 (2015)
  %doi:10.1007/JHEP11(2015)057
  [arXiv:1507.04354 [hep-ph]].
  %%CITATION = doi:10.1007/JHEP11(2015)057;%%

%\cite{Chiang:2017fjr}
\bibitem{Chiang:2017fjr} 
  C.~W.~Chiang, H.~Fukuda, M.~Takeuchi and T.~T.~Yanagida,
  %``Current Status of Top-Specific Variant Axion Model,''
  Phys.\ Rev.\ D {\bf 97}, no. 3, 035015 (2018)
  %doi:10.1103/PhysRevD.97.035015
  [arXiv:1711.02993 [hep-ph]].
  %%CITATION = doi:10.1103/PhysRevD.97.035015;%%

%\cite{Chiang:2018bnu}
\bibitem{Chiang:2018bnu} 
  C.~W.~Chiang, M.~Takeuchi, P.~Y.~Tseng and T.~T.~Yanagida,
  %``Muon $g-2$ and rare top decays in up-type specific variant axion models,''
  Phys.\ Rev.\ D {\bf 98}, no. 9, 095020 (2018)
  %doi:10.1103/PhysRevD.98.095020
  [arXiv:1807.00593 [hep-ph]].
  %%CITATION = doi:10.1103/PhysRevD.98.095020;%%

%\cite{Davis:1986xc}
\bibitem{Davis:1986xc} 
  R.~L.~Davis,
  %``Cosmic Axions from Cosmic Strings,''
  Phys.\ Lett.\ B {\bf 180}, 225 (1986).
  %doi:10.1016/0370-2693(86)90300-X
  %%CITATION = doi:10.1016/0370-2693(86)90300-X;%%

%\cite{Lyth:1991bb}
\bibitem{Lyth:1991bb} 
  D.~H.~Lyth,
  %``Estimates of the cosmological axion density,''
  Phys.\ Lett.\ B {\bf 275}, 279 (1992).
  %doi:10.1016/0370-2693(92)91590-6
  %%CITATION = doi:10.1016/0370-2693(92)91590-6;%%

%\cite{Hiramatsu:2012gg}
\bibitem{Hiramatsu:2012gg} 
  T.~Hiramatsu, M.~Kawasaki, K.~Saikawa and T.~Sekiguchi,
  %``Production of dark matter axions from collapse of string-wall systems,''
  Phys.\ Rev.\ D {\bf 85}, 105020 (2012)
  Erratum: [Phys.\ Rev.\ D {\bf 86}, 089902 (2012)]
  %doi:10.1103/PhysRevD.86.089902, 10.1103/PhysRevD.85.105020
  [arXiv:1202.5851 [hep-ph]].
  %%CITATION = doi:10.1103/PhysRevD.86.089902, 10.1103/PhysRevD.85.105020;%%

%\cite{Fleury:2015aca}
\bibitem{Fleury:2015aca} 
  L.~Fleury and G.~D.~Moore,
  %``Axion dark matter: strings and their cores,''
  JCAP {\bf 1601}, 004 (2016)
  %doi:10.1088/1475-7516/2016/01/004
  [arXiv:1509.00026 [hep-ph]].
  %%CITATION = doi:10.1088/1475-7516/2016/01/004;%%

%\cite{Klaer:2017ond}
\bibitem{Klaer:2017ond} 
  V.~B.~Klaer and G.~D.~Moore,
  %``The dark-matter axion mass,''
  JCAP {\bf 1711}, no. 11, 049 (2017)
  %doi:10.1088/1475-7516/2017/11/049
  [arXiv:1708.07521 [hep-ph]].
  %%CITATION = doi:10.1088/1475-7516/2017/11/049;%%

%\cite{Gorghetto:2018myk}
\bibitem{Gorghetto:2018myk} 
  M.~Gorghetto, E.~Hardy and G.~Villadoro,
  %``Axions from Strings: the Attractive Solution,''
  JHEP {\bf 1807}, 151 (2018)
  %doi:10.1007/JHEP07(2018)151
  [arXiv:1806.04677 [hep-ph]].
  %%CITATION = doi:10.1007/JHEP07(2018)151;%%

\bibitem{Gorghetto:2019patras}
  M.~Gorghetto, 
  ``Axions from Strings",
  Talk at the 15th Patras Workshop on Axions, WIMPs and WISPs,
  Albert-Ludwigs-University Freiburg, 4 June 2019.
  
\bibitem{Saikawa:2019patras}
  K.~Saikawa, 
  ``Production of dark matter axions from global strings",
  Talk at the 15th Patras Workshop on Axions, WIMPs and WISPs,
  Albert-Ludwigs-University Freiburg, 5 June 2019.

%\cite{Horns:2012jf}
\bibitem{Horns:2012jf} 
  D.~Horns, J.~Jaeckel, A.~Lindner, A.~Lobanov, J.~Redondo and A.~Ringwald,
  %``Searching for WISPy Cold Dark Matter with a Dish Antenna,''
  JCAP {\bf 1304}, 016 (2013)
  %doi:10.1088/1475-7516/2013/04/016
  [arXiv:1212.2970 [hep-ph]].
  %%CITATION = doi:10.1088/1475-7516/2013/04/016;%%

%\cite{Hochberg:2016ajh}
\bibitem{Hochberg:2016ajh} 
  Y.~Hochberg, T.~Lin and K.~M.~Zurek,
  %``Detecting Ultralight Bosonic Dark Matter via Absorption in Superconductors,''
  Phys.\ Rev.\ D {\bf 94}, no. 1, 015019 (2016)
  %doi:10.1103/PhysRevD.94.015019
  [arXiv:1604.06800 [hep-ph]].
  %%CITATION = doi:10.1103/PhysRevD.94.015019;%%

%\cite{Marsh:2018dlj}
\bibitem{Marsh:2018dlj}
  D.~J.~E.~Marsh, K.~C.~Fong, E.~W.~Lentz, L.~Smejkal and M.~N.~Ali,
  %``Proposal to Detect Dark Matter using Axionic Topological Antiferromagnets,''
  Phys.\ Rev.\ Lett.\  {\bf 123} (2019) no.12,  121601
  %doi:10.1103/PhysRevLett.123.121601
  [arXiv:1807.08810 [hep-ph]].
  %%CITATION = doi:10.1103/PhysRevLett.123.121601;%%

%\cite{Arvanitaki:2014dfa}
\bibitem{Arvanitaki:2014dfa} 
  A.~Arvanitaki and A.~A.~Geraci,
  %``Resonantly Detecting Axion-Mediated Forces with Nuclear Magnetic Resonance,''
  Phys.\ Rev.\ Lett.\  {\bf 113}, no. 16, 161801 (2014)
  %doi:10.1103/PhysRevLett.113.161801
  [arXiv:1403.1290 [hep-ph]].
  %%CITATION = doi:10.1103/PhysRevLett.113.161801;%%
  
%\cite{Moody:1984ba}
\bibitem{Moody:1984ba} 
  J.~E.~Moody and F.~Wilczek,
  %``New Macroscopic Forces?,''
  Phys.\ Rev.\ D {\bf 30}, 130 (1984).
  %doi:10.1103/PhysRevD.30.130
  %%CITATION = doi:10.1103/PhysRevD.30.130;%%

%\cite{Jaeckel:2018mbn}
\bibitem{Jaeckel:2018mbn} 
  J.~Jaeckel and L.~J.~Thormaehlen,
  %``Distinguishing Axion Models with IAXO,''
  JCAP {\bf 1903}, no. 03, 039 (2019)
  %doi:10.1088/1475-7516/2019/03/039
  [arXiv:1811.09278 [hep-ph]].
  %%CITATION = doi:10.1088/1475-7516/2019/03/039;%%
  
%\cite{Dafni:2018tvj}
\bibitem{Dafni:2018tvj} 
  T.~Dafni {\it et al.},
  %``Weighing the solar axion,''
  Phys.\ Rev.\ D {\bf 99}, no. 3, 035037 (2019)
  %doi:10.1103/PhysRevD.99.035037
  [arXiv:1811.09290 [hep-ph]].
  %%CITATION = doi:10.1103/PhysRevD.99.035037;%%

%\cite{Cowan}
\bibitem{Cowan}
  G.~Cowan,
  ``Statistics,'' in Review of Particle Physics,
  Phys.\ Rev.\ D {\bf 98} (2018) no.3,  030001.
  %doi:10.1103/PhysRevD.98.030001

%\cite{Stump:2001gu}
\bibitem{Stump:2001gu}
  D.~Stump, J.~Pumplin, R.~Brock, D.~Casey, J.~Huston, J.~Kalk, H.~L.~Lai and W.~K.~Tung,
  %``Uncertainties of predictions from parton distribution functions. 1. The Lagrange multiplier method,''
  Phys.\ Rev.\ D {\bf 65} (2001) 014012
  %doi:10.1103/PhysRevD.65.014012
  [hep-ph/0101051].
  %%CITATION = doi:10.1103/PhysRevD.65.014012;%%

%\cite{Fogli:2002pt}
\bibitem{Fogli:2002pt}
  G.~L.~Fogli, E.~Lisi, A.~Marrone, D.~Montanino and A.~Palazzo,
  %``Getting the most from the statistical analysis of solar neutrino oscillations,''
  Phys.\ Rev.\ D {\bf 66} (2002) 053010
  %doi:10.1103/PhysRevD.66.053010
  [hep-ph/0206162].
  %%CITATION = doi:10.1103/PhysRevD.66.053010;%%

%\cite{Aoki:2009ha}
\bibitem{Aoki:2009ha} 
  M.~Aoki, S.~Kanemura, K.~Tsumura and K.~Yagyu,
  %``Models of Yukawa interaction in the two Higgs doublet model, and their collider phenomenology,''
  Phys.\ Rev.\ D {\bf 80}, 015017 (2009)
  %doi:10.1103/PhysRevD.80.015017
  [arXiv:0902.4665 [hep-ph]].
  %%CITATION = doi:10.1103/PhysRevD.80.015017;%%

\end{thebibliography}
\end{document}